\documentclass[twocolumn,tighten,times]{aastex63}
\usepackage[T1]{fontenc}
\usepackage{times}
\usepackage{boondox-cal}
\usepackage{amsthm}
\usepackage{blindtext}
\usepackage{booktabs}
\usepackage{amsthm,amsmath,amssymb}
\usepackage{mathrsfs}
\usepackage{longtable} 

\newcommand\bhm{M_{\bullet}}

\defcitealias{Du2018a}{I}
\defcitealias{Brotherton2020}{II}

\graphicspath{{pic/}}

\begin{document}

\title{\large Monitoring AGNs with H$\beta$ Asymmetry. III. \\
Long-term Reverberation Mapping Results of 15 Palomar-Green Quasars}

\correspondingauthor{Michael S. Brotherton and Pu Du}
\email{mbrother@uwyo.edu, dupu@ihep.ac.cn}

\author{Dong-Wei Bao}
\affiliation{Key Laboratory for Particle Astrophysics, Institute of High Energy Physics, Chinese Academy of Sciences, 19B Yuquan Road, Beijing 100049, People's Republic of China}
\affiliation{School of Astronomy and Space Science, University of Chinese Academy of Sciences, 19A Yuquan Road, Beijing 100049, People’s Republic of China}

\author{Michael S. Brotherton}
\affiliation{Department of Physics and Astronomy, University of Wyoming, Laramie, WY 82071, USA}

\author{Pu Du}
\affiliation{Key Laboratory for Particle Astrophysics, Institute of High Energy Physics, Chinese Academy of Sciences, 19B Yuquan Road, Beijing 100049, People's Republic of China}

\author{Jacob N. McLane}
\affiliation{Department of Physics and Astronomy, University of Wyoming, Laramie, WY 82071, USA}

\author{T. E. Zastrocky}
\affiliation{Department of Physics and Astronomy, University of Wyoming, Laramie, WY 82071, USA}
\affiliation{Physics and Astronomy Department, Regis University, Denver, CO 80212, USA}

\author{Kianna A. Olson}
\affiliation{Department of Physics and Astronomy, University of Wyoming, Laramie, WY 82071, USA}

\author{Feng-Na Fang}
\affiliation{Key Laboratory for Particle Astrophysics, Institute of High Energy Physics, Chinese Academy of Sciences, 19B Yuquan Road, Beijing 100049, People's Republic of China}
\affiliation{School of Astronomy and Space Science, University of Chinese Academy of Sciences, 19A Yuquan Road, Beijing 100049, People’s Republic of China}

\author{Shuo Zhai}
\affiliation{Key Laboratory for Particle Astrophysics, Institute of High Energy Physics, Chinese Academy of Sciences, 19B Yuquan Road, Beijing 100049, People's Republic of China}
\affiliation{School of Astronomy and Space Science, University of Chinese Academy of Sciences, 19A Yuquan Road, Beijing 100049, People’s Republic of China}

\author{Zheng-Peng Huang}
\affiliation{Key Laboratory for Particle Astrophysics, Institute of High Energy Physics, Chinese Academy of Sciences, 19B Yuquan Road, Beijing 100049, People's Republic of China}

\author{Kai Wang}
\affiliation{Key Laboratory for Particle Astrophysics, Institute of High Energy Physics, Chinese Academy of Sciences, 19B Yuquan Road, Beijing 100049, People's Republic of China}

\author{Bi-Xuan Zhao}
\affiliation{Shanghai Observatory, Chinese Academy of Sciences, 80 Nandan Road, Shanghai 200030, People's Republic of China}

\author{Sha-Sha Li}
\affiliation{Yunnan Observatories, Chinese Academy of Sciences, Kunming 650011, People's Republic of China}

\author{Sen Yang}
\affiliation{Key Laboratory for Particle Astrophysics, Institute of High Energy Physics, Chinese Academy of Sciences, 19B Yuquan Road, Beijing 100049, People's Republic of China}
\affiliation{School of Astronomy and Space Science, University of Chinese Academy of Sciences, 19A Yuquan Road, Beijing 100049, People’s Republic of China}

\author{Yong-Jie Chen}
\affiliation{Key Laboratory for Particle Astrophysics, Institute of High Energy Physics, Chinese Academy of Sciences, 19B Yuquan Road, Beijing 100049, People's Republic of China}
\affiliation{School of Astronomy and Space Science, University of Chinese Academy of Sciences, 19A Yuquan Road, Beijing 100049, People’s Republic of China}

\author{Jun-Rong Liu}
\affiliation{Key Laboratory for Particle Astrophysics, Institute of High Energy Physics, Chinese Academy of Sciences, 19B Yuquan Road, Beijing 100049, People's Republic of China}
\affiliation{School of Astronomy and Space Science, University of Chinese Academy of Sciences, 19A Yuquan Road, Beijing 100049, People’s Republic of China}

\author{Zhu-Heng Yao}
\affiliation{Key Laboratory for Particle Astrophysics, Institute of High Energy Physics, Chinese Academy of Sciences, 19B Yuquan Road, Beijing 100049, People's Republic of China}
\affiliation{School of Astronomy and Space Science, University of Chinese Academy of Sciences, 19A Yuquan Road, Beijing 100049, People’s Republic of China}

\author{Yue-Chang Peng}
\affiliation{Key Laboratory for Particle Astrophysics, Institute of High Energy Physics, Chinese Academy of Sciences, 19B Yuquan Road, Beijing 100049, People's Republic of China}
\affiliation{School of Astronomy and Space Science, University of Chinese Academy of Sciences, 19A Yuquan Road, Beijing 100049, People’s Republic of China}

\author{Wei-Jian Guo}
\affiliation{Key Laboratory for Particle Astrophysics, Institute of High Energy Physics, Chinese Academy of Sciences, 19B Yuquan Road, Beijing 100049, People's Republic of China}
\affiliation{School of Astronomy and Space Science, University of Chinese Academy of Sciences, 19A Yuquan Road, Beijing 100049, People’s Republic of China}

\author{Yu-Yang Songsheng}
\affiliation{Key Laboratory for Particle Astrophysics, Institute of High Energy Physics, Chinese Academy of Sciences, 19B Yuquan Road, Beijing 100049, People's Republic of China}
\affiliation{School of Astronomy and Space Science, University of Chinese Academy of Sciences, 19A Yuquan Road, Beijing 100049, People’s Republic of China}

\author{Yan-Rong Li}
\affiliation{Key Laboratory for Particle Astrophysics, Institute of High Energy Physics, Chinese Academy of Sciences, 19B Yuquan Road, Beijing 100049, People's Republic of China}

\author{Bo-Wei jiang}
\affiliation{Key Laboratory for Particle Astrophysics, Institute of High Energy Physics, Chinese Academy of Sciences, 19B Yuquan Road, Beijing 100049, People's Republic of China}
\affiliation{School of Astronomy and Space Science, University of Chinese Academy of Sciences, 19A Yuquan Road, Beijing 100049, People’s Republic of China}

\author{David H. Kasper}
\affiliation{Department of Physics and Astronomy, University of Wyoming, Laramie, WY 82071, USA}

\author{William T. Chick}
\affiliation{Department of Physics and Astronomy, University of Wyoming, Laramie, WY 82071, USA}

\author{My L. Nguyen}
\affiliation{Department of Physics and Astronomy, University of Wyoming, Laramie, WY 82071, USA}

\author{Jaya Maithil}
\affiliation{Department of Physics and Astronomy, University of Wyoming, Laramie, WY 82071, USA}

\author{H. A. Kobulnicky}
\affiliation{Department of Physics and Astronomy, University of Wyoming, Laramie, WY 82071, USA}

\author{D. A. Dale}
\affiliation{Department of Physics and Astronomy, University of Wyoming, Laramie, WY 82071, USA}

\author{Derek Hand}
\affiliation{Department of Physics and Astronomy, University of Wyoming, Laramie, WY 82071, USA}

\author{C. Adelman}
\affiliation{Department of Physics and Astronomy, University of Wyoming, Laramie, WY 82071, USA}
\affiliation{Department of Physics \& Astronomy, Cal Poly Pomona, Pomona, CA 91768, USA}

\author{Z. Carter}
\affiliation{Department of Physics and Astronomy, University of Wyoming, Laramie, WY 82071, USA}
\affiliation{Department of Physics and Astronomy, Trinity University, San Antonio, TX 78212, USA}

\author{A. M. Murphree}
\affiliation{Department of Physics and Astronomy, University of Wyoming, Laramie, WY 82071, USA}
\affiliation{Department of Physics, Rhodes College, Memphis, TN 38112, USA}

\author{M. Oeur}
\affiliation{Department of Physics and Astronomy, University of Wyoming, Laramie, WY 82071, USA}
\affiliation{Department of Physics and Astronomy, State Long Beach, Long Beach, CA 90840, USA}

\author{S. Schonsberg}
\affiliation{Department of Physics and Astronomy, University of Wyoming, Laramie, WY 82071, USA}
\affiliation{Department of Physics and Astronomy, University of Montana, Missoula, MT 59812, USA}

\author{T. Roth} 
\affiliation{Department of Physics and Astronomy, University of Wyoming, Laramie, WY 82071, USA}
\affiliation{Department of Physics \& Astronomy, California State University, Sacramento, CA 95747, USA}

\author{Hartmut Winkler}
\affiliation{Department of Physics, University of Johannesburg, P.O. Box 524, 2006 Auckland Park, South Africa}

\author{Paola Marziani}
\affiliation{Istituto Nazionale di Astrofisica (INAF), Osservatorio Astronomico di Padova, I-35122 Padova, Italy}

\author[0000-0001-6441-9044]{Mauro D'Onofrio} 
\affiliation{Dipartimento di Fisica \& Astronomia ``Galileo Galilei'', Universit\`{a} di Padova, Padova, Italy}

\author{Chen Hu}
\affiliation{Key Laboratory for Particle Astrophysics, Institute of High Energy Physics, Chinese Academy of Sciences, 19B Yuquan Road, Beijing 100049, People's Republic of China}

\author{Ming Xiao}
\affiliation{Key Laboratory for Particle Astrophysics, Institute of High Energy Physics, Chinese Academy of Sciences, 19B Yuquan Road, Beijing 100049, People's Republic of China}

\author{Suijian Xue}
\affiliation{Key Laboratory of Optical Astronomy, National Astronomical Observatories, Chinese Academy of Sciences, Beijing 100012, People’s Republic of China}

\author{Bo{\.z}ena Czerny}
\affiliation{Center for Theoretical Physics, Polish Academy of Sciences, Al. Lotnikow 32/46, 02-668 Warsaw, Poland}

\author{Jes\'us Aceituno}
\affil{Centro Astronomico Hispano Alem\'an, Sierra de los filabres sn, 04550 gergal.  Almer\'ia, Spain}
\affil{Instituto de Astrof\'isica de Andaluc\'ia (CSIC), Glorieta de la astronom\'ia sn, 18008 Granada, Spain}

\author{Luis C. Ho}
\affiliation{Kavli Institute for Astronomy and Astrophysics, Peking University, Beijing 100871, People's Republic of China}
\affiliation{Department of Astronomy, School of Physics, Peking University, Beijing 100871, People's Republic of China}

\author{Jin-Ming Bai}
\affiliation{Yunnan Observatories, Chinese Academy of Sciences, Kunming 650011, People's Republic of China}

\author{Jian-Min Wang}\altaffiliation{PI of the MAHA Project}
\affiliation{Key Laboratory for Particle Astrophysics, Institute of High Energy Physics, Chinese Academy of Sciences, 19B Yuquan Road, Beijing 100049, People's Republic of China}
\affiliation{National Astronomical Observatories of China, Chinese Academy of Sciences, 20A Datun Road, Beijing 100020, People’s Republic of China}
\affiliation{School of Astronomy and Space Science, University of Chinese Academy of Sciences, 19A Yuquan Road, Beijing 100049, People’s Republic of China}

\collaboration{99}{(MAHA collaboration)}

\begin{abstract}

In this third paper of the series reporting on the reverberation mapping (RM)
campaign of active galactic nuclei with asymmetric H$\beta$ emission-line
profiles, we present results for 15 Palomar-Green (PG) quasars using spectra
obtained between the end of 2016 to May 2021. This campaign combines long time
spans with relatively high cadence. For 8 objects, both the time lags obtained
from the entire light curves and the measurements from individual observing
seasons are provided. Reverberation mapping of 9 of our targets has been
attempted for the first time, while the results for 6 others can be compared
with previous campaigns. We measure the H$\beta$ time lags over periods of years
and estimate their black hole masses. The long duration of the campaign enables
us to investigate their broad line region (BLR) geometry and kinematics for
different years by using velocity-resolved lags, which demonstrate signatures of
diverse BLR geometry and kinematics. The BLR geometry and kinematics of
individual objects are discussed. In this sample, the BLR kinematics of
Keplerian/virialized motion and inflow is more common than outflow. 

\end{abstract}


\section{INTRODUCTION}
\label{sec:intro}

The broad emission lines of active galactic nuclei (AGNs), are the primary
features in their UV/optical spectra, and arise from the photoionization of gas
in the broad-line regions (BLRs) by the continuum emission from the accretion
disks around the central supermassive black holes (SMBHs). Although the profiles
of the broad Balmer emission lines (e.g, H$\alpha$, H$\beta$, H$\gamma$) in AGNs
are sometimes well approximated by Gaussian or Lorentzian functions, a fraction
of them are more complex and possess significant asymmetries (redward, blueward,
or double-peaked) sometimes with systematic velocity shifts of their peaks
\citep[e.g.,][]{DeRobertis1985, Sulentic1989, Marziani2003atlas, eracleous2012}.
The physical origin of the profile asymmetries of broad emission lines is far
from fully understood, but it is likely that the asymmetries are connected with
the kinematics of BLRs or opacity effects.

In past decades, observational studies often focused on emission-line profiles
and their correlations with other AGN properties. For example,
\cite{Boroson1992} discovered that the H$\beta$ profile tends to be red
asymmetric if the Fe {\sc ii} emissions are weak and the [O {\sc iii}] lines are
strong (the main variations in the so-called Eigenvector 1). \cite{Marziani2003}
divided a sample of AGNs into several bins with different black hole (BH) masses
and Eddington ratios, and investigated the systematic properties of the median
profiles of broad H$\beta$ in each bin, showing that redward asymmetries are
observed at low Eddington ratio. \cite{netzer2007} studied AGNs from the Sloan
Digital Sky Survey (SDSS), and found that the fractional flux of the red part of
the H$\beta$ line shows a positive correlation with luminosity and a negative
correlation with the flux ratio of Fe {\sc ii}/H$\beta$. \cite{hu2008a}
discovered that the H$\beta$ line shows a more significantly red asymmetry if
the Fe {\sc ii} emission lines have stronger redshifted velocities. 

Additionally, theoretical efforts were made to understand the diversity of
emission-line profiles. \cite{Capriotti1979} proposed that the line asymmetries
could be attributed to optically thick inflowing or outflowing BLR clouds.
\cite{Ferland1979} calculated  asymmetric profiles from an expanding BLR by
taking into account Balmer self-absorption of optically thick clouds.
\cite{chen1989} and \cite{Chen1989b} found that a relativistic Keplerian disk
can explain the observed asymmetric and double-peaked profile observed in
Arp~102B. \cite{eracleous1995} suggested that a relativistic eccentric disk
could account for observed asymmetries. \cite{storchi-begmann2003} used the
spiral arms in a disk to explain the H$\alpha$ line-profile variations of NGC
1097. \cite{Wang2017} suggested that the BLR could be formed through tidal
disruption of clumps from a dusty torus, showing asymmetric profiles due to the
infall of the captured gas. Asymmetries of profiles generated by this model are
generally consistent with profiles of Palomar-Green quasars. Furthermore,
supermassive binary black holes (SMBBHs) were also recently proposed to explain
double-peaked profiles \cite[e.g.,][]{shen2010, bon2012, Li2016Binary, ji2021}. 

The reverberation mapping (RM) technique \citep{Blandford1982, Peterson1993} is
a powerful tool to investigate the geometry and kinematics of BLRs, by
monitoring the delayed response of the broad emission lines with respect to the
continuum variation, and has been carried out for more than a hundred AGNs over
the past several decades. Before 2000, investigations focused on, e.g.,  
bright but heterogeneous samples of Seyfert 1 galaxies \citep{Peterson1998},
Palomar-Green (PG) quasars \citep{Kaspi2000}, or intensive studies of some
individual objects \citep[e.g., International AGN Watch project,
see][]{Clavel1991, Peterson2002}. These efforts established a general
understanding of the RM properties of AGNs. Since 2000, significant progress has
been made by dedicated RM projects with different goals. For example, the Lick
AGN Monitoring Project \citep[LAMP, see, e.g., ][]{Bentz2008, Barth2015, U2021}
resolved the BLR kinematics of some local Seyfert galaxies. The super-Eddington
accreting massive black holes (SEAMBHs) project \citep[e.g.,][]{Du2014, Du2015,
du2016V, Du2018b} focuses on the AGNs with the highest accretion rates and found
shortened time lags compared to other objects of similar luminosity.
Industrial-scale RM campaigns like the Sloan Digital Sky Survey RM project
\citep[SDSS-RM, e.g., ][]{Shen2016, Grier2017a} and the Australian Dark Energy
Survey (OzDES) RM program \citep{Yu2021} use fiber-fed instruments and can
obtain the time delays of multiple objects in the field-of-view simultaneously.
\cite{Barth2013} and \cite{Hu2015} measured the time lags of Fe {\sc ii} lines.  
\cite{Rafter2011} and \cite{Woo2019} monitored intermediate-mass black holes,
while \cite{Rakshit2019} and \cite{Li2021} observed luminous nearby quasars
(e.g., 5100\AA\, luminosity  $\gtrsim 10^{45}\,{\rm erg\,s^{-1}}$). Some
long-term projects aim to measure C {\sc iv} or C {\sc iii}] emission lines in
high-redshift quasars in a time span of decades \citep{Kaspi2017, Kaspi2021,
Lira2018}. There are also many recent campaigns for small samples of (or
individual) interesting AGNs \citep[e.g.,][]{Denney2009, Grier2012, Lu2016,
fausnaugh2017, DeRosa2018, Zhang2019, Czerny2019, Hu2020a, Zajacek2020,
Zajacek2021, oknyansky2021}.

\begin{deluxetable*}{ccccccccc}
    \tablecaption{Basic Information of the 15 PG Targets\label{tab:basic_info}}
    \footnotesize
    \tablewidth{0pt}
    \tablehead{ \colhead{Name} & \colhead{Other Names} &  \colhead{RA} &
    \colhead{Dec} & \colhead{$z$}  & \colhead{$A$} & \colhead{$A$ (Boroson+92)}  &
    \colhead{Observatories} & \colhead{Previous RM} }
    \startdata
    PG 0007$+$106 & Mrk 1501, III Zw 2 & 00:10:31.0 & $+$10:58:29 & 0.0872 & $-$0.022 &  $-$0.046 & WIRO, Asiago, SAAO     & (1)       \\
    PG 0049$+$171 & Mrk 1148           & 00:51:54.7 & $+$17:25:59 & 0.0645 & $-$0.063 &  $-$0.047 & WIRO                   &           \\
    PG 0923$+$129 & Mrk 705, Ark 202   & 09:26:03.3 & $+$12:44:03 & 0.0288 & $-$0.072 &  $-$0.031 & WIRO                   &           \\
    PG 0947$+$396 &                    & 09:50:48.4 & $+$39:26:51 & 0.2055 & $-$0.116 &  $-$0.148 & WIRO                   &           \\
    PG 1001$+$054 &                    & 10:04:20.1 & $+$05:13:00 & 0.1601 & $+$0.065 &  $+$0.082 & Lijiang                &           \\
    PG 1048$+$342 &                    & 10:51:43.8 & $+$33:59:27 & 0.1671 & $-$0.226 &  $+$0.045 & WIRO                   & (2)$^*$   \\
    PG 1100$+$772 & 3C 249.1           & 11:04:13.6 & $+$76:58:58 & 0.3115 & $-$0.106 &  $-$0.097 & WIRO \& Asiago         &           \\
    PG 1202$+$281 & GQ Com             & 12:04:42.1 & $+$27:54:12 & 0.1650 & $-$0.095 &  $-$0.298 & WIRO \& Asiago         &           \\
    PG 1211$+$143 &                    & 12:14:17.6 & $+$14:03:13 & 0.0809 & $+$0.039 &  $-$0.003 & Lijiang \& CAHA        & (2)       \\
    PG 1310$-$108 &                    & 13:13:05.7 & $-$11:07:42 & 0.0343 & $-$0.112 &  $-$0.075 & WIRO                   &           \\
    PG 1351$+$640 &                    & 13:53:15.8 & $+$63:45:46 & 0.0882 & $-$0.139 &  $+$0.136 & WIRO                   & (2)$^*$   \\
    PG 1351$+$695 & Mrk 279            & 13:53:03.4 & $+$69:18:29 & 0.0305 & $-$0.043 &           & WIRO                   & (3,4,5)   \\
    PG 1501$+$106 & Mrk 841            & 15:04:01.2 & $+$10:26:16 & 0.0364 & $-$0.071 &  $-$0.039 & WIRO, Asiago, SAAO     & (6)       \\
    PG 1534$+$580 & Mrk 290            & 15:35:52.3 & $+$57:54:09 & 0.0302 & $-$0.109 &  $+$0.044 & WIRO                   & (7)       \\
    PG 1613$+$658 & Mrk 876            & 16:13:57.1 & $+$65:43:10 & 0.1211 & $-$0.183 &  $-$0.207 & WIRO                   & (2,8)
    \enddata
    \tablecomments{$A$ is a dimensionless parameter to describe the asymmetry, which
    is measured from our campaign (see Section \ref{subsec:Targets}). $A$
    (Boroson+92) is the asymmetry parameter listed in \cite{Boroson1992}.
    References: (1) \cite{Grier2012}, (2) \cite{Kaspi2000}, (3) \cite{Maoz1990}, (4)
    \cite{Santos-Lleo2001}, (5) \cite{Barth2015}, (6) \cite{Brotherton2020}, (7)
    \cite{denney2010}, (8) \cite{Minezaki2019}. $^*$ means that the previous RM
    campaign did not successfully measure the time lag of H$\beta$.}
    \end{deluxetable*}

To understand the kinematics associated with the asymmetries of emission-line
profiles, and to explore the evolution of BLR gas, we initiated a dedicated RM
campaign in 2016 named the ``Monitoring AGNs with H$\beta$ Asymmetry'' (MAHA)
project. We focus on AGNs with asymmetric (or double-peaked) emission lines,
which are more likely connected with complicated BLR geometry or kinematics.
Another goal of the MAHA project is to search for SMBBH candidates from transfer
functions (also called ``velocity-delay maps'') produced by RM \citep{wang2018,
songsheng2020,Kovacevic2020}. 

We have previously published the RM results of seven Seyfert galaxies observed
from the end of 2016 to May 2017 \cite{Du2018a} (hereafter Paper
\citetalias{Du2018a}) and \cite{Brotherton2020} (hereafter Paper
\citetalias{Brotherton2020}). Some of the objects show very complicated
signatures in their velocity-resolved lags (e.g., Ark~120 and Mrk~6) or
velocity-delay maps (e.g., Mrk~79), which are difficult to interpret as simple
inflow, outflow, or virialized motions (see Papers \citetalias{Du2018a} and
\citetalias{Brotherton2020}). The discovery of the diverse BLR kinematics in
Seyfert galaxies with asymmetric line profiles (Papers \citetalias{Du2018a} and
\citetalias{Brotherton2020}) motivates us to consider whether the BLR geometry
and kinematics are also complex in more luminous quasars with asymmetric
H$\beta$.

The Palomar-Green (PG) sample of objects with ultraviolet excesses
\citep{Schmidt1983, Boroson1992} includes subsamples of quasars that have been
extensively studied in almost all wavelengths of the electromagnetic spectrum
and some have already been spectroscopically monitored for RM
\citep[e.g.,][]{Kaspi2000, Bentz2009, Grier2012, Barth2015, Zhang2019, Hu2020a,
Hu2020b}. The asymmetries of their emission-line profiles have been investigated
using single-epoch spectra \citep{Boroson1992, Marziani2003}, but not
systematically in the time domain. It is valuable to investigate the geometry
and kinematics of their BLRs for the PG quasars with significantly asymmetric
emission lines by the velocity-resolved lags \citep[e.g.,][]{Bentz2009,
denney2010, Du2016VI} or velocity-delay maps \citep[e.g.,][]{Grier2013,
Xiao2018ApJ, Horne2021}. As the third paper of the MAHA series, we report here
the RM observations of 15 PG quasars, most with significantly asymmetric
H$\beta$ emission lines. 

The paper is organized as follows. The target selection and the observations are
given in Section \ref{sec:observation}. The analyses are provided in Section
\ref{sec:analysis}, including the mean and root-mean-square (rms) spectra, the
light curves, the line widths, the time lag measurements, the black hole masses,
and the velocity-resolved time lags. The discussion of individual objects is in
Section \ref{sec:discussion}. Finally, in Section \ref{sec:summary}, we briefly
summarize the present paper.

\section{OBSERVATIONS}
\label{sec:observation}

\subsection{Targets}
\label{subsec:Targets}

The primary goal of the MAHA project is to monitor the AGNs showing current or
historical asymmetric emission-line profiles in order to investigate BLR
geometry and kinematics, their evolution, and the possible presence of SMBBHs.
\cite{Boroson1992} adopted the line asymmetry parameter 
\begin{equation}
A = \frac{[\lambda_{\rm c} (3/4) - \lambda_{\rm c} (1/4)]}{\Delta\lambda(1/2)}, 
\end{equation} 
defined by \cite{DeRobertis1985}, and measured the asymmetries of the H$\beta$
emission lines in PG quasars, where $\lambda_{\rm c} (3/4)$ and $\lambda_{\rm c}
(1/4)$ are the central wavelengths where the profile is 3/4 and 1/4 of the peak
value respectively, and $\Delta\lambda(1/2)$ is the FWHM of emission line. $A<0$
indicates that the emission line has a profile with a more pronounced red wing,
while $A>0$ means the line has a stronger blue wing (see Figure 1 in Paper
\citetalias{Du2018a}). \cite{Boroson1992} demonstrated that the $A$ parameter is
positively correlated with the relative strength of Fe {\sc ii} with respect to
H$\beta$ in PG quasars and some of them have strong asymmetries with $A \lesssim
-0.1$ or $A \gtrsim 0.1$. Based on the asymmetry measurements of
\cite{Boroson1992}, we selected 5 PG quasars (PG~0947+396, PG~1100+772,
PG~1202+281, PG~1310$-$108, and PG~1613+658) with significant red asymmetries
($A \approx -0.08$ -- $-0.3$) and 4 PG quasars (PG~1001+054, PG~1048+342,
PG~1351+640, and PG~1534+580) with moderate to significant blue asymmetries ($A
\approx 0.05$ -- $0.15$) as our MAHA targets from the PG sample in
\cite{Boroson1992}. It is intriguing that the H$\beta$ profile of PG~1048+342
has changed to red asymmetry in our observations (see its $A$ parameter
measurements from our campaign in Table \ref{tab:basic_info}). 

We also selected an additional 6 PG quasars as  RM targets: PG~0007+106,
PG~0049+171, PG~0923+129, PG~1211+143, PG~1351+695, and PG~1501+106. The
radio-emission variability of PG~0007+106 demonstrates
quasi-periodicity/periodicity (with a period of $\sim5$ years, see
\citealt{Terasranta2005}, \citealt{Li2010}) which is potentially caused by jet
precession. SMBBHs are a possible cause of jet precession \citep{begelman1980,
romero2000}, thus we chose this object as our target. The line profile of
PG~1211+143 was almost symmetric in \cite{Boroson1992}, but showed mild blue
asymmetry recently (see Table \ref{tab:basic_info}). PG~1351+695 displayed
significant blue asymmetry in 2011 \citep{Barth2015, williams2018}. PG~1501+106
showed weak red asymmetry in \cite{Boroson1992}; however, this asymmetry became
stronger in 2017-2020 (see Table \ref{tab:basic_info}). The H$\beta$
emission-line profiles of PG~0049+171 and PG~0923+129 were only weakly
asymmetric \citep{Boroson1992}, but we included them in our target list as they
fit well into our program schedule (showing stronger red asymmetry in our
campaign). Table \ref{tab:basic_info} provides for each target the coordinates,
redshifts, asymmetries measured in our campaign (from an individual exposure
with high S/N ratio) and from \cite{Boroson1992}, and the specific telescopes
used. The mean spectra of our targets are displayed in Figure
\ref{fig:mean_spec}.

\begin{deluxetable}{ccc}
    \tablecaption{RA and DEC of the comparison stars for spectroscopy of two Lijiang targets\label{tab:comparison stars}}
    \setlength{\tabcolsep}{20pt}
    \tablehead{
    \colhead{Target} &  \colhead{RA$_{\rm comp}$}   & \colhead{Dec$_{\rm comp}$} 
    }
    \startdata
    PG~1001+054 & 10:04:24 & +05:15:29 \\
    PG~1211+143 & 12:13:59 & +14:05:16 
    \enddata
    \end{deluxetable}

\subsection{Spectroscopy}
\label{subsec:Spectroscopy}

The spectroscopic observations were carried out using the 2.3 m telescope of
Wyoming Infrared Observatory (WIRO) in the United States, the Lijiang 2.4 m
telescope of the Yunnan Observatories of the Chinese Academy of Sciences in
China, the 2.2 m telescope of Calar Alto Astronomical Observatory of Centro
Astron$\acute{\rm o}$mico Hispano-Alem$\acute{\rm a}$n (CAHA) in Spain, the
Copernico 1.82 m telescope of the Italian National Institute for Astrophysics
(INAF) at Mount Ekar in Italy, and the Sutherland 1.9 m telescope at South
African Astronomical Observatory (SAAO) in South Africa. The sites at which
individual objects were observed are listed in Table \ref{tab:basic_info}.
Observations for some objects date back to December 2016, and continued until
the northern spring of 2021 for all targets except PG~1211+143 (for which
observations concluded in July 2017). We monitored most of the objects for more
than one year. To investigate the potential changes of the BLRs in different
years, we usually divided the data for each target into observing seasons
bounded by the periods when objects were inaccessible. We did not divide the
observations of PG~1100+772, PG~1351+640, PG~1351+695, PG~1534+580 into segments
because the seasonal gaps were small or their variation time scales are too long
to get reliable lag measurements from individual seasons (see figures in the
following sections). We divided the data of PG~1613+658 into only two seasons
because of its relatively long variation time scale. Our observations of
PG~0923+129, PG~1211+143, and PG~1310$-$108 span only one season. The detailed
beginning and end dates, spectroscopic epochs, and cadences for different
seasons are listed in Table \ref{tab:lc_info}. The spectra obtained from the 5
telescopes were all reduced using standard procedures (including the corrections
of bias and flat field, and the wavelength calibration) using IRAF v2.16. Here
we briefly introduce the settings of the instruments, apertures, and the
calibration of the observations at these telescopes.

\subsubsection{WIRO data}
\label{sec:wiro}

We performed RM  at WIRO using a 900 l/mm grating which provides a dispersion of
1.49 \AA\ pixel$^{-1}$ and a wavelength range of $\sim$4000-7000 \AA. To
minimize slit losses and their influence on the flux calibration, a 5" wide slit
oriented north-south was adopted (wider than the typical seeing of
\textasciitilde 2"-3"). Spectrophotometric standard stars (usually
BD+$28^{\circ}$4211, G191B2B, Feige~34, and Hz~44) were used for flux
calibration. We used an extraction aperture from $-$6".84 to 6".84, with
background windows [$-$15".2, $-$7".6] and [7".6, 15".2] relative to the
object's nuclear position. We adopted the [O {\sc iii}]-based technique
\citep[e.g.,][]{vanGroningen1992, fausnaugh2017b} to perform the relative flux
calibration. Where necessary, the spectra of the targets are artificially
broadened to achieve the same spectral resolution throughout, and then scaled
according to their [O {\sc iii}] fluxes (see more details in Paper
\citetalias{Du2018a}). The fiducial [O {\sc iii}] fluxes were determined using
the spectra taken in photometric conditions. The [O {\sc iii}]$\lambda$4959
lines of PG~1202+281, PG~1351+640, and PG~1351+695 overlap with their [O {\sc
iii}]$\lambda$5007 because of their broad line widths (please note that, during
the [O {\sc iii}]-based flux calibration, the original spectra were broadened).
We used both of the [O {\sc iii}] lines to do the flux calibration in these
cases. 

During each night, we took 3 to 4 consecutive exposures in order to both improve
the S/N ratios and evaluate the calibration accuracy by  checking their
differences. The spectra taken during the same night (after the [O {\sc iii}]
calibration) were combined to produce the spectrum for that epoch. In addition
to Poisson noise, the difference between the consecutive exposures during the
night is caused by the varying weather conditions, seeing variations, or
tracking variations during the exposures. This systematic uncertainty was
estimated by comparing the fluxes of the exposures in a wide range of
wavelengths (4740-5125\AA, effectively eliminating the contribution from Poisson
noise), and was added to the error of the continuum and emission-line fluxes of
the corresponding epoch using quadratic summation (see more details in Paper
\citetalias{Du2018a}).

\subsubsection{Lijiang data}
\label{sec:lijiang}

We used the Yunnan Faint Object Spectrograph and Camera (YFOSC) installed in
Lijiang 2.4 m telescope, which is an instrument both for low-resolution
spectroscopy and imaging. Grism 14 (with a resolution of 1.8\AA\ pixel$^{-1}$
and a wavelength range of 3800\AA - 7200\AA) and a 2".5-wide slit were adopted
in the campaign. The spectra were extracted in an aperture of $\pm$4".25 around
the nuclear position, with background windows [$-$14".15, $-$7".36] and [7".36,
14".15]. The field de-rotator of the telescope makes it easy to rotate the slit
accurately, thus we perform the flux calibration by placing a comparison star
simultaneously in the slit (\citealt{Maoz1990, Kaspi2000}, see more details in
\citealt{Du2014}). The advantage of the comparison-star-based calibration
technique is that it can accurately correct for the changes of the
wavelength-dependent atmosphere extinction in different nights. The information
of the comparison stars are listed in Table \ref{tab:comparison stars}. The
fiducial spectra of the comparison stars were generated from the data in
good-weather conditions (calibrated by the spectrophotometric standard stars
Feige 34 and Hiltner 600). The target spectra were flux-corrected by scaling the
comparison stars to standard values. In order to ensure that the comparison
stars were not variable during observations, we performed differential
photometry using several field stars. The standard deviations of the photometric
light curves of the comparison stars are $\sim$1\% and much smaller than the
variation amplitudes of the targets, which means that they can be treated as
calibration standards. Similar to the WIRO observations, we took 2 to 3
consecutive exposures each epoch. We combined them to obtain the
individual-night spectra. In addition, we corrected the small
wavelength-calibration uncertainties of the spectra according to their [O {\sc
iii}] emission lines before producing the light curves.

\subsubsection{CAHA data}

Several spectra of PG~1211+143 were taken using the CAHA 2.2 m telescope from
May to August in 2017 using the Calar Alto Faint Object Spectrograph (CAFOS). We
took the spectra using Grism G-200 and a 3".0-wide long slit. The wavelength
coverage is from 4000\AA\ to 8000\AA\ (with a dispersion of 4.47\AA\
pixel$^{-1}$). The spectra were extracted in an aperture of $\pm$5".58, with
background windows [$-$23".85, $-$6".30] and [23".85, 6".30]. Similar to the
observations at Lijiang, we also adopted the comparison-star-based calibration
technique. The coordinates of the comparison star are listed in Table
\ref{tab:comparison stars}. The calibration procedures are the same as for the
Lijiang data (see Section \ref{sec:lijiang}).

\subsubsection{Asiago data}

For PG~0007+106, PG~1100+772, PG~1202+281, and PG~1501+106, some of the data
points come from the Asiago 1.82 m telescope. The spectra were taken using the
Asiago Faint Object Spectrograph and Camera (AFOSC), which is a focal reducer
instrument similar to YFOSC and CAFOS, with a 4".2 slit. For PG~0007+106 and
PG~1501+106, Grism VPH7 was used, with a wavelength coverage of 3200\AA\ to
7000\AA\ and a dispersion of 2.95\AA\ pixel$^{-1}$. For PG~1100+772 and
PG~1202+281, Grism VPH6 was used, with a wavelength coverage of 4500\AA\ to
10000\AA\ with a dispersion of 2.95\AA\ pixel$^{-1}$.  We also adopted the [O
{\sc iii}]-based calibration, similar to that in the WIRO data reduction. We
extracted the spectra using a window of $\pm$30 pixels (corresponding to
$\pm$7.8"). The background was determined using the windows [$-$13", $-$6.76"]
and [6.76", 13"] on both sides of the objects.

\subsubsection{SAAO data}

We also took spectra using the SAAO 1.9 m telescope for PG~0007+106 and
PG~1501+106. The 600 lines mm$^{-1}$ grating and a slit width of 4".04 were
used. The flux calibration was also performed using the [O {\sc iii}]-based
technique. More details of the observations and data reduction can be found in
\cite{Winkler2017}. We extracted the spectra using a window of $\pm6$ pixels
(corresponding to $\pm$8.16"). The background was determined using the windows
[$-$20.4", $-$10.9"] and [10.9", 20.4"] on both sides of the objects.

\subsection{Photometry} 
\label{subsec:Photometry}

The YFOSC and CAFOS instruments can also perform imaging observations. For
PG~1001+054 and PG~1211+143, we took Johnson V-band images and carried out
differential photometry for the targets and the in-slit comparison stars using
several other stars in the same fields. The purpose was (1) to make sure that
the in-slit comparison stars were not variable during our campaign and (2) to
check the flux calibration accuracy of the spectroscopic observations. The
fluxes of the targets and comparison stars were extracted using circular
apertures with radii of 5".66 and 5".30 for YFOSC and CAFOS, respectively. The
typical exposure times were 20 to 50 seconds. For PG~1001+054, the small scatter
of the photometric light curve of the comparison star is at the level of
$\sim$1-2\%, which is stable enough for calibrations. While the comparison star
of PG~1211+143 is not in the field of view for photometry, the consistency
between its photometric and spectroscopic continuum light curves indicates that
its comparison star did not vary significantly and our calibration procedures
appear accurate.

To improve the cadence and extend the temporal coverage of the continuum light
curves, we also employ archival time-domain photometric data from the All-Sky
Automated Survey for
SuperNovae\footnote{http://www.astronomy.ohio-state.edu/asassn/index.shtml}
(ASAS-SN) and the Zwicky Transient
Facility\footnote{https://www.ztf.caltech.edu/} (ZTF). The ASAS-SN project
\citep{Shappee2014, Kochanek2017} started in 2013 to identify transients and
variable sources. Objects with magnitudes between 8 mag and 17 mag in the whole
sky are monitored. The details of the data reduction are provided in
\cite{Shappee2014} and \cite{Kochanek2017}. ZTF makes use of the Palomar 48-inch
Schmidt telescope and provides high-quality photometric light curves for objects
with magnitudes $\lesssim20$ \citep{Masci2019}. As of May 2021, there were 6
data releases in ZTF. We employ the light curves from ASAS-SN (g and V bands)
and ZTF (g and r bands) to supplement our photometric and spectroscopic
continuum light curves. Considering that the scatter in the ASAS-SN light curves
is larger than that of our spectroscopic continuum and the ZTF light curves, we
adopted the ASAS-SN data only if they can significantly lengthen the continuum
light curves or supplement their temporal coverage (PG~0049+171, PG~0923+129,
PG~1211+143, PG~1351+695, PG~1501+106, PG~1534+580, and PG~1613+658). Otherwise,
the ZTF light curves are used in the present work.

    \setlength{\tabcolsep}{0.5mm}
    \startlongtable
    \begin{deluxetable*}{ccccccccccccccccc}
    \tablecaption{Basic Information of Light Curves \label{tab:lc_info}}
    \tablewidth{0pt}
    \tablehead{ & & & & \multicolumn{5}{c}{Spectroscopy} & & \multicolumn{3}{c}{Continuum} & & \multicolumn{3}{c}{H$\beta$} \\
    \cmidrule(r){5-9} \cmidrule(r){11-13} \cmidrule(r){15-17} \\
    \colhead{Name} & & \colhead{Season} & & \colhead{Duration} & & 
    \colhead{Epochs} & & \colhead{Cadence} & & \colhead{$\it F_{\rm var}$} & & \colhead{Flux density}  & & \colhead{$\it F_{\rm var}$}  & & \colhead{Flux} \\
        & & & & & & & & (days) & &  (\%) & &  & &  (\%) & & \\
    (1) & & (2) & & (3) & & (4) & & (5) & & (6) & & (7) & & (8) & & (9) }
    \startdata
    PG 0007+106 & & All & & 2017.10--2021.01 & & 132 & & 8.9 & & 10.4$\pm$0.7 & & 0.95$\pm$0.10  & & 11.3$\pm$0.7 & & 1.51$\pm$0.17   \\ 
     & & 1 & & 2017.10--2018.01 & & 23 & & 4.3 & & 7.5$\pm$1.2 & & 1.08$\pm$0.08 & &  5.5$\pm$0.9 & & 1.61$\pm$0.09  \\ 
     & & 2 & & 2018.08--2019.02 & & 47 & & 3.4 & & 3.8$\pm$0.5 & & 0.95$\pm$0.04 & &  7.8$\pm$0.9 & & 1.64$\pm$0.13  \\ 
     & & 3 & & 2019.06--2020.02 & & 35 & & 6.5 & & 11.3$\pm$1.4 & & 0.93$\pm$0.11 & &  8.2$\pm$1.1 & & 1.40$\pm$0.12  \\ 
     & & 4 & & 2020.08--2021.01 & & 27 & & 5.1 & & 4.8$\pm$0.9 & & 0.86$\pm$0.05 & &  7.3$\pm$1.1 & & 1.35$\pm$0.10  \\ 
    PG 0049+171 & & All & & 2017.10--2021.02 & & 160 & & 7.5 & & 13.4$\pm$0.8 & & 1.77$\pm$0.24  & & 5.9$\pm$0.4 & & 1.88$\pm$0.12   \\ 
     & & 1 & & 2017.10--2018.02 & & 28 & & 4.5 & & 6.4$\pm$0.9 & & 1.89$\pm$0.13 & &  7.2$\pm$1.1 & & 2.01$\pm$0.15  \\ 
     & & 2 & & 2018.08--2019.02 & & 48 & & 3.7 & & 5.4$\pm$0.6 & & 1.50$\pm$0.08 & &  2.2$\pm$0.4 & & 1.78$\pm$0.05  \\ 
     & & 3 & & 2019.06--2020.02 & & 44 & & 5.5 & & 8.9$\pm$1.0 & & 2.00$\pm$0.18 & &  3.3$\pm$0.5 & & 1.92$\pm$0.08  \\ 
     & & 4 & & 2020.08--2021.02 & & 40 & & 4.3 & & 6.5$\pm$0.8 & & 1.74$\pm$0.12 & &  2.9$\pm$0.5 & & 1.85$\pm$0.07  \\ 
    PG 0923+129 & & All & & 2020.10--2021.05 & & 41 & & 5.0 & & 6.3$\pm$0.8 & & 4.41$\pm$0.30  & & 8.5$\pm$1.0 & & 1.90$\pm$0.17   \\ 
    PG 0947+396 & & All & & 2017.10--2021.05 & & 83 & & 15.7 & & 6.9$\pm$0.6 & & 0.57$\pm$0.04  & & 6.3$\pm$0.6 & & 0.49$\pm$0.03   \\ 
     & & 1 & & 2017.10--2018.05 & & 26 & & 7.8 & & 7.5$\pm$1.3 & & 0.55$\pm$0.05 & &  9.7$\pm$1.4 & & 0.48$\pm$0.05  \\ 
     & & 2 & & 2018.11--2019.06 & & 22 & & 10.2 & & 8.2$\pm$1.3 & & 0.57$\pm$0.05 & &  3.3$\pm$0.8 & & 0.50$\pm$0.02  \\ 
     & & 3 & & 2020.02--2020.05 & & 16 & & 7.2 & & 1.1$\pm$0.7 & & 0.60$\pm$0.01 & &  4.3$\pm$0.9 & & 0.50$\pm$0.02  \\ 
     & & 4 & & 2020.11--2021.05 & & 19 & & 10.4 & & 4.6$\pm$1.0 & & 0.56$\pm$0.03 & &  3.2$\pm$0.8 & & 0.48$\pm$0.02  \\ 
    PG 1001+054 & & All & & 2017.10--2021.05 & & 102 & & 12.8 & & 7.8$\pm$0.6 & & 0.81$\pm$0.06  & & 5.5$\pm$0.4 & & 0.81$\pm$0.05   \\ 
     & & 1 & & 2017.10--2018.04 & & 31 & & 5.8 & & 2.6$\pm$0.4 & & 0.80$\pm$0.02 & &  2.4$\pm$0.4 & & 0.83$\pm$0.02  \\ 
     & & 2 & & 2018.10--2019.06 & & 34 & & 6.6 & & 2.8$\pm$0.4 & & 0.75$\pm$0.02 & &  1.8$\pm$0.5 & & 0.77$\pm$0.02  \\ 
     & & 3 & & 2019.11--2020.05 & & 23 & & 8.1 & & 4.3$\pm$0.7 & & 0.90$\pm$0.04 & &  3.5$\pm$0.6 & & 0.85$\pm$0.03  \\ 
     & & 4 & & 2020.11--2021.05 & & 14 & & 14.1 & & 3.5$\pm$0.9 & & 0.85$\pm$0.03 & &  5.9$\pm$1.4 & & 0.86$\pm$0.06  \\ 
    PG 1048+342 & & All & & 2017.11--2021.05 & & 87 & & 14.4 & & 11.1$\pm$0.9 & & 0.58$\pm$0.07  & & 8.0$\pm$0.7 & & 0.52$\pm$0.04   \\ 
     & & 1 & & 2017.11--2018.05 & & 23 & & 7.3 & & 5.8$\pm$1.2 & & 0.49$\pm$0.03 & &  3.4$\pm$0.7 & & 0.46$\pm$0.02  \\ 
     & & 2 & & 2018.11--2019.06 & & 36 & & 6.2 & & 3.5$\pm$0.6 & & 0.63$\pm$0.03 & &  2.6$\pm$0.6 & & 0.56$\pm$0.02  \\ 
     & & 3 & & 2019.11--2020.04 & & 13 & & 11.9 & & 2.3$\pm$1.8 & & 0.60$\pm$0.03 & &  1.3$\pm$1.2 & & 0.54$\pm$0.02  \\ 
     & & 4 & & 2020.12--2021.05 & & 15 & & 10.2 & & 6.5$\pm$1.5 & & 0.55$\pm$0.04 & &  2.1$\pm$0.8 & & 0.54$\pm$0.02  \\ 
    PG 1100+772 & & All & & 2018.11--2021.04 & & 42 & & 20.9 & & 8.9$\pm$1.0 & & 1.37$\pm$0.13  & & 1.6$\pm$0.5 & & 2.32$\pm$0.06   \\ 
    PG 1202+281 & & All & & 2016.12--2021.04 & & 101 & & 15.5 & & 9.2$\pm$0.7 & & 0.58$\pm$0.05  & & 7.4$\pm$0.6 & & 0.37$\pm$0.03   \\ 
     & & 1 & & 2016.12--2017.05 & & 26 & & 5.7 & & 6.4$\pm$1.0 & & 0.59$\pm$0.04 & &  3.7$\pm$0.7 & & 0.34$\pm$0.02  \\ 
     & & 2 & & 2018.01--2018.05 & & 22 & & 5.5 & & 5.4$\pm$0.9 & & 0.60$\pm$0.03 & &  1.0$\pm$0.8 & & 0.35$\pm$0.01  \\ 
     & & 3 & & 2018.12--2019.07 & & 27 & & 7.8 & & 10.1$\pm$1.4 & & 0.58$\pm$0.06 & &  2.6$\pm$0.6 & & 0.40$\pm$0.01  \\ 
     & & 4 & & 2020.01--2020.05 & & 21 & & 6.4 & & 6.2$\pm$1.0 & & 0.54$\pm$0.03 & &  4.4$\pm$0.8 & & 0.36$\pm$0.02  \\ 
     & & 5 & & 2020.12--2021.04 & & 5 & & 23.8 & & 6.3$\pm$2.1 & & 0.47$\pm$0.03 & &  5.3$\pm$1.8 & & 0.36$\pm$0.02  \\ 
    PG 1211+143 & & All & & 2016.12--2017.07 & & 52 & & 4.2 & & 12.1$\pm$1.3 & & 4.66$\pm$0.58  & & 10.3$\pm$1.1 & & 4.61$\pm$0.50   \\ 
    PG 1310-108 & & All & & 2021.01--2021.05 & & 17 & & 7.7 & & 3.3$\pm$0.9 & & 1.76$\pm$0.08  & & 3.3$\pm$0.7 & & 1.14$\pm$0.04   \\ 
    PG 1351+640 & & All & & 2016.12--2021.02 & & 109 & & 13.8 & & 14.1$\pm$1.0 & & 3.63$\pm$0.51  & & 4.7$\pm$0.4 & & 1.22$\pm$0.06   \\ 
    PG 1351+695 & & All & & 2019.06--2021.04 & & 108 & & 6.2 & & 12.2$\pm$0.9 & & 3.80$\pm$0.47  & & 26.9$\pm$1.9 & & 1.75$\pm$0.48   \\ 
    PG 1501+106 & & All & & 2017.02--2020.06 & & 136 & & 8.9 & & 13.2$\pm$0.8 & & 5.18$\pm$0.69  & & 7.6$\pm$0.5 & & 3.61$\pm$0.28   \\ 
     & & 1 & & 2017.02--2017.05 & & 17 & & 6.1 & & 8.6$\pm$1.6 & & 6.06$\pm$0.53 & &  6.9$\pm$1.2 & & 3.67$\pm$0.26  \\ 
     & & 2 & & 2018.01--2018.05 & & 28 & & 4.3 & & 2.9$\pm$0.5 & & 5.64$\pm$0.18 & &  2.7$\pm$0.5 & & 3.69$\pm$0.11  \\ 
     & & 3 & & 2019.02--2019.10 & & 57 & & 4.3 & & 9.0$\pm$0.9 & & 5.20$\pm$0.47 & &  10.0$\pm$1.0 & & 3.64$\pm$0.37  \\ 
     & & 4 & & 2020.01--2020.06 & & 34 & & 3.9 & & 7.2$\pm$0.9 & & 4.34$\pm$0.32 & &  4.1$\pm$0.5 & & 3.47$\pm$0.15  \\ 
    PG 1534+580 & & All & & 2020.02--2021.05 & & 83 & & 5.5 & & 5.8$\pm$0.5 & & 3.90$\pm$0.24  & & 6.0$\pm$0.6 & & 1.85$\pm$0.12   \\ 
    PG 1613+658 & & All & & 2016.12--2021.04 & & 200 & & 7.9 & & 11.9$\pm$0.6 & & 2.62$\pm$0.32  & & 5.1$\pm$0.3 & & 3.55$\pm$0.20   \\ 
     & & 1 & & 2016.12--2018.05 & & 55 & & 9.2 & & 12.2$\pm$1.2 & & 2.84$\pm$0.35  & & 3.7$\pm$0.5 & & 3.78$\pm$0.17   \\ 
     & & 2 & & 2018.12--2021.04 & & 145 & & 6.0 & & 10.0$\pm$0.6 & & 2.53$\pm$0.26  & & 3.0$\pm$0.3 & & 3.46$\pm$0.13   
    \enddata
    \tablecomments{Column (1) is the name of object. Column (2) is the season for
    the measurement. Columns (3-5) are the duration, epoch and cadence of the
    spectroscopy. Columns (6-7) are the variation amplitude and mean flux for the
    continuum light curve. The uncertainty range of the mean flux is the standard
    deviation of the light curve. The unit for mean flux is $\rm
    10^{-15}~erg~s^{-1}~cm^{-2}\mathring{A}^{-1}$.  Columns (8-9) are the variation
    amplitude and mean flux for the H$\beta$ light curve. The unit for mean flux is
    $\rm 10^{-13}~erg~s^{-1}~cm^{-2}$.
    }
    \end{deluxetable*}

\section{Analysis} 
\label{sec:analysis}

\subsection{Mean and RMS spectra}
\label{sec:meanrms}

To check the general H$\beta$ profiles, evaluate their variation amplitudes, and
investigate their changes in different seasons, we calculated the mean and rms
spectra of the objects for the whole campaign as well as for individual seasons
(see Figures \ref{0007lc} through \ref{1613lc})  using
\begin{equation}
\bar{F}_{\lambda}=\frac{1}{N}\sum_{i=1}^NF_{\lambda}^i,
\end{equation}
and 
\begin{equation}
S_{\lambda}=\left[\frac{1}{N}\sum_{i=1}^N\left(F_{\lambda}^i-\bar{F}_{\lambda}\right)^2\right]^{1/2},
\end{equation}
respectively. Here $F_{\lambda}^i$ is the $i$-th spectrum of the object and $N$
is the number of its spectra. The narrow [O {\sc iii}] emission lines in the rms
spectra are extremely weak or negligible compared to the mean spectra of the
same objects, which indicates that our calibration procedure works well. Only
the [O {\sc iii}] lines in the rms spectrum of PG~1211+143 have some residual
signals. This is caused by the variation of spectral resolution in its exposures
with different seeing rather than the flux variations of the [O {\sc iii}]
lines. We took the spectra at Lijiang/CAHA (see Table \ref{tab:basic_info}) and
performed the flux calibration based on the comparison star (see Section
\ref{sec:lijiang}) for this object. The variable spectral resolution was not
corrected. We measure the standard deviation of [O {\sc iii}]$\lambda$5007 flux
to be $\sim$3\%, which indicates the reliability of our calibration procedures.

\begin{figure*}[ht!]
    \centering
    \includegraphics[width=0.8\textwidth]{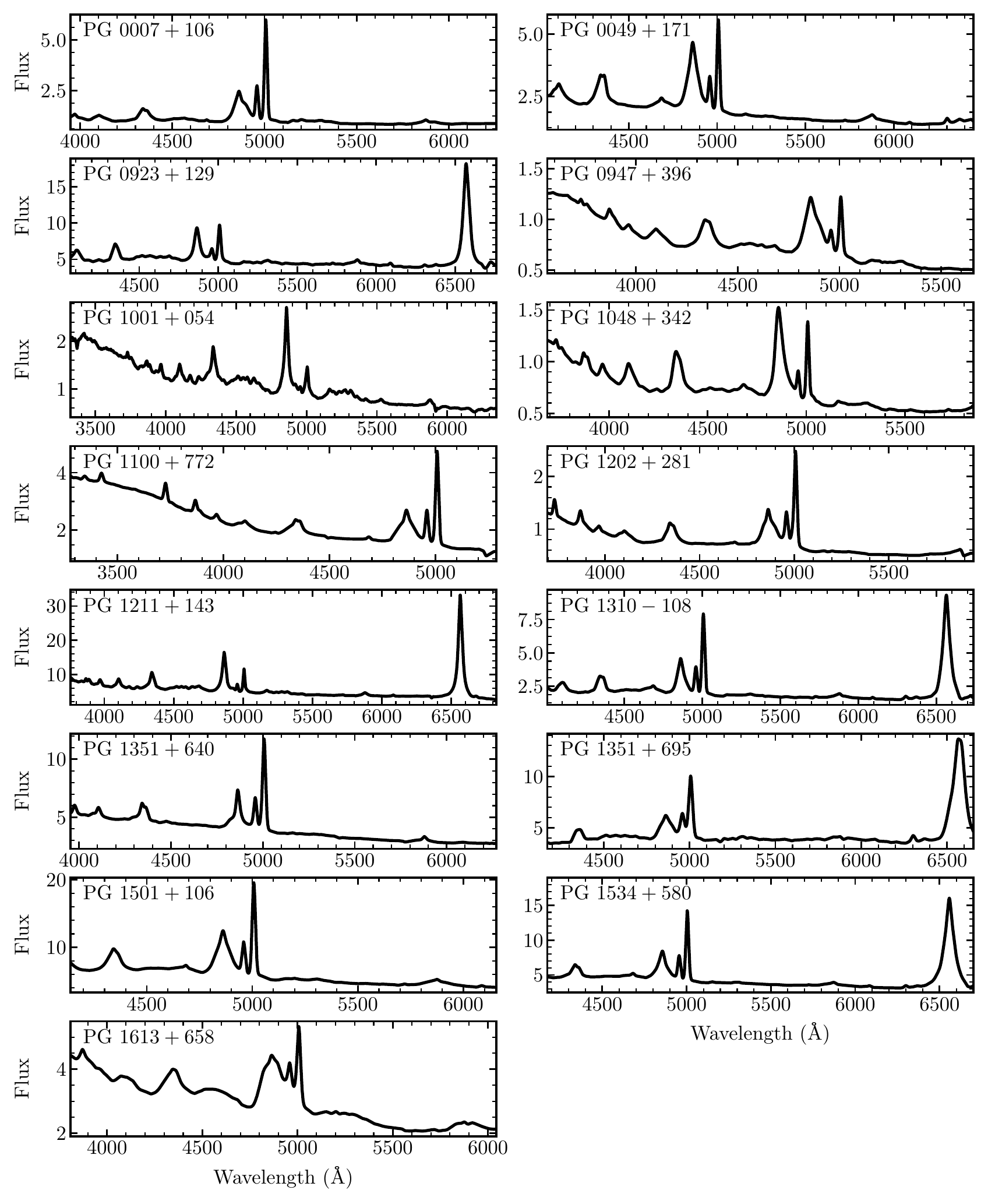} 
    \caption{Mean spectra (observed flux density vs. rest-frame wavelength) of the objects. 
    Flux units are $\rm 10^{-15}\ erg\ s^{-1}$ $\rm cm^{-2}$ $\rm \mathring{A}^{-1}$. \label{fig:mean_spec}}
    \end{figure*}

The rms spectra of several objects in some seasons only show weak H$\beta$
emission lines or even what appear to be ``absorption lines.'' There are two
main reasons for this: (1) the variation amplitudes of the H$\beta$ fluxes in
the corresponding periods are significantly smaller than those in other seasons
(e.g., Season 4 in PG~0947+396, Season 2 in PG~1202+281 and PG~1501+106, see
Figures \ref{0947lc}, \ref{1202lc}, and \ref{1501lc}) and (2) the variations of
their H$\beta$ light curves show reverse modulation with respect to the
continuum light curves -- in other words, the peaks (troughs) of the H$\beta$
fluxes happen to appear during the troughs (peaks) of the continuum fluxes (see,
e.g., the light curves of PG~0947+396 in Season 2 and PG~1202+281 in Seasons 1
and 3, in Figures \ref{0947lc} and \ref{1202lc}). To check if the contribution
from the reverse variations of the continuum can really weaken the emission-line
signals in the rms spectra, we subtracted the continuum beneath the H$\beta$
lines, determined from the linear interpolation between two continuum windows on
both sides, from each individual spectrum before calculating the rms spectra for
those objects in which the rms spectra showed very weak or ``absorption-like''
H$\beta$ signals. The continuum-cleaned rms spectra of PG~0049+171, PG~0947+396,
PG~1202+281, PG~1351+640, and PG~1501+106 are plotted in Figures \ref{0049lc},
\ref{0947lc}, \ref{1202lc}, \ref{1351640lc}, and \ref{1501lc}, respectively. The
continuum-cleaned rms spectra have much stronger H$\beta$ emission lines than
the original rms spectra, consistent with the idea that the apparent absorption
effect  is due the continuum contributions to the emission lines in the rms
spectra. 

\begin{deluxetable}{ccccccc}
    \tablecaption{Systematic errors of Light Curves\label{tab:syserr_info}}
    \tablewidth{0pt}
    \renewcommand\arraystretch{0.7}
    \tablehead{
    \colhead{Target} & & \colhead{Duration} & & \colhead{$\sigma_{\rm sys}$ (conti)} & & \colhead{$\sigma_{\rm sys}$ (H$\beta$)}
    }
    \startdata
      PG 0007+106 & & 2017.10--2018.01 & & 0.019 & & 0.026 \\ 
      PG 0007+106 & & 2018.08--2019.02 & & 0.015 & & 0.031 \\ 
      PG 0007+106 & & 2019.06--2020.02 & & 0.022 & & 0.034 \\ 
      PG 0007+106 & & 2020.08--2021.01 & & 0.022 & & 0.026 \\ 
      PG 0049+171 & & 2017.10--2018.02 & & 0.037 & & 0.042 \\ 
      PG 0049+171 & & 2018.08--2019.02 & & 0.014 & & 0.028 \\ 
      PG 0049+171 & & 2019.06--2020.02 & & 0.030 & & 0.043 \\ 
      PG 0049+171 & & 2020.08--2021.02 & & 0.021 & & 0.038 \\ 
      PG 0923+129 & & 2020.10--2021.05 & & 0.000 & & 0.000 \\ 
      PG 0947+396 & & 2017.10--2018.05 & & 0.017 & & 0.009 \\ 
      PG 0947+396 & & 2018.11--2019.06 & & 0.003 & & 0.000 \\ 
      PG 0947+396 & & 2020.02--2020.05 & & 0.000 & & 0.000 \\ 
      PG 0947+396 & & 2020.11--2021.05 & & 0.009 & & 0.000 \\ 
      PG 1001+054 & & 2017.10--2018.04 & & 0.010 & & 0.011 \\ 
      PG 1001+054 & & 2018.10--2019.06 & & 0.007 & & 0.013 \\ 
      PG 1001+054 & & 2019.11--2020.05 & & 0.009 & & 0.013 \\ 
      PG 1001+054 & & 2020.11--2021.05 & & 0.011 & & 0.020 \\ 
      PG 1048+342 & & 2017.11--2018.05 & & 0.017 & & 0.007 \\ 
      PG 1048+342 & & 2018.11--2019.06 & & 0.005 & & 0.007 \\ 
      PG 1048+342 & & 2019.11--2020.04 & & 0.022 & & 0.010 \\ 
      PG 1048+342 & & 2020.12--2021.05 & & 0.013 & & 0.004 \\ 
      PG 1100+772 & & 2018.11--2019.09 & & 0.000 & & 0.000 \\ 
      PG 1100+772 & & 2019.10--2020.05 & & 0.000 & & 0.000 \\ 
      PG 1100+772 & & 2020.11--2021.04 & & 0.017 & & 0.000 \\ 
      PG 1202+281 & & 2016.12--2017.05 & & 0.014 & & 0.007 \\ 
      PG 1202+281 & & 2018.01--2018.05 & & 0.012 & & 0.006 \\ 
      PG 1202+281 & & 2018.12--2019.07 & & 0.009 & & 0.007 \\ 
      PG 1202+281 & & 2020.01--2020.05 & & 0.008 & & 0.006 \\ 
      PG 1202+281 & & 2020.12--2021.04 & & 0.007 & & 0.004 \\ 
      PG 1211+143 & & 2016.12--2017.07 & & 0.148 & & 0.146 \\ 
      PG 1310-108 & & 2021.01--2021.05 & & 0.038 & & 0.000 \\ 
      PG 1351+640 & & 2016.12--2017.05 & & 0.000 & & 0.020 \\ 
      PG 1351+640 & & 2017.12--2018.05 & & 0.020 & & 0.021 \\ 
      PG 1351+640 & & 2019.01--2019.11 & & 0.049 & & 0.021 \\ 
      PG 1351+640 & & 2020.02--2020.05 & & 0.002 & & 0.025 \\ 
      PG 1351+640 & & 2020.08--2021.02 & & 0.036 & & 0.030 \\ 
      PG 1351+695 & & 2019.06--2020.05 & & 0.061 & & 0.036 \\ 
      PG 1351+695 & & 2020.08--2021.04 & & 0.108 & & 0.136 \\ 
      PG 1501+106 & & 2017.02--2017.05 & & 0.098 & & 0.043 \\ 
      PG 1501+106 & & 2018.01--2018.05 & & 0.056 & & 0.048 \\ 
      PG 1501+106 & & 2019.02--2019.10 & & 0.041 & & 0.046 \\ 
      PG 1501+106 & & 2020.01--2020.06 & & 0.000 & & 0.019 \\ 
      PG 1534+580 & & 2020.02--2020.05 & & 0.037 & & 0.029 \\ 
      PG 1534+580 & & 2020.08--2021.05 & & 0.070 & & 0.051 \\ 
      PG 1613+658 & & 2016.12--2017.05 & & 0.022 & & 0.083 \\ 
      PG 1613+658 & & 2018.01--2018.05 & & 0.000 & & 0.000 \\ 
      PG 1613+658 & & 2018.12--2020.05 & & 0.027 & & 0.058 \\ 
      PG 1613+658 & & 2020.08--2021.04 & & 0.027 & & 0.048 
    \enddata
    \tablecomments{These are the systematic errors of the spectroscopy data in
    separate seasons. The systematic error of ``0.00'' means that it can be ignored.
    For PG~1100+772, PG~1351+640, PG~1351+695, PG~1534+580, and PG~1613+658, we did
    not divide their light curves into different seasons  
    according to their gaps in the campaign because of the long variation time
    scales (see details in Section \ref{subsec:Spectroscopy}). However, their
    systematic uncertainties for continuum and H$\beta$ used in the time-series
    analysis in Section \ref{sec:analysis} are estimated in light of the gaps in the
    campaign (if necessary). The unit for continuum systematic errors is $\rm
    10^{-15}~erg~s^{-1}~cm^{-2}\mathring{A}^{-1}$. The unit for H$\beta$ systematic
    errors is $\rm 10^{-13}~erg~s^{-1}~cm^{-2}$.}
    \end{deluxetable}

\subsection{Light Curves} 
\label{sec:light curves}

The H$\beta$ light curves can be measured by the direct integration method 
\citep[e.g.,][]{Peterson1998, Kaspi2000, Bentz2009, Grier2012, Du2014}
or spectral fitting methods \citep[e.g.,][]{Barth2013, Hu2015}. Paper
\citetalias{Du2018a} has described the advantages and disadvantages of these two
methods and explained the reason why we decided to use the direct integration
method in our MAHA campaign (see Section 3.1 there). As in Papers
\citetalias{Du2018a} and \citetalias{Brotherton2020}, we adopted the integration
method to measure the fluxes of the H$\beta$ emission lines. The H$\beta$ fluxes
are measured after subtracting the underlying continuum. The continuum and the
integration windows are selected according to the emission-line signals in the
rms spectra, but also to avoid the possible influence of the He {\sc ii} line
and [O {\sc iii}] lines as much as possible. The narrow-line fluxes remaining in
the integration windows are also included in the H$\beta$ light curves. The
5100\AA\ continuum light curves are obtained by measuring the median fluxes
density from 5075\AA\ to 5125\AA. The measurement windows for the continuum and
H$\beta$ are marked in the mean and rms spectra in Figures \ref{0007lc} --
\ref{1613lc} for individual objects in different seasons. The light curves are
provided in Table \ref{tab:data for light curves} and shown in Figures
\ref{0007lc} -- \ref{1613lc}.

For some objects, the uncertainties described in Section \ref{sec:observation}
are still smaller than the apparent scatter in the light curves. This indicates
that the changes of the weather, pointing, and tracking conditions on different
nights have introduced some extra systematic uncertainties. We estimate these
systematic uncertainties using the median-filter method (see more details in
\citealt{Du2014} or Paper \citetalias{Du2018a}), and are provided in Table
\ref{tab:syserr_info} as needed. In the following analysis, these systematic
uncertainties are also included in the calculations by quadratic summation.

\subsection{Inter-calibration of Light Curves}
\label{sec:intercalibration}

Because of the different apertures used for the telescopes in our campaign (as
well as ASAS-SN and ZTF), and the correspondingly different contributions from
the host galaxies, we need to take care to properly inter-calibrate the
photometric and spectroscopic light curves. The inter-calibration is performed
by the Bayesian-based package PyCALI\footnote{PyCALI is available at:
https://github.com/LiyrAstroph/PyCALI} \citep{Li2014}. It assumes that the light
curves can be described by a damped random walk model and determines the best
multiplicative and additive factors by exploring the posterior probability
distribution with a diffusive nested sampling algorithm \citep{Brewer2011}. The
5100\AA\ continuum and H$\beta$ light curves from different telescopes are
inter-calibrated and then combined by averaging the observations during the same
nights. The inter-calibrated and combined light curves are shown in Figures
\ref{0007lc} -- \ref{1613lc}. The light curves from different telescopes are
generally quite consistent with each other. Several severely deviant data points
differing from adjacent epochs and the MICA reconstruction (see below) or
possessing significantly larger error bars are not included in the following
time-series analysis in Figures \ref{0007lc} -- \ref{1613lc}. 

In principle, the emission-line contributions (e.g., H$\beta$, H$\gamma$, He
{\sc ii}, Fe {\sc ii}) in the broad bands of photometric light curves may
slightly influence the lag measurements. However, the broad-band photometric and
spectroscopic (at 5100\AA) continuum light curves are all well consistent with
each other in the present paper (see Figures \ref{0007lc}-\ref{1613lc}), which
means these influences can be ignored given the current uncertainties of the
light curves. This is very natural because the integrated emission-line fluxes
in these broad bands are roughly smaller than 10\% of the continuum fluxes and
the emission-line variation amplitudes are generally smaller than those of the
continuum (see Table \ref{tab:lc_info}).

The average fluxes and variability of the continuum and H$\beta$ light curves are provided 
in Table \ref{tab:lc_info}. The variability and its uncertainty of a light curve have been 
defined \citep{Rodriguez-Pascual1997, Edelson2002} as
\begin{equation}
    F_{\rm var} = \frac{(\sigma^2 - \Delta^2)^{1/2}}{\langle F \rangle}
\end{equation}
and
\begin{equation}
    \sigma_{\rm Fvar} = \frac{1}{{(2N)}^{1/2} F_{\rm var}}\frac{\sigma^2}{{\langle F \rangle}^2},
\end{equation}
where $\sigma$ is the mean square root of the variance, $\Delta^2$ is the mean
square value of the flux uncertainties,  $\langle F \rangle$ is the average
flux, and $N$ is the number of epochs.

\begin{deluxetable}{ccccccccc}
    \tablecaption{Light Curves\label{tab:data for light curves}}
    \tablehead{
    \colhead{Target} & & \colhead{Telescope} & & \colhead{Data} & & \colhead{JD}  & & \colhead{Flux}
    }
    \startdata
    PG~0007 & & WIRO & & Conti    & & 1046.674 & & $0.982\pm0.004$\\ 
    PG~0007 & & WIRO & & H$\beta$ & & 1046.674 & & $1.501\pm0.007$\\ 
    PG~0007 & & WIRO & & Conti    & & 1049.699 & & $0.963\pm0.004$\\ 
    PG~0007 & & WIRO & & H$\beta$ & & 1049.699 & & $1.519\pm0.006$\\ 
    PG~0007 & & WIRO & & Conti    & & 1050.721 & & $0.985\pm0.005$\\ 
    PG~0007 & & WIRO & & H$\beta$ & & 1050.721 & & $1.473\pm0.008$
    \enddata
    \tablecomments{This table is available in its entirety online. The uncertainty doesn't include the systematic errors measured from 
    median filter method (see Section \ref{sec:light curves}). The Julian dates are from 2,457,000. The units for continuum 
    and $\rm H\beta$ are $\rm 10^{-15}\ erg\ s^{-1}$ $\rm cm^{-2}$ $\rm \mathring{A}^{-1}$ and 
    $\rm 10^{-13}\ erg\ s^{-1}$ $\rm cm^{-2}$, respectively. }
    \end{deluxetable}

    \begin{deluxetable*}{cccccccccccccccccccccc}
        \setlength{\tabcolsep}{0.5mm}
        \tablecaption{Line Widths, Time Lags in Rest Frame, and 5100$\rm \mathring{A}$ Luminosity \label{tab:linewidthtimelags}}
        \tablewidth{0pt}
        \tablehead{
        & & & & \multicolumn{3}{c}{Mean Spectra} & & \multicolumn{3}{c}{RMS} & & \multicolumn{3}{c}{ICCF} & & \colhead{MICA} & & \colhead{$\chi^2$} & & & \\
        \cmidrule(r){5-7} \cmidrule(r){9-11} \cmidrule(r){13-15} 
        \colhead{Target} & & \colhead{Season} & & \colhead{FWHM} & & \colhead{$\sigma_{\rm line}$} & & \colhead{FWHM} & & \colhead{$\sigma_{\rm line}$} & & \colhead{$\tau_{\rm cent}$} & & \colhead{$\tau_{\rm peak}$} & & \colhead{$\tau_{\rm cent}$} 
        & & \colhead{$\tau_{\rm peak}$} & & \colhead{$\lambda L_{\lambda}(5100\AA$)}& \\
         & &  & &($\rm km\ s^{-1}$) & & ($\rm km\ s^{-1}$) & & ($\rm km\ s^{-1}$) & & ($\rm km\ s^{-1}$) & & (days) & & (days) & & (days) & & (days) & & ($\times {\rm 10^{44}\ \rm erg\ s^{-1}}$)}
        \startdata
        PG 0007+106 & & All & & $5301_{-28}^{+33}$ & & $2424_{-41}^{+39}$ & & $4832_{-11}^{+10}$ & & $1766_{-9}^{+11}$ & & $30.9_{-2.4}^{+2.5}$ & & $25.1_{-4.4}^{+1.9}$ & & $25.8_{-1.1}^{+1.2}$ & & $24.2_{-3.2}^{+3.8}$ & & $1.61\pm0.17$  \\ 
        & &  1  & & $5365_{-28}^{+28}$ & & $2524_{-44}^{+43}$ & & $5396_{-24}^{+25}$ & & $1895_{-21}^{+21}$ & & $22.1_{-5.7}^{+7.8}$ & & $19.5_{-4.0}^{+24.3}$ & & $23.3_{-4.0}^{+6.2}$ & & $51.2_{-32.9}^{+7.8}$ & & $1.84\pm0.14$  \\ 
        & &  2  & & $5244_{-30}^{+33}$ & & $2347_{-34}^{+34}$ & & $4621_{-6}^{+7}$ & & $1881_{-14}^{+13}$ & & $34.7_{-4.3}^{+4.0}$ & & $32.0_{-7.6}^{+5.6}$ & & $19.7_{-13.7}^{+9.3}$ & & $15.0_{-2.7}^{+4.4}$ & & $1.62\pm0.07$  \\ 
        & &  3  & & $5433_{-29}^{+44}$ & & $2365_{-37}^{+41}$ & & $4471_{-29}^{+29}$ & & $1750_{-24}^{+24}$ & & $15.6_{-11.1}^{+15.4}$ & & $14.4_{-6.0}^{+4.5}$ & & $14.3_{-2.6}^{+2.5}$ & & $4.6_{-150.2}^{+19.3}$ & & $1.58\pm0.18$  \\ 
        & &  4  & & $5176_{-24}^{+37}$ & & $2558_{-53}^{+56}$ & & $4686_{-10}^{+9}$ & & $1558_{-12}^{+14}$ & & $25.1_{-6.7}^{+10.4}$ & & $20.7_{-3.8}^{+6.9}$ & & $20.8_{-2.0}^{+2.1}$ & & $98.2_{-12.9}^{+9.8}$ & & $1.45\pm0.08$  \\ 
        PG 0049+171 & & All & & $4262_{-38}^{+411}$ & & $2272_{-32}^{+38}$ & & $2873_{-4}^{+7}$ & & $1193_{-13}^{+13}$ & & $34.7_{-4.5}^{+3.8}$ & & $28.1_{-5.5}^{+11.4}$ & & $39.5_{-2.6}^{+3.3}$ & & $84.9_{-31.4}^{+43.7}$ & & $1.27\pm0.17$  \\ 
        & &  1  & & $4131_{-48}^{+394}$ & & $2005_{-46}^{+45}$ & & $2804_{-17}^{+9}$ & & $1109_{-21}^{+17}$ & & $51.2_{-3.8}^{+3.7}$ & & $54.2_{-9.9}^{+5.2}$ & & $41.8_{-6.3}^{+7.2}$ & & $31.8_{-20.6}^{+20.9}$ & & $1.36\pm0.09$  \\ 
        & &  2  & & $4426_{-95}^{+399}$ & & $2309_{-42}^{+37}$ & & $2919_{-8}^{+13}$ & & $1370_{-13}^{+14}$ & & $23.7_{-4.4}^{+14.0}$ & & $22.2_{-4.6}^{+9.4}$ & & $30.9_{-7.6}^{+3.7}$ & & $27.1_{-9.5}^{+36.8}$ & & $1.07\pm0.06$  \\ 
        & &  3  & & $4296_{-47}^{+311}$ & & $2333_{-32}^{+36}$ & & $1896_{-7}^{+9}$ & & $988_{-19}^{+21}$ & & $33.0_{-9.3}^{+6.7}$ & & $34.6_{-11.4}^{+5.1}$ & & $46.3_{-8.6}^{+64.3}$ & & $28.2_{-10.3}^{+13.3}$ & & $1.44\pm0.13$  \\ 
        & &  4  & & $4222_{-42}^{+478}$ & & $2327_{-36}^{+52}$ & & $3425_{-13}^{+12}$ & & $1223_{-23}^{+19}$ & & $20.4_{-3.4}^{+4.0}$ & & $20.6_{-3.6}^{+4.4}$ & & $22.6_{-2.2}^{+2.7}$ & & $20.3_{-5.2}^{+7.7}$ & & $1.25\pm0.08$  \\ 
        PG 0923+129 & & All & & $2461_{-34}^{+34}$ & & $1711_{-48}^{+51}$ & & $2138_{-13}^{+17}$ & & $1215_{-19}^{+20}$ & & $4.6_{-4.8}^{+3.4}$ & & $5.5_{-3.7}^{+1.6}$ & & $6.2_{-1.8}^{+3.2}$ & & $5.0_{-90.5}^{+4.5}$ & & $0.56\pm0.04$  \\ 
        PG 0947+396 & & All & & $5440_{-76}^{+24}$ & & $2872_{-27}^{+33}$ & & $3292_{-33}^{+28}$ & & $2021_{-32}^{+27}$ & & $34.4_{-4.9}^{+4.5}$ & & $36.3_{-8.9}^{+8.4}$ & & $39.5_{-1.7}^{+3.8}$ & & $41.4_{-10.8}^{+6.8}$ & & $5.85\pm0.43$  \\ 
        & &  1  & & $5222_{-24}^{+44}$ & & $2797_{-33}^{+36}$ & & $3757_{-37}^{+28}$ & & $1653_{-29}^{+26}$ & & $18.4_{-6.8}^{+6.0}$ & & $17.4_{-14.1}^{+15.5}$ & & $15.9_{-7.8}^{+7.0}$ & & $24.8_{-18.9}^{+44.7}$ & & $5.68\pm0.47$  \\ 
        & &  2  & & $5156_{-24}^{+68}$ & & $2816_{-27}^{+33}$ & & $5002_{-16}^{+15}$ & & $1673_{-14}^{+12}$ & & $38.5_{-5.8}^{+5.6}$ & & $41.1_{-8.0}^{+6.2}$ & & $41.8_{-1.1}^{+1.3}$ & & $44.8_{-7.5}^{+8.1}$ & & $5.89\pm0.51$  \\ 
        & &  3  & & $5590_{-93}^{+24}$ & & $2906_{-22}^{+20}$ & & $4216_{-19}^{+21}$ & & $1470_{-17}^{+15}$ & & $57.1_{-9.5}^{+10.7}$ & & $48.5_{-20.4}^{+17.0}$ & & $32.6_{-18.2}^{+16.1}$ & & $34.9_{-16.6}^{+59.4}$ & & $6.17\pm0.13$  \\ 
        & &  4  & & $5783_{-34}^{+97}$ & & $2984_{-22}^{+28}$ & & $6181_{-28}^{+46}$ & & $1861_{-32}^{+35}$ & & $46.9_{-11.2}^{+6.0}$ & & $48.9_{-18.0}^{+5.5}$ & & $29.4_{-4.0}^{+10.0}$ & & $48.2_{-18.1}^{+12.4}$ & & $5.78\pm0.31$  \\ 
        PG 1001+054 & & All & & $1688_{-21}^{+21}$ & & $1325_{-18}^{+18}$ & & $1933_{-20}^{+31}$ & & $1370_{-62}^{+44}$ & & $99.4_{-27.2}^{+15.8}$ & & $68.0_{-16.0}^{+11.2}$ & & $65.5_{-3.9}^{+5.6}$ & & $63.3_{-11.2}^{+17.4}$ & & $4.46\pm0.35$  \\ 
        & &  3  & & $1666_{-11}^{+11}$ & & $1318_{-11}^{+11}$ & & $917_{-50}^{+60}$ & & $840_{-185}^{+134}$ & & $57.6_{-13.5}^{+18.9}$ & & $63.0_{-28.1}^{+22.6}$ & & $64.7_{-6.9}^{+11.2}$ & & $117.0_{-62.5}^{+11.5}$ & & $4.93\pm0.22$  \\ 
        PG 1048+342 & & All & & $2905_{-59}^{+27}$ & & $1797_{-16}^{+16}$ & & $2147_{-7}^{+8}$ & & $1175_{-14}^{+16}$ & & $24.8_{-8.7}^{+10.4}$ & & $32.6_{-32.6}^{+13.3}$ & & $36.8_{-3.4}^{+2.4}$ & & $31.8_{-35.5}^{+77.2}$ & & $3.52\pm0.41$  \\ 
        & &  1  & & $3004_{-45}^{+34}$ & & $1840_{-14}^{+16}$ & & $3043_{-34}^{+33}$ & & $1449_{-31}^{+36}$ & & $26.2_{-8.2}^{+8.6}$ & & $25.5_{-8.2}^{+13.7}$ & & $28.0_{-4.8}^{+5.6}$ & & $33.6_{-11.9}^{+11.1}$ & & $3.02\pm0.21$  \\ 
        PG 1100+772 & & All & & $5733_{-21}^{+32}$ & & $3449_{-30}^{+31}$ & & $11229_{-23}^{+29}$ & & $4002_{-110}^{+87}$ & & $44.9_{-30.8}^{+30.5}$ & & $37.4_{-27.5}^{+45.3}$ & & $55.9_{-1.4}^{+3.0}$ & & $48.6_{-120.9}^{+46.6}$ & & $41.95\pm3.85$  \\ 
        PG 1202+281 & & All & & $5199_{-22}^{+24}$ & & $2035_{-4}^{+5}$ & & $4255_{-17}^{+23}$ & & $1301_{-24}^{+18}$ & & $98.5_{-30.1}^{+28.2}$ & & $66.4_{-13.4}^{+65.7}$ & & $66.3_{-1.9}^{+2.3}$ & & $116.3_{-31.4}^{+39.4}$ & & $3.42\pm0.32$  \\ 
        & &  1  & & $4891_{-19}^{+19}$ & & $3412_{-12}^{+12}$ & & $3825_{-70}^{+38}$ & & $1597_{-19}^{+17}$ & & $50.0_{-4.6}^{+6.6}$ & & $50.4_{-5.1}^{+7.9}$ & & $48.7_{-3.9}^{+3.5}$ & & $74.4_{-27.8}^{+8.8}$ & & $3.52\pm0.24$  \\ 
        & &  3  & & $4863_{-16}^{+22}$ & & $3260_{-17}^{+20}$ & & $3814_{-747}^{+399}$ & & $1540_{-63}^{+53}$ & & $71.6_{-9.5}^{+15.6}$ & & $68.7_{-7.1}^{+36.6}$ & & $69.5_{-4.4}^{+4.8}$ & & $133.0_{-68.8}^{+28.2}$ & & $3.45\pm0.36$  \\ 
        & &  4  & & $4949_{-16}^{+16}$ & & $3738_{-24}^{+25}$ & & $3658_{-29}^{+26}$ & & $1428_{-33}^{+29}$ & & $53.3_{-8.5}^{+10.9}$ & & $46.5_{-11.1}^{+22.4}$ & & $63.3_{-12.9}^{+13.3}$ & & $79.4_{-22.1}^{+37.4}$ & & $3.22\pm0.21$  \\ 
        PG 1211+143 & & All & & $1918_{-95}^{+95}$ & & $1499_{-49}^{+49}$ & & $1358_{-11}^{+14}$ & & $697_{-19}^{+18}$ & & $33.0_{-5.5}^{+5.6}$ & & $47.5_{-8.9}^{+10.8}$ & & $53.0_{-5.8}^{+5.1}$ & & $43.7_{-17.5}^{+18.2}$ & & $4.94\pm0.56$  \\ 
        PG 1310-108 & & All & & $3613_{-781}^{+812}$ & & $1978_{-65}^{+87}$ & & $2425_{-19}^{+15}$ & & $1092_{-54}^{+36}$ & & $13.2_{-2.8}^{+3.8}$ & & $12.5_{-2.1}^{+3.6}$ & & $12.8_{-1.7}^{+1.7}$ & & $12.6_{-3.5}^{+6.8}$ & & $0.33\pm0.01$  \\ 
        PG 1351+640 & & All & & $7625_{-81}^{+95}$ & & $3114_{-50}^{+65}$ & & $2154_{-13}^{+18}$ & & $1527_{-21}^{+23}$ & & $68.6_{-20.7}^{+20.4}$ & & $61.6_{-27.6}^{+81.3}$ & & $74.8_{-2.3}^{+2.3}$ & & $-31.5_{-37.2}^{+77.8}$ & & $4.87\pm0.69$  \\ 
        PG 1351+695 & & All & & $5297_{-10}^{+10}$ & & $1871_{-6}^{+6}$ & & $4478_{-3}^{+3}$ & & $1583_{-10}^{+12}$ & & $18.6_{-2.0}^{+2.3}$ & & $16.7_{-1.9}^{+4.1}$ & & $19.9_{-1.0}^{+1.0}$ & & $11.7_{-5.3}^{+6.0}$ & & $0.50\pm0.06$  \\ 
        PG 1501+106 & & All & & $5006_{-32}^{+50}$ & & $2490_{-47}^{+47}$ & & $4152_{-5}^{+8}$ & & $1986_{-17}^{+14}$ & & $26.0_{-2.2}^{+2.0}$ & & $24.0_{-6.1}^{+4.6}$ & & $22.0_{-0.4}^{+0.5}$ & & $113.7_{-66.2}^{+10.7}$ & & $1.03\pm0.14$  \\ 
        & &  1  & & $5081_{-44}^{+26}$ & & $2528_{-43}^{+50}$ & & $3855_{-14}^{+13}$ & & $2291_{-19}^{+18}$ & & $24.1_{-8.8}^{+10.8}$ & & $6.1_{-3.2}^{+17.9}$ & & $5.0_{-1.4}^{+1.3}$ & & $117.0_{-43.7}^{+7.5}$ & & $1.20\pm0.11$  \\ 
        & &  3  & & $5002_{-53}^{+34}$ & & $2454_{-44}^{+47}$ & & $4141_{-5}^{+5}$ & & $1384_{-12}^{+10}$ & & $24.8_{-1.4}^{+1.6}$ & & $24.9_{-1.6}^{+1.8}$ & & $42.9_{-4.7}^{+9.9}$ & & $25.9_{-1.8}^{+3.0}$ & & $1.03\pm0.09$  \\ 
        & &  4  & & $4718_{-41}^{+43}$ & & $2378_{-45}^{+56}$ & & $2945_{-18}^{+17}$ & & $1156_{-14}^{+14}$ & & $32.2_{-4.0}^{+3.6}$ & & $24.8_{-7.5}^{+16.8}$ & & $43.7_{-2.7}^{+3.7}$ & & $108.8_{-23.0}^{+5.5}$ & & $0.86\pm0.06$  \\ 
        PG 1534+580 & & All & & $4217_{-155}^{+751}$ & & $3180_{-47}^{+84}$ & & $2362_{-7}^{+11}$ & & $1142_{-20}^{+20}$ & & $26.0_{-8.6}^{+5.7}$ & & $35.5_{-21.6}^{+3.5}$ & & $25.4_{-1.4}^{+2.0}$ & & $28.2_{-14.0}^{+9.3}$ & & $0.48\pm0.03$  \\ 
        PG 1613+658 & & All & & $10269_{-167}^{+773}$ & & $3927_{-16}^{+13}$ & & $6762_{-10}^{+9}$ & & $3504_{-14}^{+13}$ & & $51.2_{-6.0}^{+5.2}$ & & $55.8_{-22.9}^{+10.5}$ & & $52.4_{-2.8}^{+3.4}$ & & $27.6_{-13.0}^{+13.0}$ & & $8.62\pm1.04$  \\ 
        & &  1  & & $9866_{-132}^{+118}$ & & $3907_{-13}^{+14}$ & & $12817_{-30}^{+36}$ & & $4654_{-16}^{+18}$ & & $49.7_{-10.2}^{+9.2}$ & & $42.9_{-10.9}^{+31.1}$ & & $79.3_{-2.8}^{+5.9}$ & & $91.8_{-32.7}^{+10.0}$ & & $9.36\pm1.16$  \\ 
        & &  2  & & $10998_{-221}^{+193}$ & & $3926_{-14}^{+13}$ & & $11469_{-48}^{+25}$ & & $4196_{-29}^{+20}$ & & $46.3_{-7.3}^{+7.3}$ & & $45.9_{-13.3}^{+11.4}$ & & $48.3_{-3.8}^{+5.0}$ & & $-191.1_{-11.0}^{+223.2}$ & & $8.34\pm0.85$  \\ 
        \enddata
        \tablecomments{The line widths of the rms spectra of PG~0049+171, PG~0947+396, PG~1202+281, PG~1351+640, 
        and PG~1501+106 are measured from their continuum-cleaned rms spectra (See Section \ref{sec:meanrms}). 
        The broadening caused by the instrument and seeing has been corrected. 
        For 5100$\rm \mathring{A}$ luminosity, the galactic extinction 
        \citep{Schlafly2011} is corrected, but the host-galaxy contamination is not removed. 
        The time lags in the table are in the rest frame.}
        \end{deluxetable*}

        \begin{deluxetable*}{ccccccccccc}
            \setlength{\tabcolsep}{0.7mm}
            \tablecaption{Virial Products and Masses of the Black Holes \label{tab:BHmasses}}
            \tablehead{
            \multicolumn{1}{c}{Target} & & \multicolumn{1}{c}{Season} & & \multicolumn{1}{c}{VP (Mean)} & & \multicolumn{3}{c}{BH mass (RMS)} & & \colhead{Note} \\
            \cmidrule(r){7-9}
             & &  & &  \colhead{$R_{\rm H\beta}V_{\rm FWHM}^{2}/G$} & & \colhead{$1.12\times R_{\rm H\beta} V_{\rm FWHM}^{2}/G$} & & \colhead{$4.47\times R_{\rm H\beta} \sigma_{\rm line}^{2}/G$} & &  \\
            & & & & ($\times10^7 M_{\odot}$)& & ($\times10^7 M_{\odot}$)& & ($\times10^7 M_{\odot}$)}
            \startdata
            PG 0007+106 & &  All  & & $14.16_{-0.62}^{+0.68}$ & & $13.18_{-0.56}^{+0.61}$ & & $7.03_{-0.31}^{+0.34}$ & & \checkmark\\
            & &  1  & & $13.10_{-2.23}^{+3.50}$ & & $14.84_{-2.53}^{+3.97}$ & & $7.31_{-1.26}^{+1.96}$ & & \\  
            & &  2  & & $10.56_{-7.33}^{+5.02}$ & & $9.18_{-6.38}^{+4.37}$ & & $6.07_{-4.22}^{+2.89}$  & & \\ 
            & &  3  & & $8.27_{-1.52}^{+1.45}$ & & $6.27_{-1.15}^{+1.10}$ & & $3.84_{-0.71}^{+0.68}$  & & \\ 
            & &  4  & & $10.86_{-1.03}^{+1.11}$ & & $9.97_{-0.94}^{+1.01}$ & & $4.40_{-0.42}^{+0.45}$  & & \\ 
            PG 0049+171 & &  All  & & $14.02_{-0.94}^{+2.95}$ & & $7.13_{-0.46}^{+0.60}$ & & $4.91_{-0.34}^{+0.42}$ & & \\ 
            & &  1  & & $13.92_{-2.13}^{+3.58}$ & & $7.18_{-1.09}^{+1.24}$ & & $4.49_{-0.70}^{+0.79}$ & & \\ 
            & &  2  & & $11.83_{-2.94}^{+2.55}$ & & $5.77_{-1.41}^{+0.68}$ & & $5.07_{-1.25}^{+0.61}$ & & \\ 
            & &  3  & & $16.69_{-3.11}^{+23.30}$ & & $3.64_{-0.67}^{+5.06}$ & & $3.95_{-0.75}^{+5.49}$ & &\\ 
            & &  4  & & $7.85_{-0.78}^{+2.02}$ & & $5.79_{-0.57}^{+0.70}$ & & $2.95_{-0.31}^{+0.37}$ & & \checkmark\\ 
            PG 0923+129 & &  All  & & $0.74_{-0.21}^{+0.38}$ & & $0.62_{-0.18}^{+0.32}$ & & $0.81_{-0.23}^{+0.42}$  & & \checkmark\\ 
            PG 0947+396 & &  All  & & $22.79_{-1.18}^{+2.19}$ & & $9.35_{-0.45}^{+0.91}$ & & $14.06_{-0.76}^{+1.40}$  & & \\  
            & &  1  & & $8.47_{-4.13}^{+3.72}$ & & $4.91_{-2.40}^{+2.16}$ & & $3.80_{-1.86}^{+1.67}$  & & \\ 
            & &  2  & & $21.70_{-0.63}^{+0.90}$ & & $22.87_{-0.65}^{+0.74}$ & & $10.22_{-0.33}^{+0.36}$  & & \checkmark\\  
            & &  3  & & $19.90_{-11.13}^{+9.79}$ & & $12.68_{-7.08}^{+6.24}$ & & $6.16_{-3.44}^{+3.03}$  & & \\  
            & &  4  & & $19.16_{-2.64}^{+6.53}$ & & $24.52_{-3.37}^{+8.33}$ & & $8.87_{-1.26}^{+3.03}$  & & \\ 
            PG 1001+054 & &  All  & & $3.65_{-0.23}^{+0.32}$ & & $5.35_{-0.34}^{+0.49}$ & & $10.73_{-1.17}^{+1.15}$  & & \checkmark\\ 
            & &  3  & & $3.50_{-0.38}^{+0.61}$ & & $1.19_{-0.18}^{+0.26}$ & & $3.99_{-1.81}^{+1.45}$ & &\\ 
            PG 1048+342 & &  All  & & $6.07_{-0.61}^{+0.41}$ & & $3.71_{-0.34}^{+0.24}$ & & $4.44_{-0.42}^{+0.31}$  & & \checkmark\\ 
            & &  1  & & $4.94_{-0.86}^{+0.99}$ & & $5.68_{-0.98}^{+1.14}$ & & $5.14_{-0.90}^{+1.06}$ & &\\ 
            PG 1100+772 & &  All  & & $35.86_{-0.91}^{+1.99}$ & & $154.05_{-3.78}^{+8.39}$ & & $78.13_{-4.72}^{+5.44}$  & & \checkmark\\  
            PG 1202+281 & &  All  & & $34.99_{-1.02}^{+1.27}$ & & $26.25_{-0.76}^{+0.96}$ & & $9.80_{-0.46}^{+0.44}$  & & \checkmark\\ 
            & &  1  & & $22.76_{-1.81}^{+1.66}$ & & $15.59_{-1.36}^{+1.17}$ & & $10.85_{-0.90}^{+0.82}$ & & \\
            & &  3  & & $32.08_{-2.05}^{+2.21}$ & & $22.10_{-8.77}^{+4.87}$ & & $14.39_{-1.50}^{+1.40}$ & & \\ 
            & &  4  & & $30.26_{-6.18}^{+6.35}$ & & $18.51_{-3.79}^{+3.89}$ & & $11.26_{-2.36}^{+2.41}$ & & \\ 
            PG 1211+143 & &  All  & & $3.81_{-0.56}^{+0.53}$ & & $2.14_{-0.24}^{+0.21}$ & & $2.25_{-0.28}^{+0.25}$ & & \checkmark\\ 
            PG 1310-108 & &  All  & & $3.25_{-1.47}^{+1.53}$ & & $1.64_{-0.22}^{+0.22}$ & & $1.33_{-0.22}^{+0.20}$ & & \checkmark\\ 
            PG 1351+640 & &  All  & & $84.95_{-3.18}^{+3.38}$ & & $7.60_{-0.25}^{+0.27}$ & & $15.24_{-0.64}^{+0.66}$ & & \checkmark\\  
            PG 1351+695 & &  All  & & $10.88_{-0.55}^{+0.57}$ & & $8.71_{-0.44}^{+0.46}$ & & $4.35_{-0.23}^{+0.24}$ & & \checkmark\\ 
            PG 1501+106 & &  All  & & $10.76_{-0.25}^{+0.32}$ & & $8.29_{-0.16}^{+0.19}$ & & $7.57_{-0.20}^{+0.20}$ & & \\ 
            & &  1  & & $2.51_{-0.73}^{+0.66}$ & & $1.61_{-0.47}^{+0.42}$ & & $2.28_{-0.66}^{+0.60}$ & & \\  
            & &  3  & & $20.95_{-2.33}^{+4.84}$ & & $16.08_{-1.76}^{+3.71}$ & & $7.17_{-0.79}^{+1.66}$ & & \checkmark\\  
            & &  4  & & $18.97_{-1.22}^{+1.65}$ & & $8.28_{-0.52}^{+0.71}$ & & $5.09_{-0.34}^{+0.45}$ & & \\ 
            PG 1534+580 & &  All  & & $8.80_{-0.82}^{+3.21}$ & & $3.09_{-0.18}^{+0.24}$ & & $2.89_{-0.19}^{+0.25}$ & & \checkmark\\
            PG 1613+658 & &  All  & & $107.82_{-6.74}^{+17.70}$ & & $52.36_{-2.79}^{+3.42}$ & & $56.13_{-3.02}^{+3.69}$ & & \checkmark\\
            & &  1  & & $150.70_{-6.63}^{+11.72}$ & & $284.85_{-10.01}^{+21.14}$ & & $149.95_{-5.32}^{+11.16}$ & & \\ 
            & &  2  & & $114.07_{-10.16}^{+12.36}$ & & $138.95_{-11.11}^{+14.25}$ & & $74.24_{-5.99}^{+7.64}$ & & \\ 
            \enddata
            \tablecomments{The VP are calculated from the FWHM of mean spectra. BH masses are estimated using the FWHM and the sigma of RMS spectra. 
            The propagation errors are from line widths and time lags, and the uncertainties of $f$ factor is not considered here. The 
            last column notes the data set we preferred for the BH mass measurement. 
            }
            \end{deluxetable*}

\subsection{Line width measurements} 
\label{sec:line widths}

\begin{figure*}[ht!]
    \centering
    \includegraphics[scale=0.50]{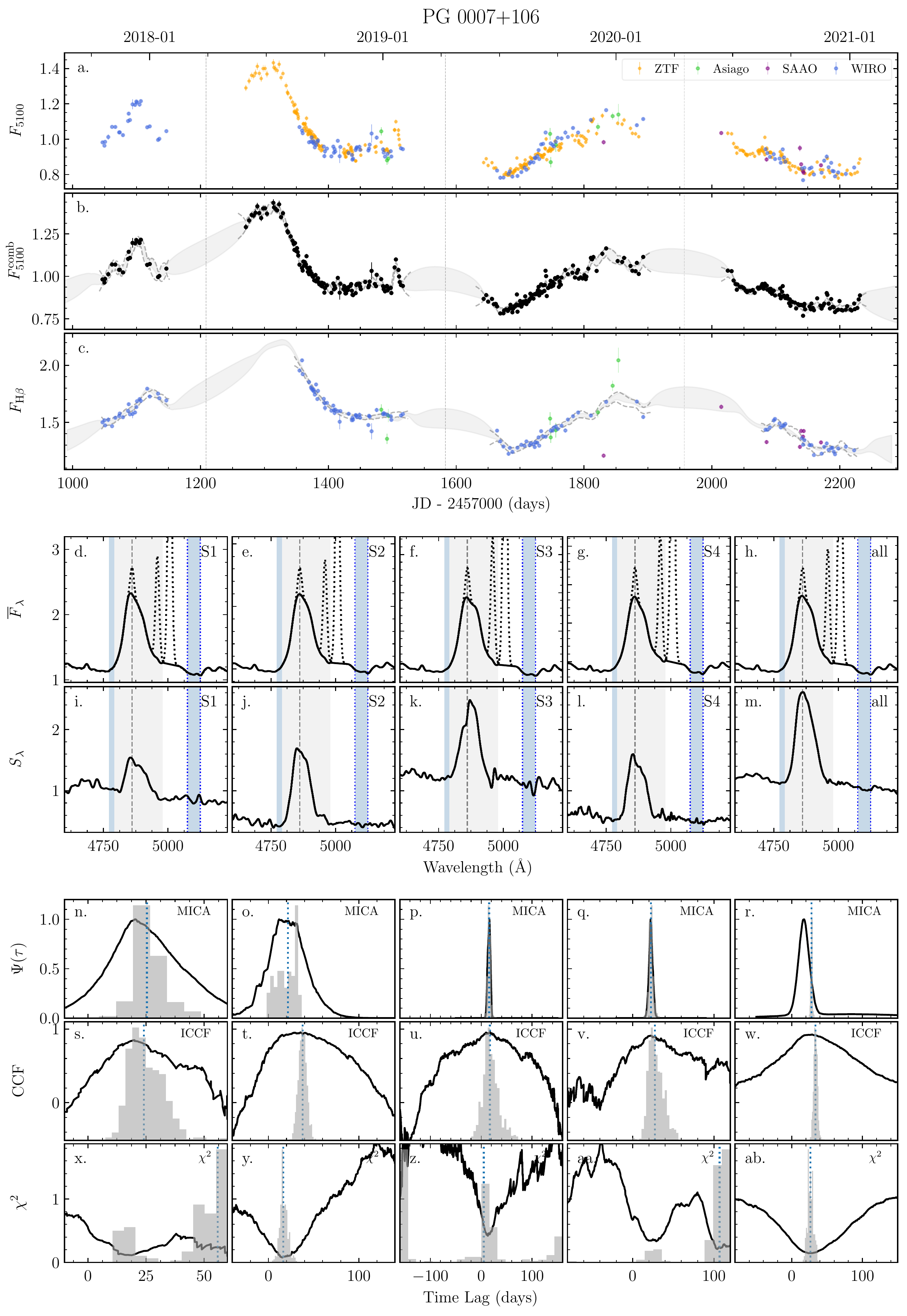} 
    \caption{Time-series analysis of PG 0007+106. 
    Panel a, b, and c are the scaled continuum, combined continuum, and
    H$\beta$ light curves. The units are 
    $\rm 10^{-15}\ erg\ s^{-1}$ $\rm cm^{-2}$ $\rm \mathring{A}^{-1}$ for Panels a and b,
    and $\rm 10^{-13}\ erg\ s^{-1}$ $\rm cm^{-2}$ for Panel c. 
    The grey dotted lines separate different seasons.
    The grey shadow and grey dashed lines are the MICA 
    reconstructions for the whole light curve and single seasons, respectively.
    Panels d -- h (i -- m) are the mean (rms) spectra 
    of the seasons and the entire light curve in the rest frames. 
    The black dashed lines are the narrow-line-subtracted mean 
    spectra. The grey and blue shades mark the integration and 
    background windows for H$\beta$ fluxes, 
    and the two blue dotted lines mark the 5100 \AA\ continuum window. 
    The units in Panels d -- h (i -- m) are 
    $\rm 10^{-15}\ erg\ s^{-1}$ $\rm cm^{-2}$ $\rm \mathring{A}^{-1}$ 
    ($\rm 10^{-16}\ erg\ s^{-1}$ $\rm cm^{-2}$ $\rm \mathring{A}^{-1}$). 
    Panels n -- ab are the MICA, ICCF and $\chi^2$ results for the corresponding 
    seasons and the entire light curve (in observed frame).
    The grey histograms are the distributions of the 
    centroid lags obtained from MICA (CCCDs from ICCF or lag distributions 
    from $\chi^2$ method) in Panels n -- r (s -- w or x -- ab).
    The blue dotted lines are the median of the distributions. 
    The error bars shown in the light curves do not include the systematic 
    uncertainties in Table \ref{tab:syserr_info} (they are used in the time-series analysis in Section \ref{sec:time_series_analysis}, 
    see also Section \ref{sec:light curves}). \label{0007lc}}
    \end{figure*}

\begin{figure*}[ht!]
    \centering
    \includegraphics[scale=0.52]{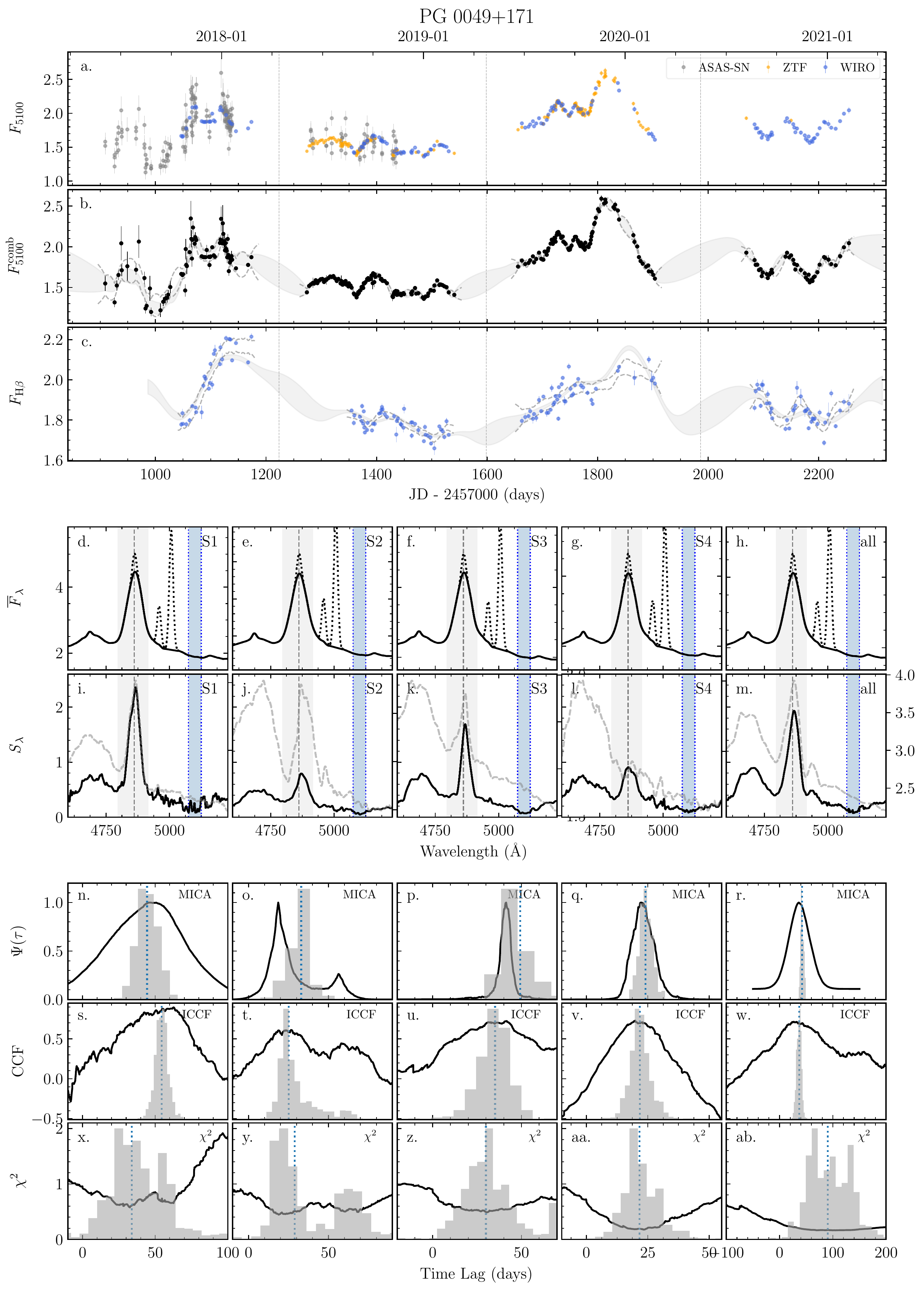} \\
    \caption{Time-series analysis of PG 0049+171. The grey and black dashed lines 
    in Panels i -- m are the original and continuum-cleaned rms spectra (see more details 
    in Section \ref{sec:meanrms}). 
    The meanings of the other panels, lines, 
    and histograms are the same as Fig.~\ref{0007lc}. \label{0049lc}}
    \end{figure*}

\begin{figure*}[ht!]
    \centering
    \includegraphics[scale=0.52]{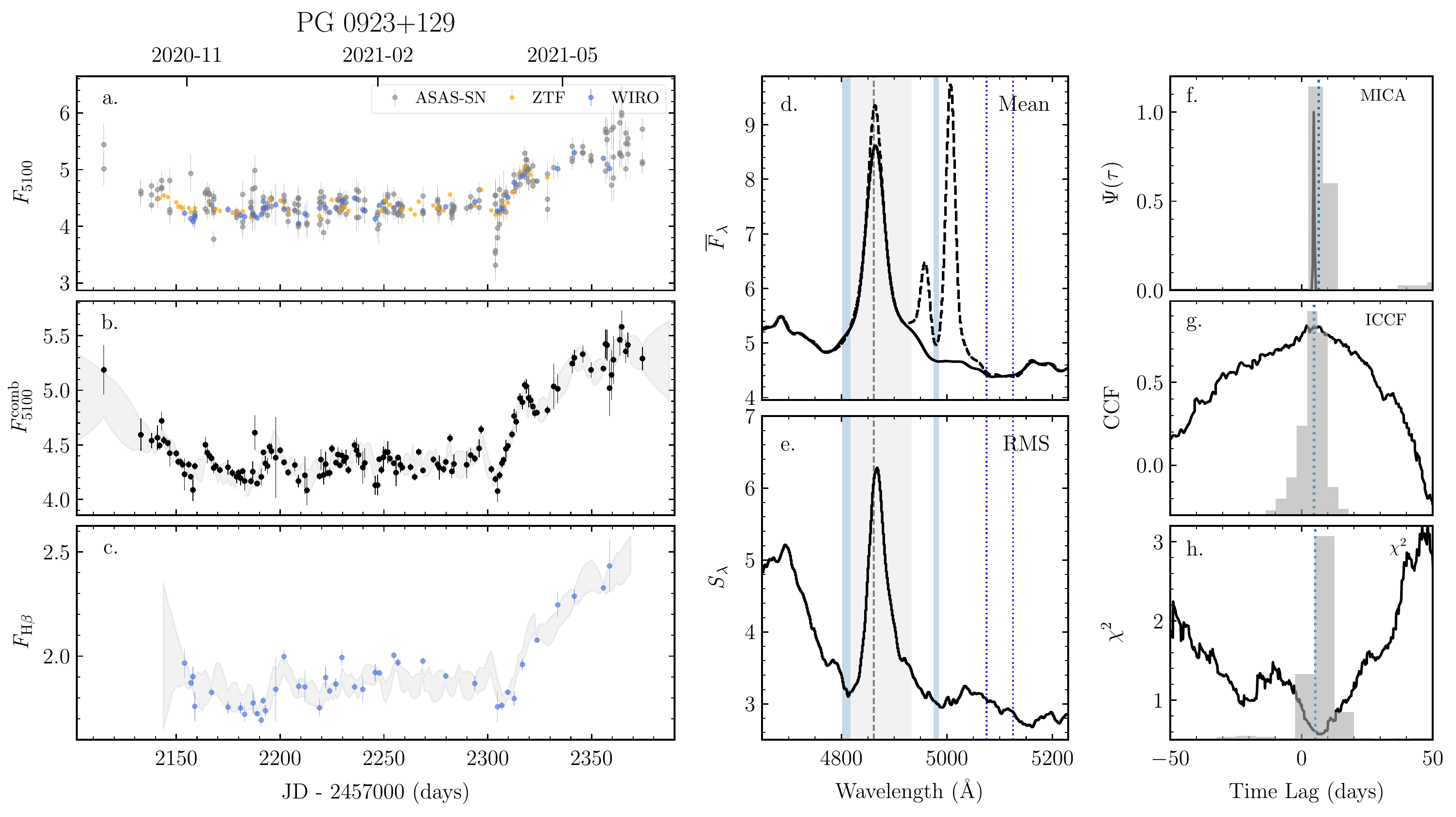}
    \caption{Time-series analysis of PG 0923+129. The meanings of the panels, lines, 
    and histograms are the same as Fig.~\ref{0007lc}. \label{0923lc}}
    \end{figure*}

\begin{figure*}[ht!]
    \centering
    \includegraphics[scale=0.52]{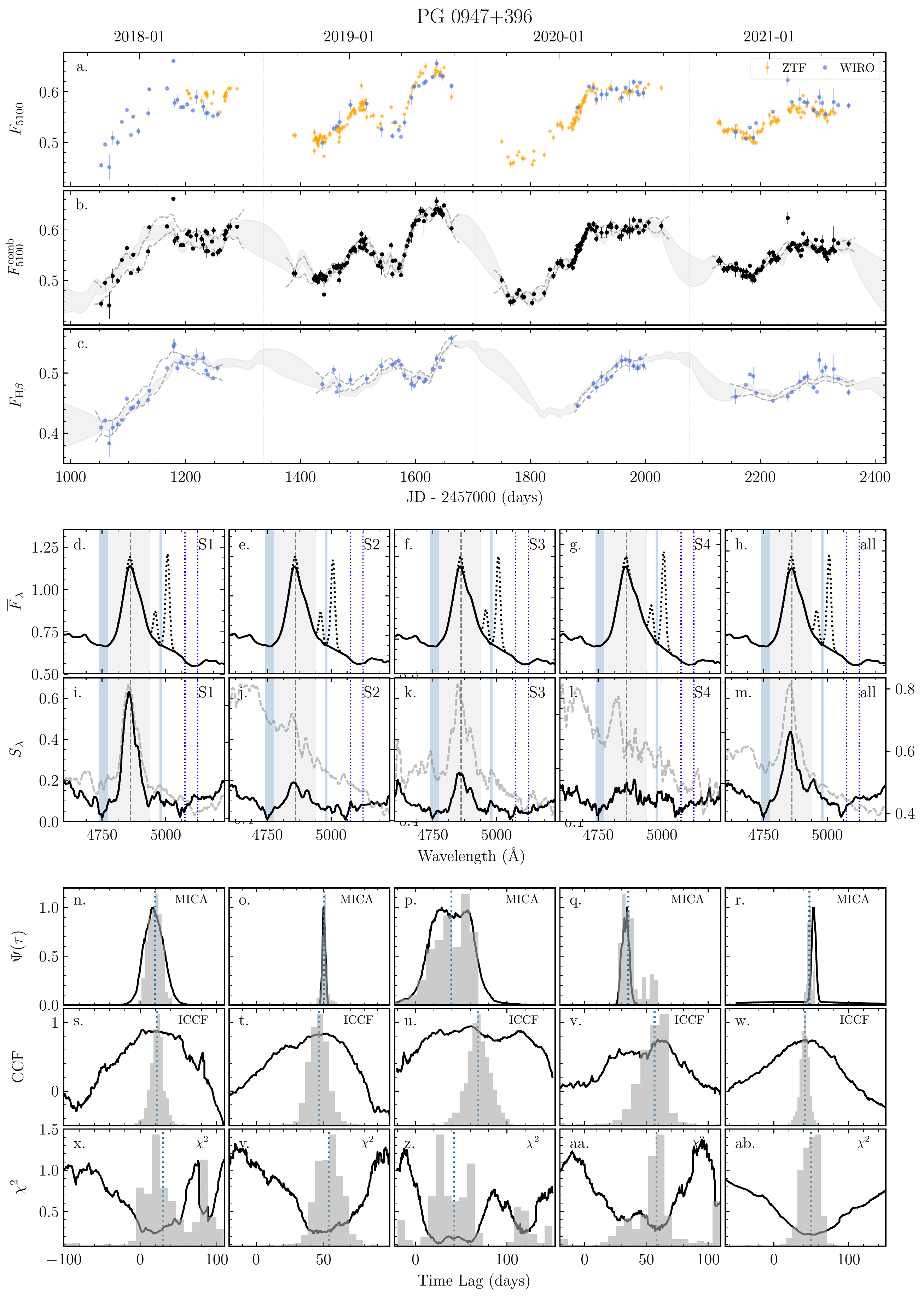}
    \caption{Time-series analysis of PG 0947+396. The grey and black dashed lines 
    in Panels i -- m are the original and continuum-cleaned rms spectra (see more details 
    in Section \ref{sec:meanrms}). 
    The meanings of the other panels, lines, 
    and histograms are the same as Fig.~\ref{0007lc}. \label{0947lc}}
    \end{figure*}

\begin{figure*}[ht!]
    \centering
    \includegraphics[scale=0.52]{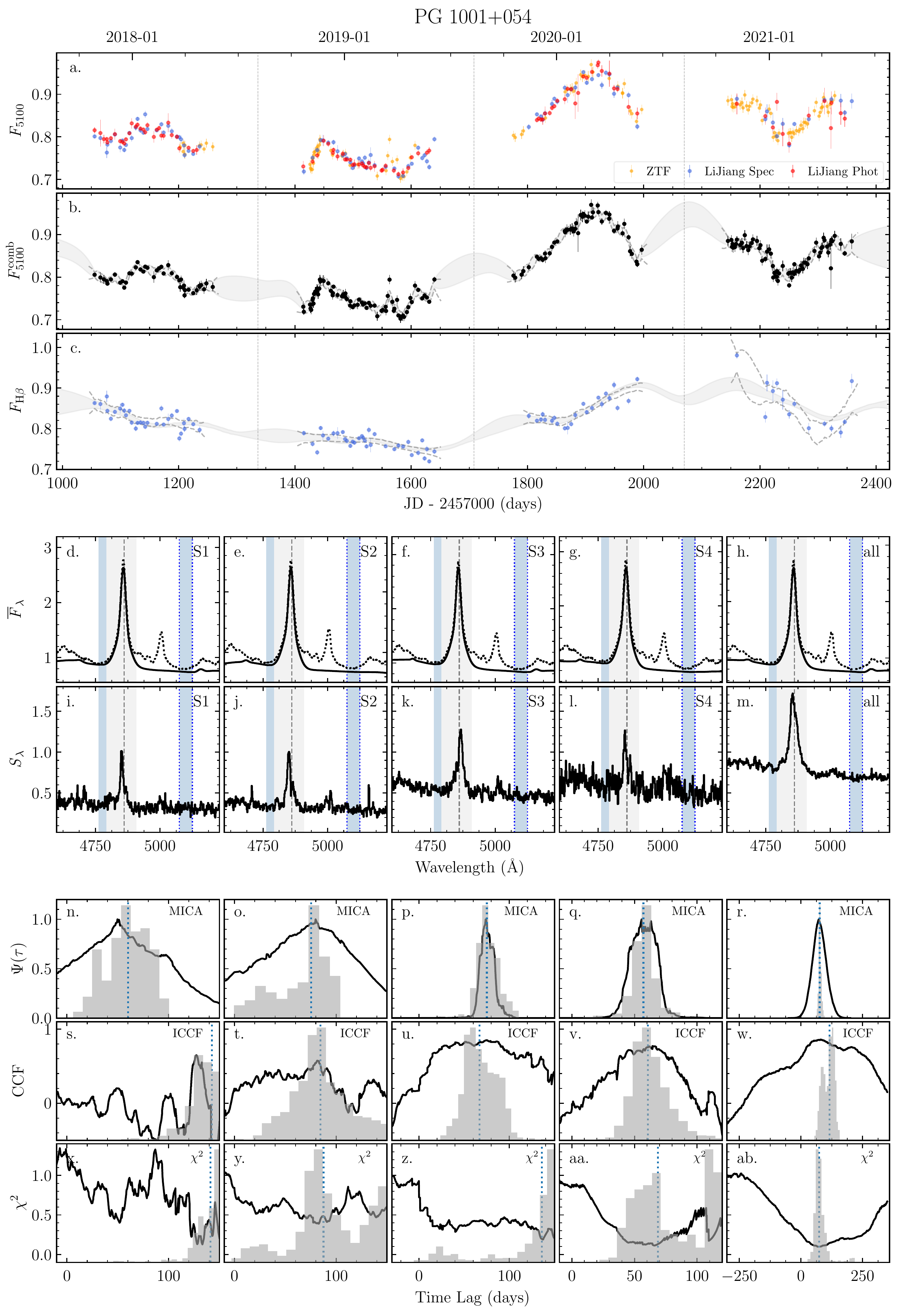} \\
    \caption{Time-series analysis of PG 1001+054. The black dotted and solid lines in Panels 
    d -- h are the original and cleaned (e.g., Fe {\sc ii}, He {\sc ii}, 
    narrow H$\beta$ and [O {\sc iii}]$\lambda\lambda$4959,5007) mean spectra. 
    The meanings of the other panels, lines, 
    and histograms are the same as Fig.~\ref{0007lc}. \label{1001lc}}
    \end{figure*}

\begin{figure*}[ht!]
    \centering
    \includegraphics[scale=0.52]{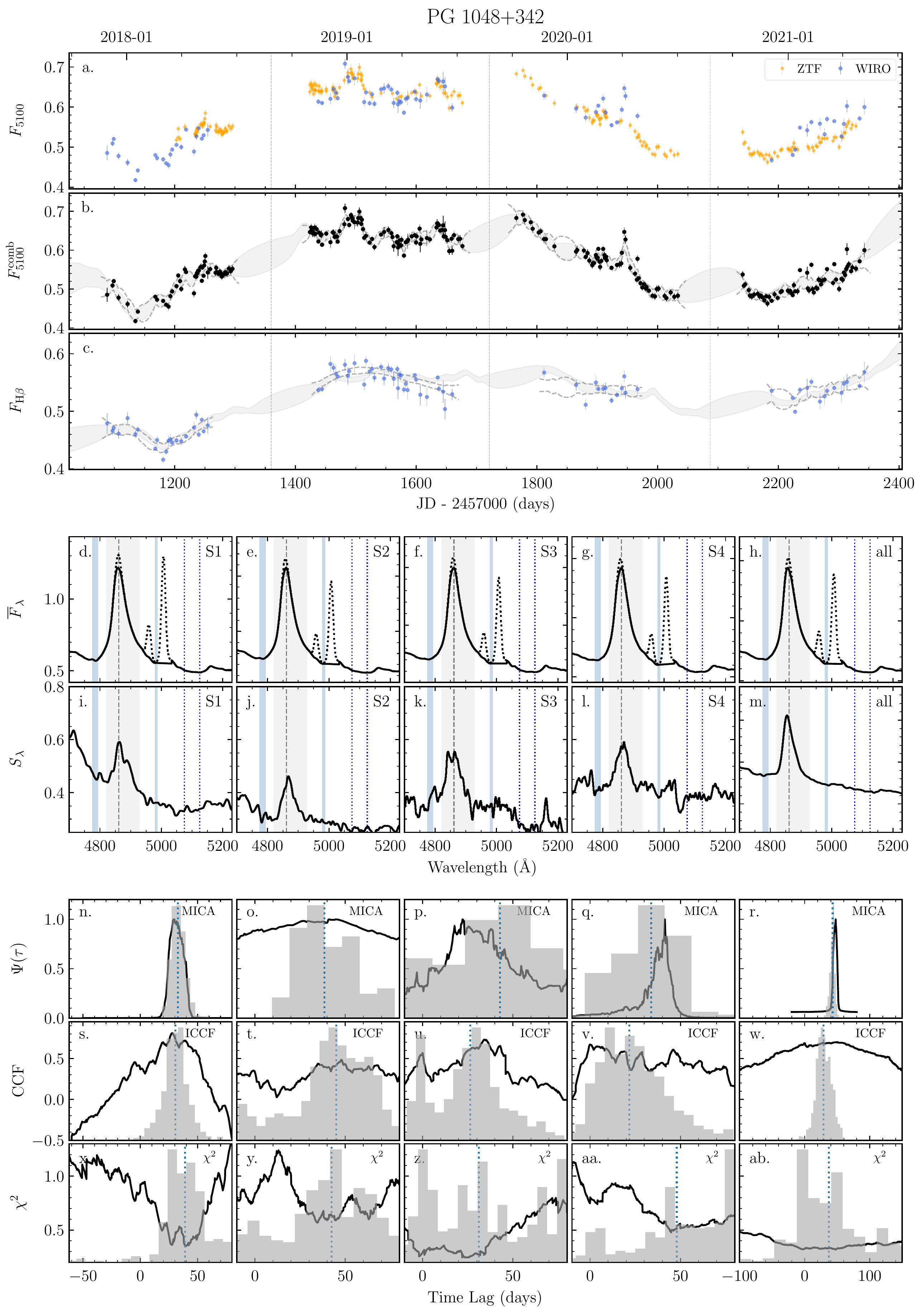}
    \caption{Time-series analysis of PG 1048+342. The meanings of the panels, lines, 
    and histograms are the same as Fig.~\ref{0007lc}. \label{1048lc}}
    \end{figure*}

\begin{figure*}[ht!]
    \centering
    \includegraphics[scale=0.52]{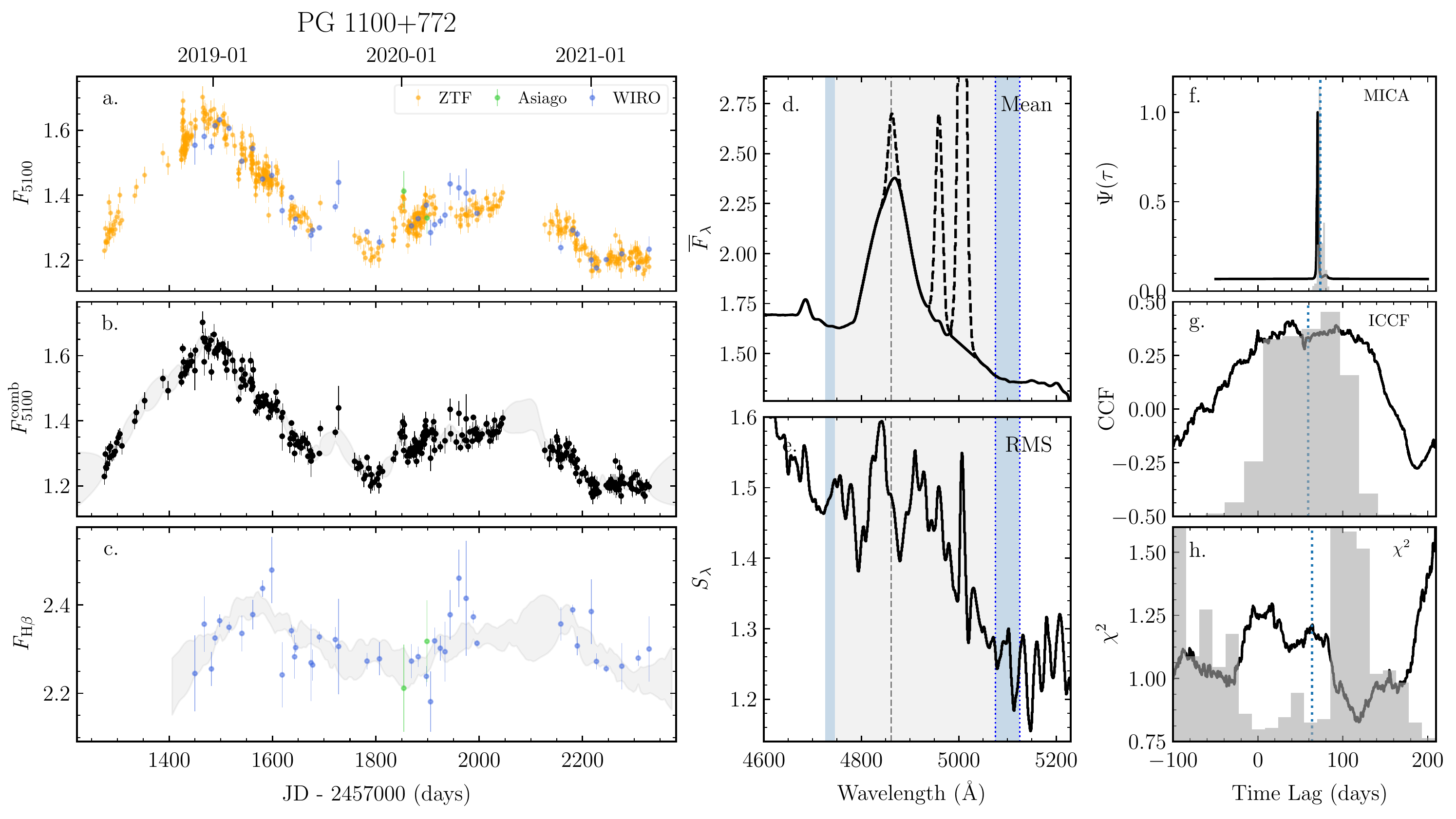} \\
    \caption{Time-series analysis of PG 1100+772. The meanings of the panels, lines, 
    and histograms are the same as Fig.~\ref{0007lc}. \label{1100lc}}
    \end{figure*}

\begin{figure*}[ht!]
    \centering
    \includegraphics[scale=0.52]{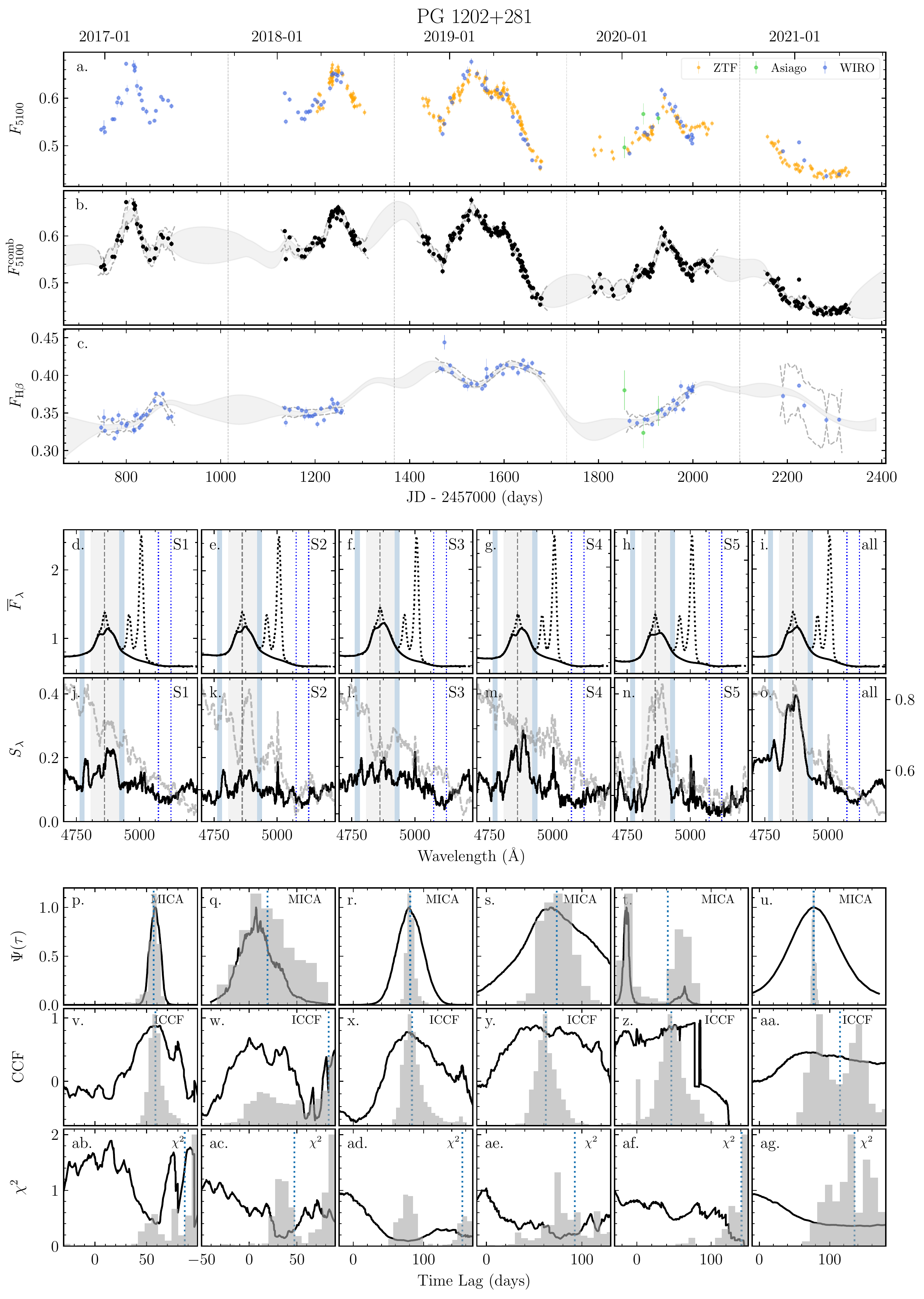} 
    \caption{Time-series analysis of PG 1202+281. 
    The meanings of the panels, lines, 
    and histograms are the same as Fig.~\ref{0007lc}.
    In the narrow-line-correct mean spectra  in 
    Panel d -- i.\label{1202lc}}
    \end{figure*}

\begin{figure*}[ht!]
    \centering
    \includegraphics[scale=0.52]{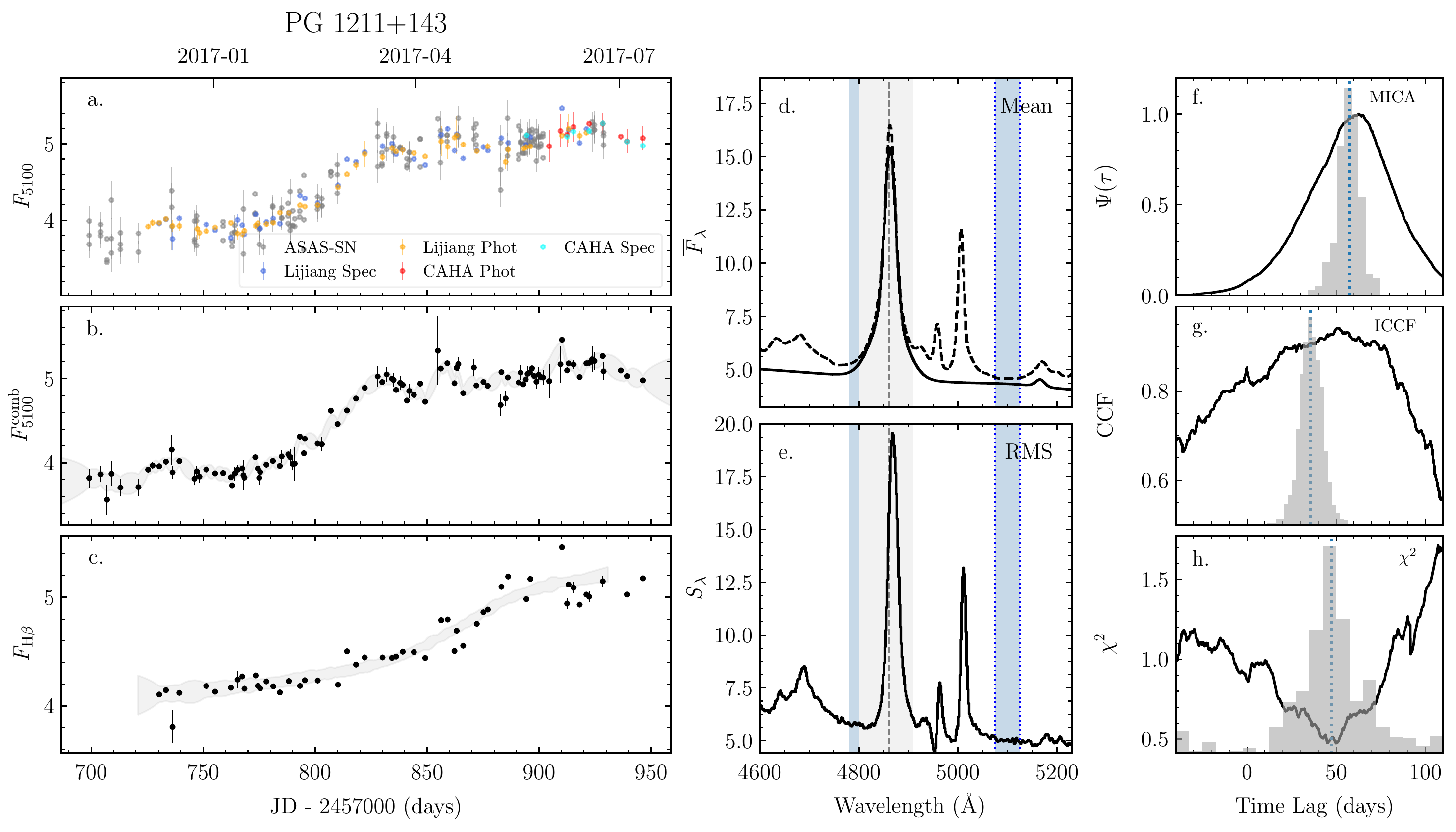} 
    \caption{Time-series analysis of PG 1211+143. The black dotted and solid lines in Panel 
    d are the original and cleaned (e.g., Fe {\sc ii}, He {\sc ii}, 
    narrow H$\beta$ and [O {\sc iii}]$\lambda\lambda$4959,5007) mean spectra. 
    The meanings of the other panels, lines, 
    and histograms are the same as Fig.~\ref{0007lc}. Panel c is the 
    combined H$\beta$ light curve from Lijiang and CAHA.\label{1211lc}}
    \end{figure*}

\begin{figure*}[ht!]
    \centering
    \includegraphics[scale=0.52]{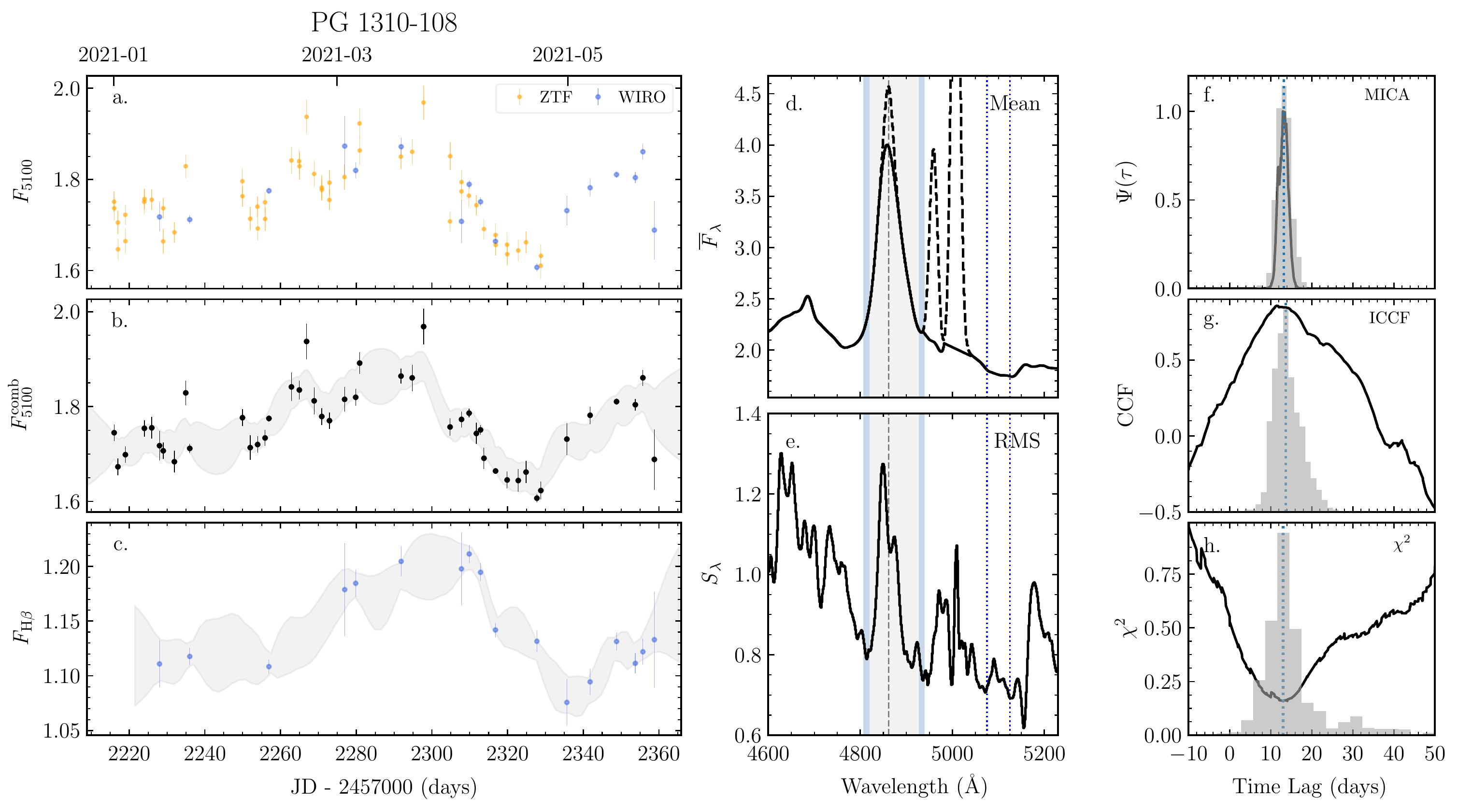} 
    \caption{Time-series analysis of PG 1310$-$108. The meanings of the panels, lines, 
    and histograms are the same as Fig.~\ref{0007lc}.\label{1310lc}}
    \end{figure*}

\begin{figure*}[ht!]
    \centering
    \includegraphics[scale=0.52]{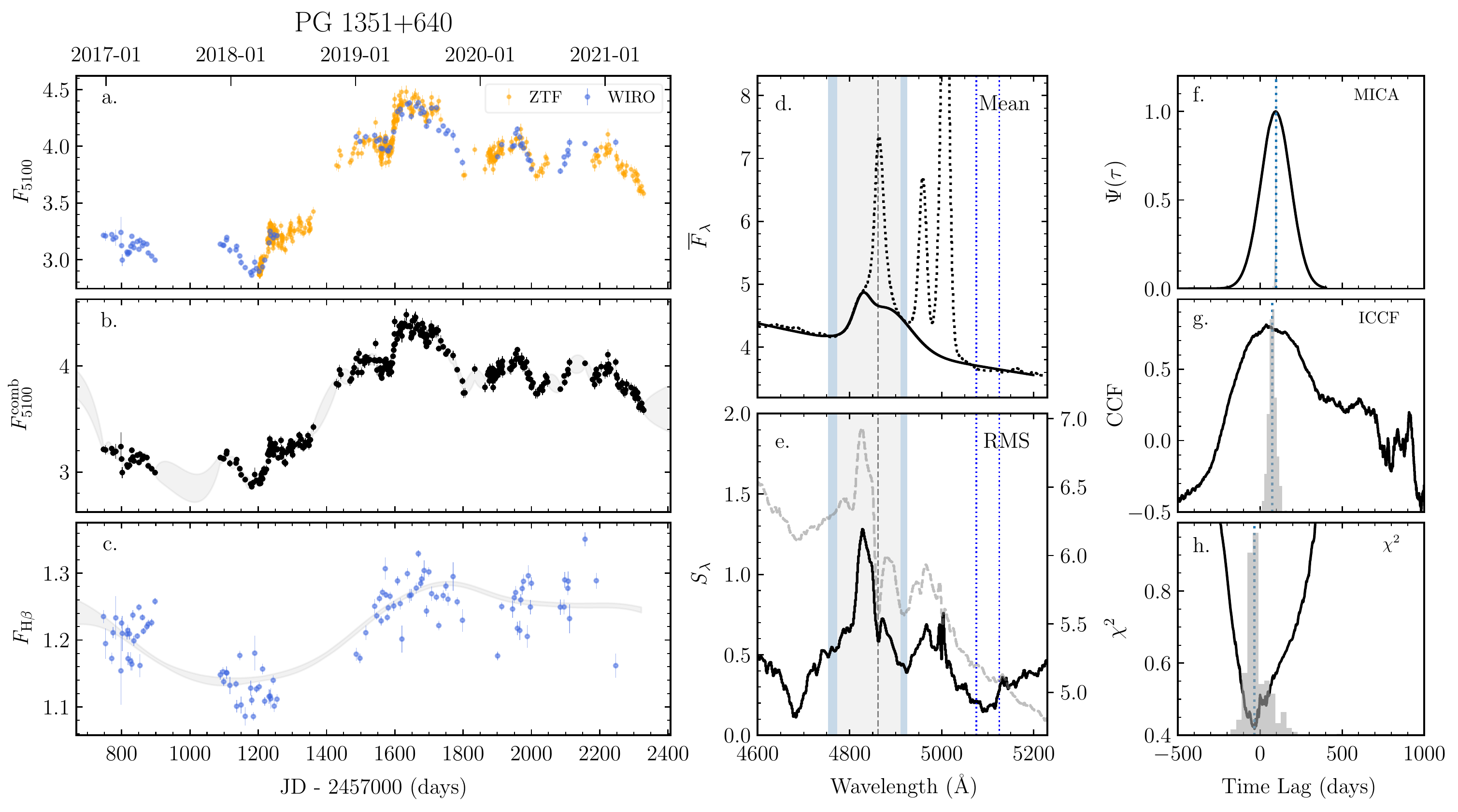} \\
    \caption{Time-series analysis of PG 1351+640. 
    The black dotted and solid lines in Panels 
    d -- h are the original and cleaned (e.g., Fe {\sc ii}, He {\sc ii}, 
    narrow H$\beta$ and [O {\sc iii}]$\lambda\lambda$4959,5007) mean spectra. 
    The meanings of the other panels, lines, 
    and histograms are the same as Fig.~\ref{0049lc}. \label{1351640lc}}
    \vspace{2cm}
    \includegraphics[scale=0.52]{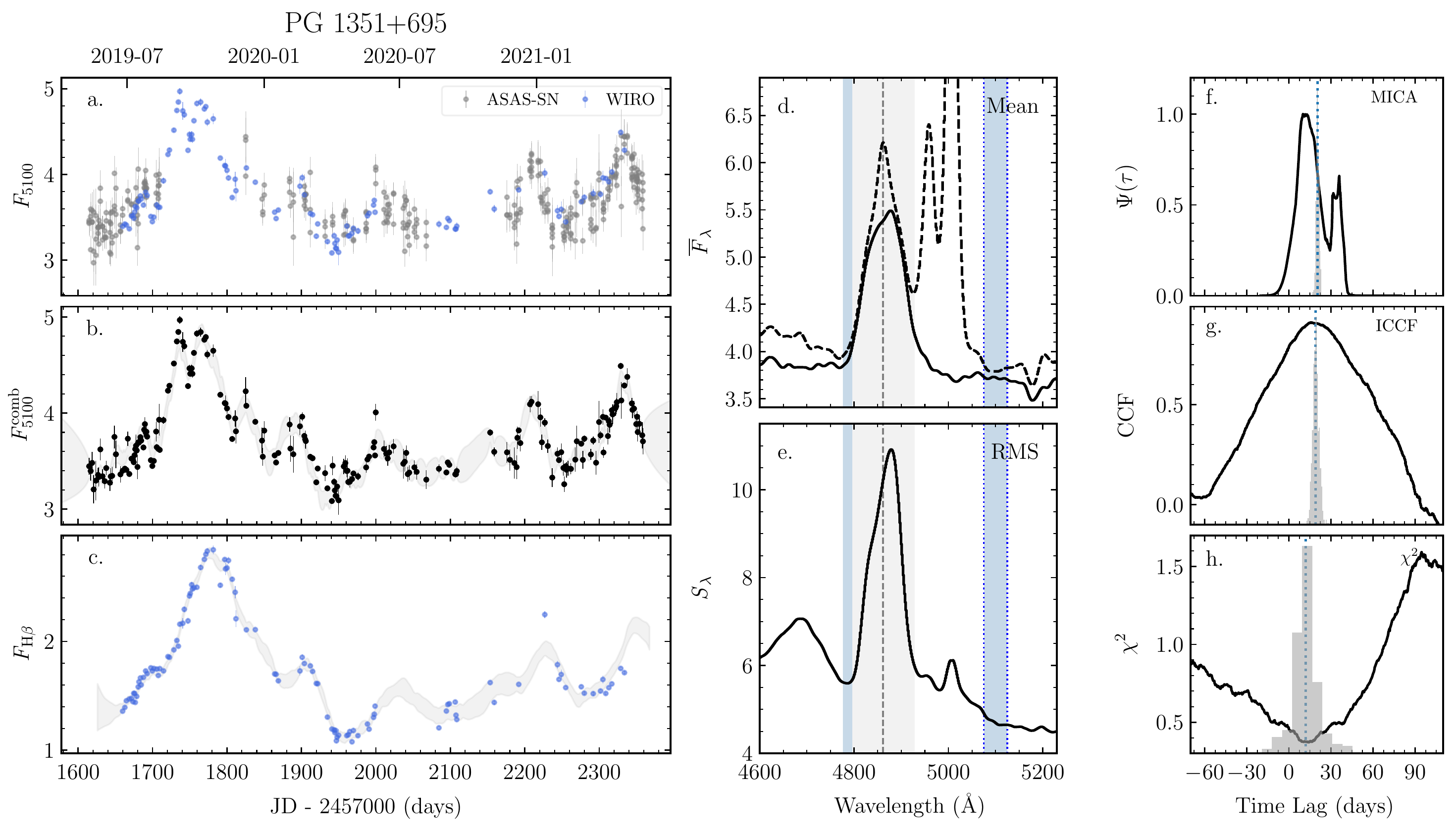} \\
    \caption{The light curves, mean spectra and RMS, time lags results of PG 1351+695. The units are the same as Fig.~\ref{0007lc}. \label{1351695lc}}
    \end{figure*}

\begin{figure*}[ht!]
    \centering
    \includegraphics[scale=0.52]{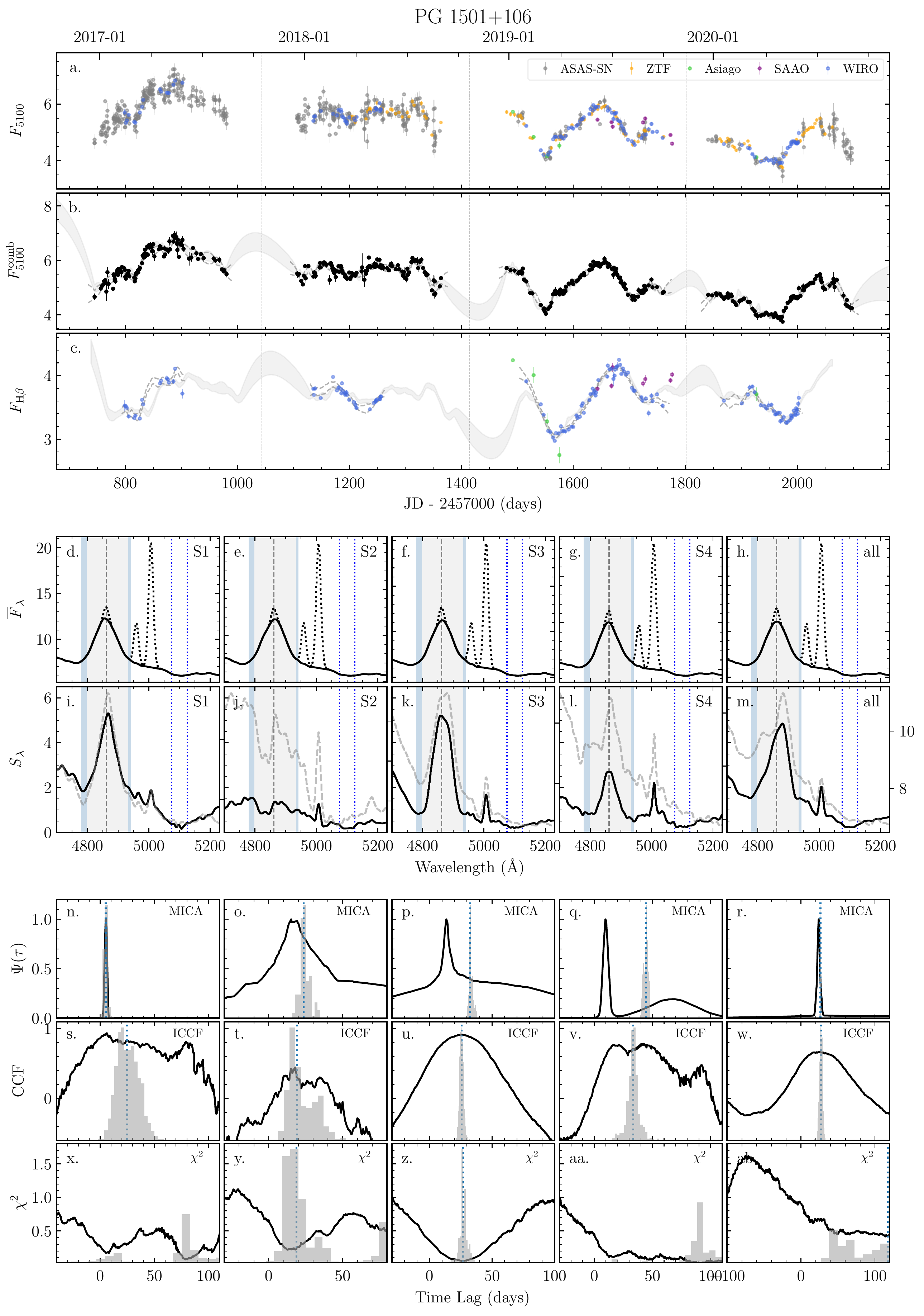} 
    \caption{Time-series analysis of PG 1501+106. The meanings of the panels, lines, 
    and histograms are the same as Fig.~\ref{0049lc}.  \label{1501lc}}
    \end{figure*}

\begin{figure*}[ht!]
    \centering
    \includegraphics[scale=0.52]{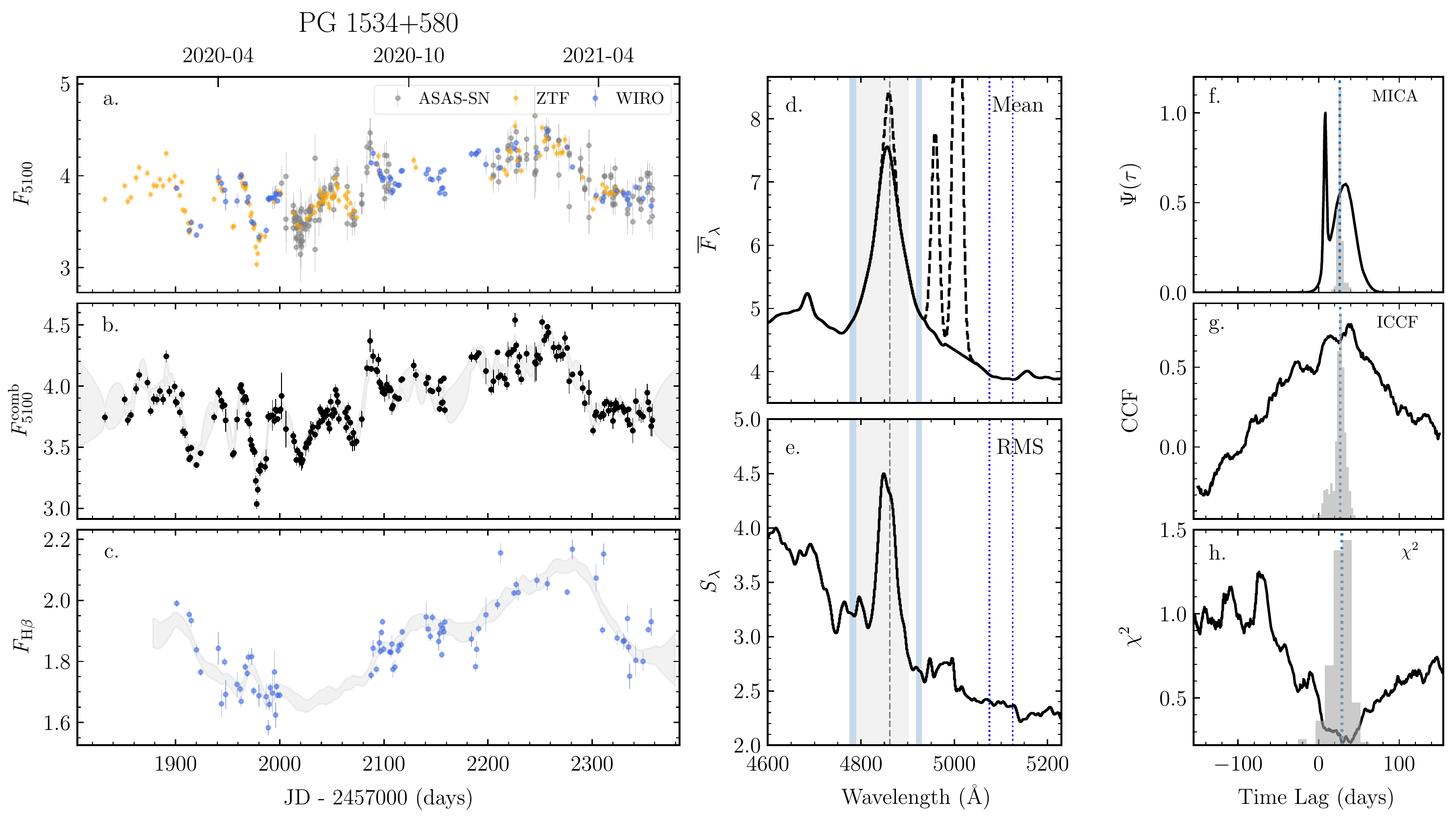} 
    \caption{Time-series analysis of PG 1534+580. The meanings of the panels, lines, 
    and histograms are the same as Fig.~\ref{0007lc}.  \label{1534lc}}
    \end{figure*}

\begin{figure*}[ht!]
    \centering
    \includegraphics[scale=0.52]{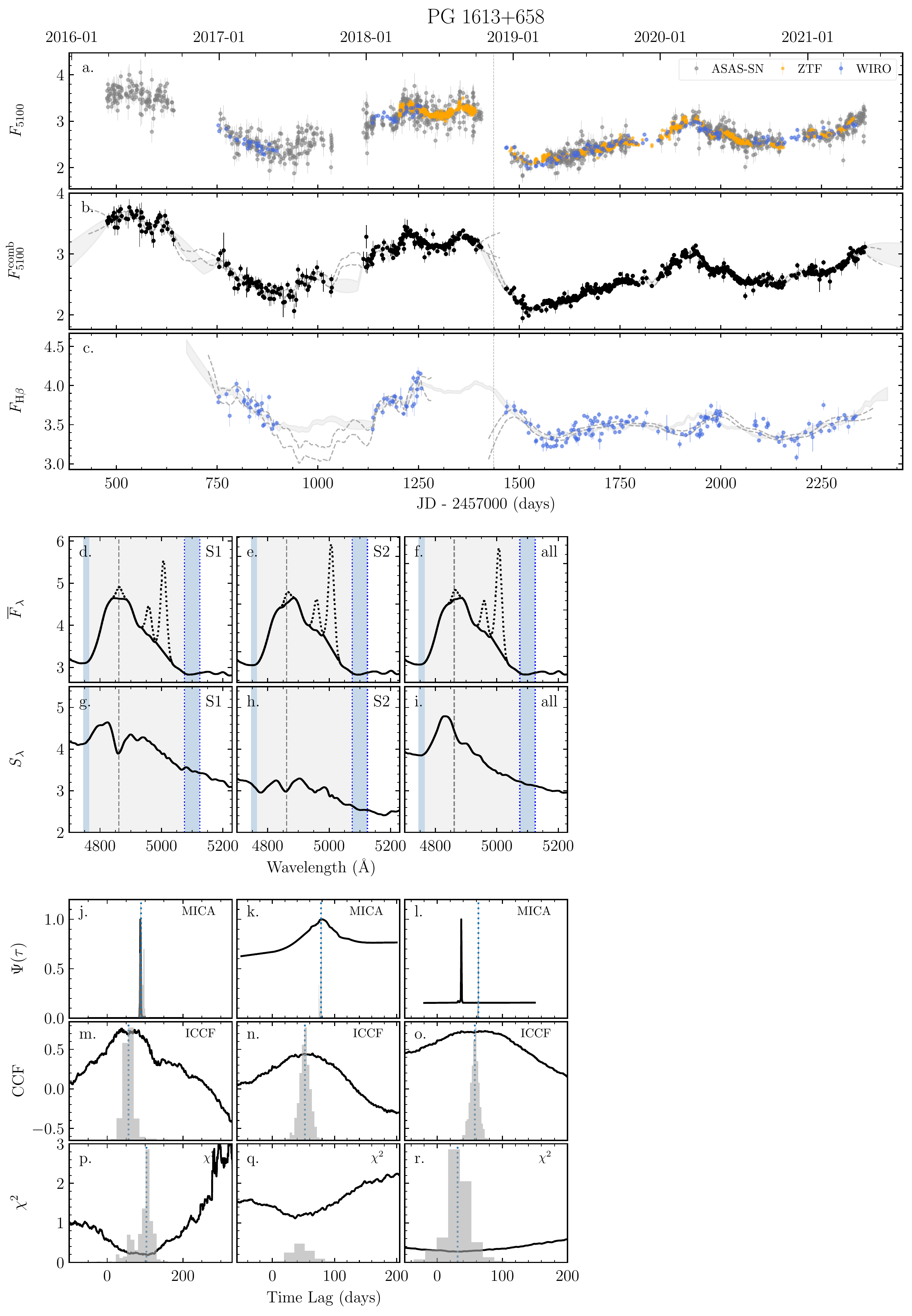}
    \caption{Time-series analysis of PG 1613+658. The meanings of the panels, lines, 
    and histograms are the same as Fig.~\ref{0007lc}. \label{1613lc}}
    \end{figure*}

The widths of the H$\beta$ emission lines are measured from both the mean and
rms spectra. Here, we use both FWHM and line dispersion $\sigma_{\rm H\beta}$ to
quantify the line widths. For the rms spectra, the narrow emission lines
(H$\beta$ and [O {\sc iii}]$\lambda\lambda$4959,5007) are generally negligible.
However, the H$\beta$ and [O {\sc iii}] narrow emission lines in the mean
spectra need to be removed before measuring the line widths of broad H$\beta$.
The narrow H$\beta$ lines were assumed to have the same profiles as the [O {\sc
iii}] lines, and were removed using the same local fitting method described in
Paper \citetalias{Du2018a}. The narrow-line subtracted spectra are shown in
Figures \ref{0007lc} -- \ref{1613lc}.

However, the [O {\sc iii}]$\lambda\lambda$4959,5007 emission lines of PG~1202+281, 
PG1351+640, and PG~1351+695 are strongly blended with each other. In addition, PG~1001+054 
and PG~1211+143 have strong Fe {\sc ii} emission lines. For these 5 objects, we make use of a 
more global fitting scheme to remove the contributions from the other emission lines 
(narrow H$\beta$, [O {\sc iii}], He {\sc ii} lines, and Fe {\sc ii} emission) before we measure the line widths 
of H$\beta$ from the mean spectra. We adopted the software DASpec\footnote{DASpec is available 
at \url{https://github.com/PuDu-Astro/DASpec}}, which is based on the Levenberg-Marquardt 
algorithm \citep{press1992}, to perform the multi-component fitting in a wide spectral range (4430\AA-5550\AA). 
The fitting included (1) a power law component for the continuum, 
(2) a template for Fe {\sc ii} emission \citep{Boroson1992}, 
(3) a simple stellar population template from \cite{bruzual2003} for the contribution from host galaxy if necessary, 
(4) two Gaussians for broad H$\beta$, 
(5) one or two Gaussians for each of the narrow emission lines (e.g., H$\beta$ and [O {\sc iii}]), and 
(6) one or two Gaussian for the He {\sc ii}$\lambda$4686 line. 
The narrow lines were assumed to have the same profiles. The narrow H$\beta$ lines 
in PG~1001+054 and PG~1211+143 are too weak to be decomposed from the broad H$\beta$. In their 
fitting, we fixed the flux of narrow H$\beta$ to be 0.1 of their [O {\sc iii}]$\lambda$5007 lines. 

We measured the line widths of the broad H$\beta$ line in the mean spectra after
removing the contributions of the other components (see Table
\ref{tab:linewidthtimelags}). The cleaned mean spectra are shown in Figures
\ref{0007lc} -- \ref{1613lc}. The uncertainties were estimated using the
bootstrap method. A subset of $N$ points were randomly extracted (with
replacement) from the original $N$ data points from the mean or rms spectrum. We
repeated this procedure 500 times and measured the FWHM and $\sigma_{\rm
H\beta}$ from the resampled spectra. The uncertainties were measured from the
generated distributions. For PG~1001+054 and PG~1211+143, we estimated the
uncertainties by assuming the flux ratio of narrow H$\beta$/[O {\sc
iii}]$\lambda$5007 to be 0.0 and 0.2 (as aforementioned, the cases with
H$\beta$/[O {\sc iii}] = 0.1 is assumed as the central value). This allowed us
to take into account the uncertainties of narrow H$\beta$ decomposition (see
more details in \citealt{Du2014}). The FWHM and $\sigma_{\rm H\beta}$ from the
mean and rms spectra are provided in Table \ref{tab:linewidthtimelags}. The
H$\beta$ signal in the rms spectra of some objects is too weak for the
line-width measurement [PG~0049+171 (Season 4), PG~0947+396 (Seasons 2 and 4),
PG~1202+281 (Seasons 1 -- 4), PG~1351+640, PG~1501+106 (Season 2)]. We measured
the FWHM and $\sigma_{\rm H\beta}$ of H$\beta$ from the aforementioned
continuum-cleaned rms spectra for all seasons. (see Section \ref{sec:meanrms}).

\subsection{Time Series Analysis}
\label{sec:time_series_analysis}

We made use of three different methods to measure the H$\beta$ time lags: the
interpolated cross-correlation function (ICCF, \citealt{Gaskell1986} and
\citealt{Gaskell1987}), the $\chi^2$ method \citep{Czerny2013}, and the MICA
algorithm which is a non-parametric approach to constrain the 1d transfer
function in RM \citep{Li2016}. Here we briefly introduce the three methods for
completeness. More details can be found in the references above.

{\it ICCF:} A commonly employed method in RM, we measured the time lags of
H$\beta$ using ICCF. In general, the time lags can be measured from the peak of
the CCF and the centroid of the CCF above a threshold (80\% of the peak), which
are marked as $\tau_{\rm peak}$ and $\tau_{\rm cent}$, respectively. The
uncertainties of the time lags were estimated using the ``flux
randomization/random subset sampling (FR/RSS)'' method \citep{Peterson1998,
Peterson2004}. In the present paper, the median and 1$\sigma$ limits of the
cross-correlation centroid distributions (CCCDs) and the cross-correlation peak
distributions generated by the FR/RSS method were adopted as the final lags and
their uncertainties.

{\it The $\chi^2$ method: } The $\chi^2$ method \citep{Czerny2013} was also
employed to measure the time lags between the continuum and H$\beta$ light
curves. \cite{Czerny2013} found that the $\chi^2$ method works better than using
ICCF for the AGNs with red-noise variability. The technique takes into account
the weights of the points in light curves through their uncertainties. After
shifting and interpolating the H$\beta$ light curves, the $\chi^2$ were
calculated by
\begin{equation}
    \chi^2 = \frac{1}{N}\sum_{i=1}^n\frac{(x_i - A_{\chi^2}y_i)^2}{\delta x_i^2 + A_{\chi^2}^2\delta y_i^2},
\end{equation}
where $x_i$ and $y_i$ are the continuum and interpolated H$\beta$ fluxes, and
$\delta x_i$ and $\delta y_i$ are their uncertainties. $A_{\chi^2}$ is a
normalized factor formulated as
\begin{equation}
    A_{\chi^2} = \frac{S_{xy} + (S_{xy}^2 + 4S_{x3y}S_{xy3})^{1/2}}{2S_{xy3}},
\end{equation}
where
\begin{equation}
\begin{aligned}
    S_{xy}  &= \sum_{i=1}^{N}(x_i^2 \delta y_i^2 - y_i^2\delta x_i^2), \\
    S_{xy3} &= \sum_{i=1}^{N}x_i y_i \delta y_i^2, \\
    S_{x3y} &= \sum_{i=1}^{N}x_i y_i\delta x_i^2.
\end{aligned}
\end{equation}
We took the minimum points in the $\chi^2$ functions as the time lag
measurements. Similar to the ICCF method, the uncertainties were generated from
FR/RSS as well. 

{\it MICA\footnote{MICA is available at https://github.com/LiyrAstroph/MICA2}:}
MICA \citep{Li2016} is a Bayesian-based non-parameteric approach to infer the 1d
transfer function from the continuum and emission-line light curves. It assumes
that the transfer function is a sum of relatively displaced Gaussians, and
employs the diffusive-nested sampling technique to obtain posterior
distributions of Gaussian parameters. For each set of parameters, we calculate
the corresponding transfer function and obtain the centroid of the transfer
function. The mean of the distribution of centroids is taken as the best
estimate of the time lag and its uncertainty by the 68.3\% confidence interval.

\begin{figure*}[ht!]
    \centering
    \includegraphics[width=\textwidth]{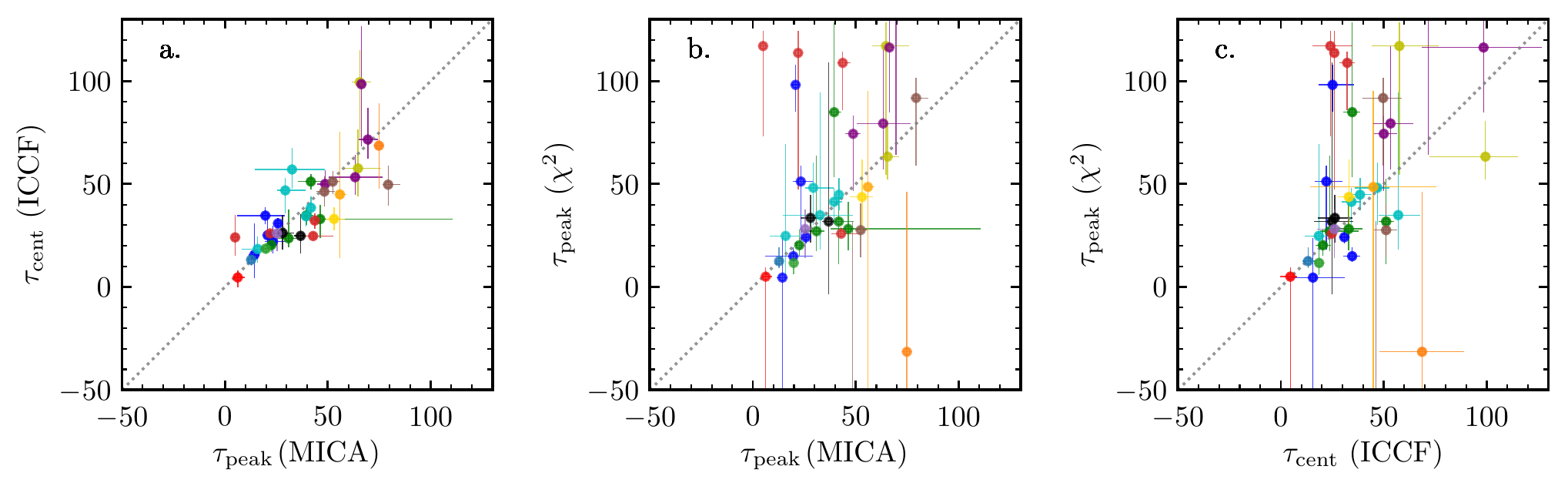}
    \caption{Pairwise correlations between the measurements from ICCF, $\chi^2$, and MICA. 
    The points in the same color are the time lags of different seasons (and the entire
    light curve) for individual objects. \label{fig:timelagsin3ways}}
    \end{figure*}

The CCFs and CCCDs, the $\chi^2$ functions and their lag distributions, and the
transfer functions and the corresponding uncertainties generated from MICA are
shown in Figures \ref{0007lc}--\ref{1613lc}. The time lags and their
uncertainties are given in Table \ref{tab:linewidthtimelags}. For the light
curves with clear variations and statistically significant time delays (e.g.,
Season 4 of PG~0049+171, Season 2 of PG~0947+396, PG~1310$-$108), the
measurements of the three methods are generally consistent with each other. The
pairwise comparison between the lag measurements of the methods are demonstrated
in Figure \ref{fig:timelagsin3ways}. The results from ICCF and MICA have the
best consistency, while the $\chi^2$ method generally gives larger scatter
compared to the other two methods. Considering that MICA takes advantage of a
damped random walk model \citep{Li2016} and can give better constraints to the
light-curve reconstruction  by incorporating the continuum and H$\beta$
variations, especially across larger gaps, we adopted the time lags from MICA
for the BH mass measurements in the following Section \ref{subsec:BHmass}.

\subsection{Black Hole Masses}
\label{subsec:BHmass}

Given the time lag and the line width measurements, the BH masses $\bhm$ can be
determined by the formula
\begin{equation}
\bhm  =  f_{\rm BLR}\frac{R_{\rm BLR} V^2}{G}
\end{equation}
where $R_{\rm BLR} =  c \tau_{\rm BLR}$ is the responsivity-weighted radius of
the BLR, $\tau_{\rm BLR}$ is the time lag, $c$ is the speed of light, $V$ is
$\sigma_{\rm H\beta}$ or FWHM of the H$\beta$ line from the mean or rms spectra,
$G$ is the gravitational constant, and $f_{\rm BLR}$ is a scaling factor. 

The average value of $f_{\rm BLR}$ for AGNs as a sample can be determined by
calibration against the $\bhm$ -- $\sigma_*$ or $\bhm$ -- $M_*$ relationships of
inactive galaxies \citep[e.g.,][]{onken2004, Woo2010, Woo2015, Ho2014}, where
$\sigma_*$ and $M_*$ are the stellar velocity dispersion and stellar mass of the
galactic bulge. However, the specific values of $f_{\rm BLR}$ in individual
objects are likely to have a significant scatter around the average
\citep[e.g.,][]{Pancoast2014, grier2017mcmc, Li2018, williams2018}. Here we
adopt the calibrated average $f_{\rm BLR}$ from \cite{Woo2015} (1.12 for FWHM
and 4.47 for $\sigma_{\rm H\beta}$) in our $\bhm$ calculations, as we did for
Papers \citetalias{Du2018a} and \citetalias{Brotherton2020}.  

It has been suggested that the line widths in rms spectra and time lags are more
consistent with the virial relationship ($\tau \propto V^{-2}$) than the mean
spectra \citep[e.g.,][]{Peterson2004, DallaBonta2020}. Therefore, we calculated
the BH masses using the line widths from the rms spectra. But for completeness,
we provided the ``virial products (VP)''  measured from the FWHM of the mean
spectra ($R_{\rm H\beta}V^2_{\rm FWHM}/G$). We divided the light curves of the
objects according to their seasonal gaps and measured the time lags for
different seasons (Section \ref{subsec:Spectroscopy}) as well as for combined
seasons. Table \ref{tab:BHmasses} gives the corresponding VP measured from the
mean spectra, as well as BH masses ($f_{\rm BLR} R_{\rm BLR} V^2/G$) measured
from the FWHM and $\sigma_{\rm H\beta}$ of the rms spectra; seasons with very
poor lag measurements are ignored in the mass determinations. For completeness,
the monochromatic luminosity at 5100\AA\ is available in Table
\ref{tab:linewidthtimelags}. It should be noted that we have corrected for the
Galactic extinction, but host galaxy contamination is present in these
measurements. We will investigate the location of our targets on the
radius-luminosity plane \citep[e.g.,][]{Kaspi2000, Bentz2013, Du2019} in a
future paper. The cosmological parameters used to calculate the luminosity are
$H_0 = 67 {\rm \ km\ s^{-1} Mpc^{-1}}, \Omega_M = 0.32, \Omega_{\Lambda} = 0.68$
\citep{Planck2014, Planck2020}. 

The time lags can vary over time, for instance if there are strong luminosity
changes. In principle, it should not make a difference to the measurement of BH
mass if we use the time lag from the light curves of any individual season or
from the entire campaign, as the BH mass cannot change on short timescales.  In
practice, some seasons have stronger variations and better sampling than others.
However, if the BLR kinematics is complex or variable, the BH masses measured
from individual seasons or the whole light curves can perhaps differ. BLR
dynamical modeling \citep[e.g.,][]{Pancoast2011, Pancoast2014, Li2013, Li2018}
can, in principle, give more reliable BH mass measurements if the BLRs deviate
from Keplerian/virialized motion, but this discussion stretches beyond the scope
of the current paper. The best data sets here may be good enough to allow
dynamical modeling, which we shall investigate in future work.

For each object, we list the measurements from the whole light curve and from
all of the seasons (except for the very poor ones) in Table \ref{tab:BHmasses},
and mark the preferred values with a ``\checkmark'' (the ones calculated from
$\sigma_{\rm H\beta}$ in the rms spectra). We prefer to adopt the results with
the smallest measurement uncertainties. They are usually the values measured
from the whole light curves, except for those objects for which the lag
measurements of individual seasons have comparable or significantly smaller
measurement uncertainties.

\subsection{Velocity-resolved Results} 
\label{sec:velocityresolved}

To investigate the BLR geometry and kinematics and their potential changes over
time for the present sample, we calculated velocity-resolved lags
\citep[e.g.,][]{Bentz2009, denney2010, Grier2013, Du2016VI, Hu2020_0026} as a
first step. We divided the emission lines into several bins, determined by the
flux ranges in the rms spectra, and measured their time lags with respect to the
continuum using ICCF.  
The lags as functions of velocity are shown in Figure \ref{fig:velocity
resolved}. Similar to the BH mass measurements in Section \ref{subsec:BHmass},
because of the limitations of variation amplitudes and S/N ratios, we cannot
obtain the velocity-resolved lag measurements for all of the objects for all
individual seasons. We did not calculate the velocity-resolved lags for the
seasons with poor data or no clear variations. For the objects with very weak
H$\beta$ signals in the rms spectra (see Section \ref{sec:meanrms}), we instead
determined their velocity bins using the mean spectra. It has been demonstrated
that using mean or rms spectra to determine the velocity bins does not usually
change the results significantly (see more details in Paper
\citetalias{Du2018a}).

\begin{figure*}
    \includegraphics[width=0.3\textwidth, height=0.21\textheight]{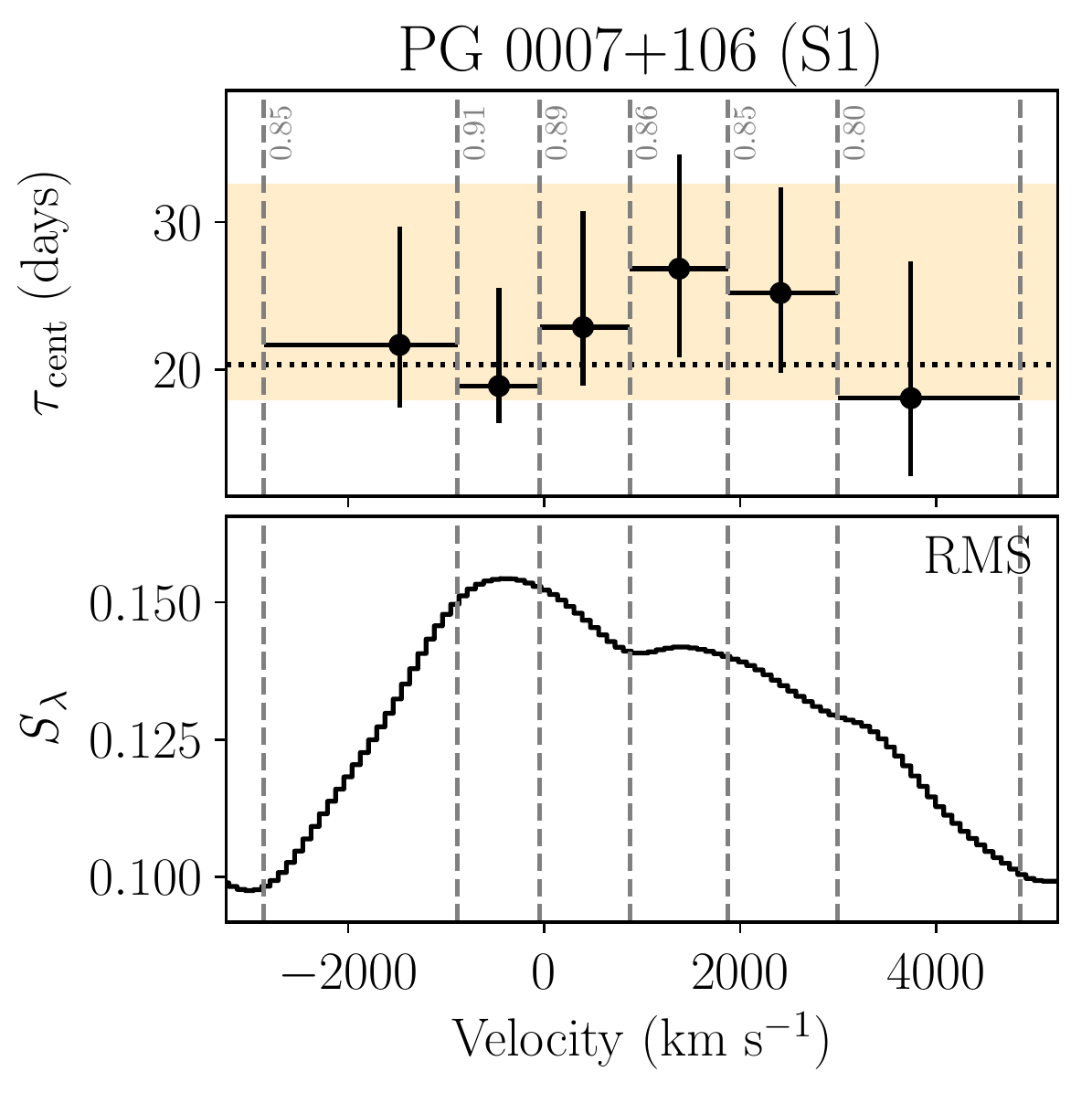}%
    \hspace{0.02in}
    \includegraphics[width=0.3\textwidth, height=0.21\textheight]{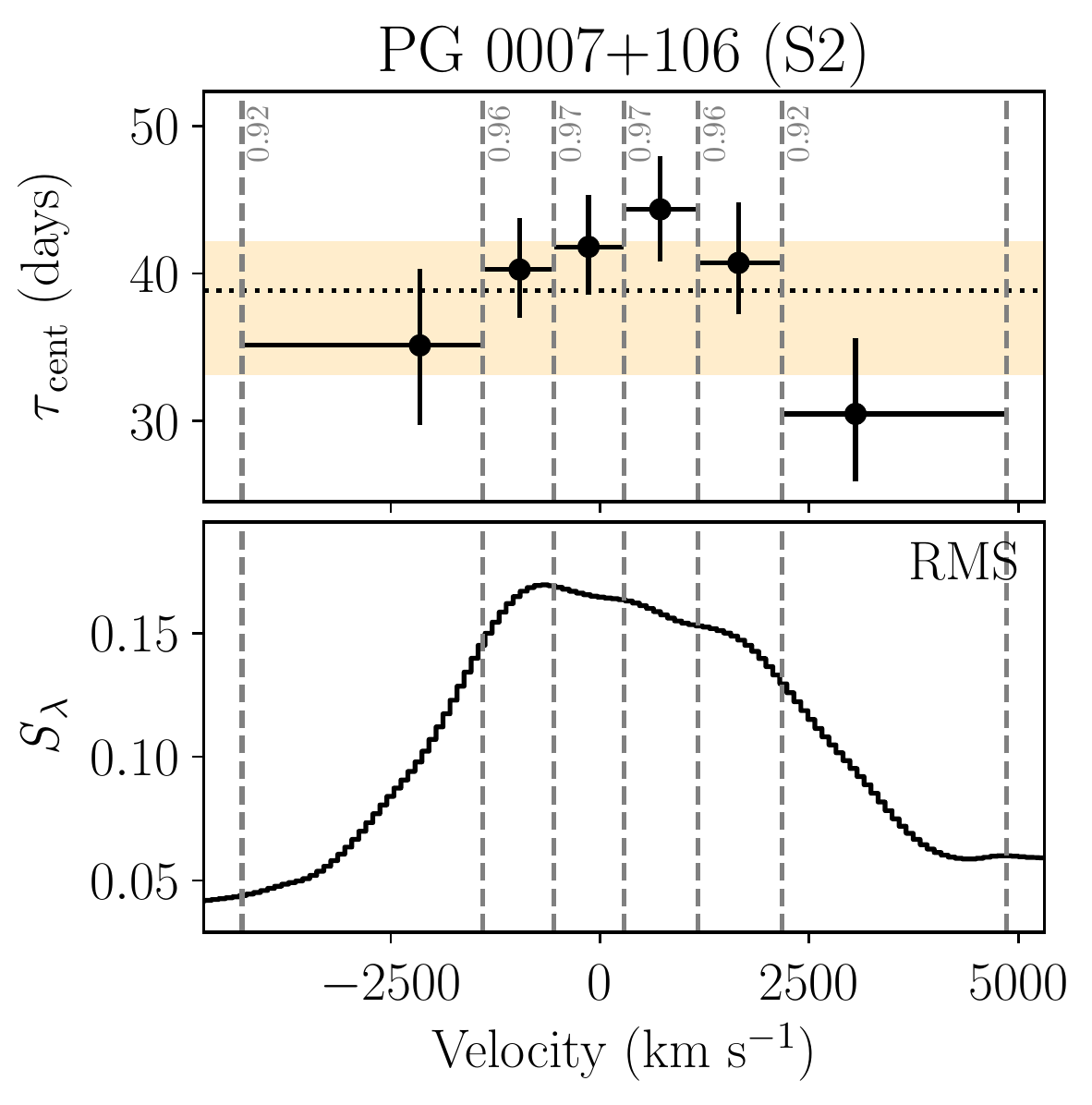}%
    \hspace{0.02in}
    \includegraphics[width=0.3\textwidth, height=0.21\textheight]{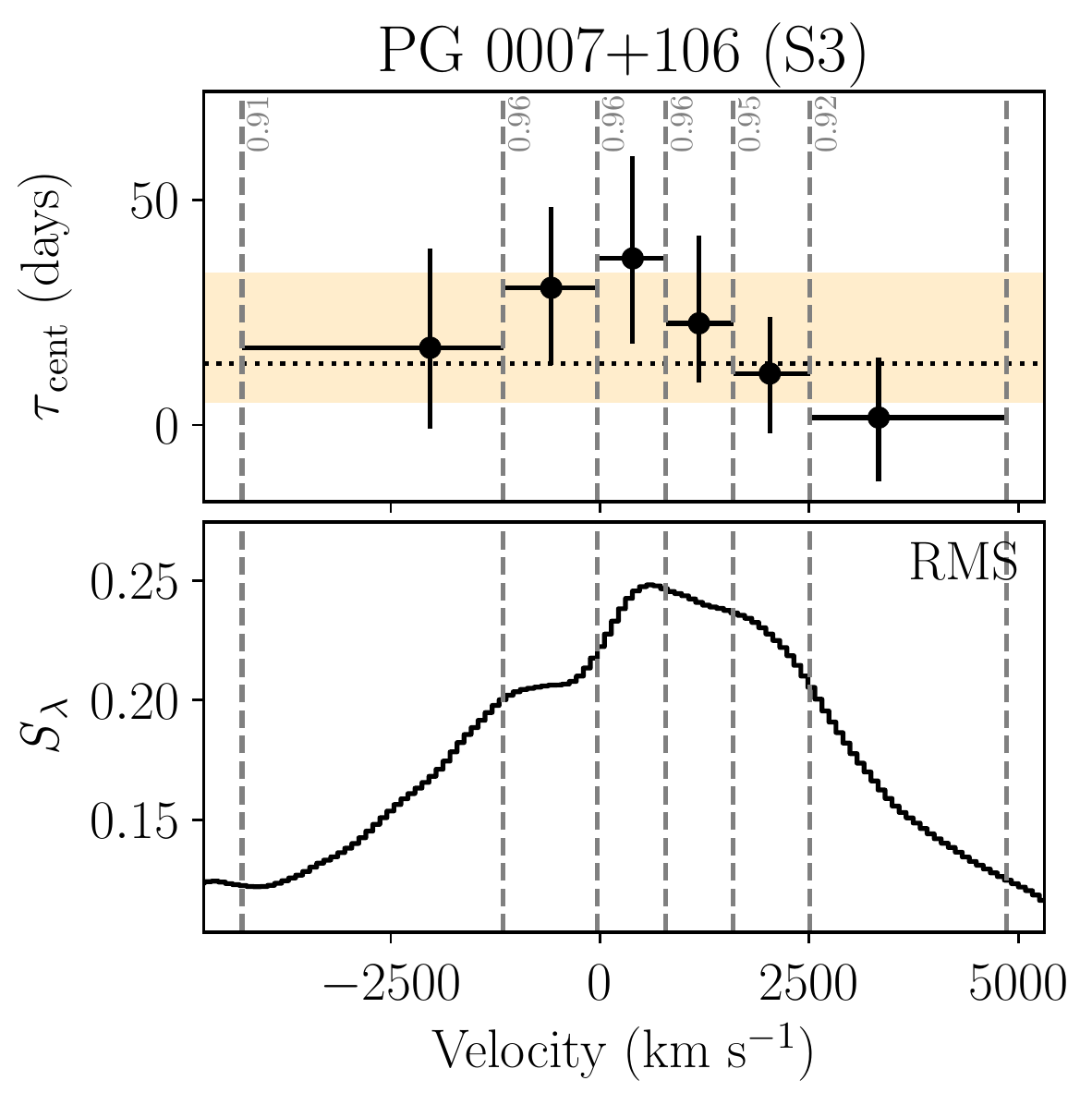}
    \quad \\
    \includegraphics[width=0.3\textwidth, height=0.21\textheight]{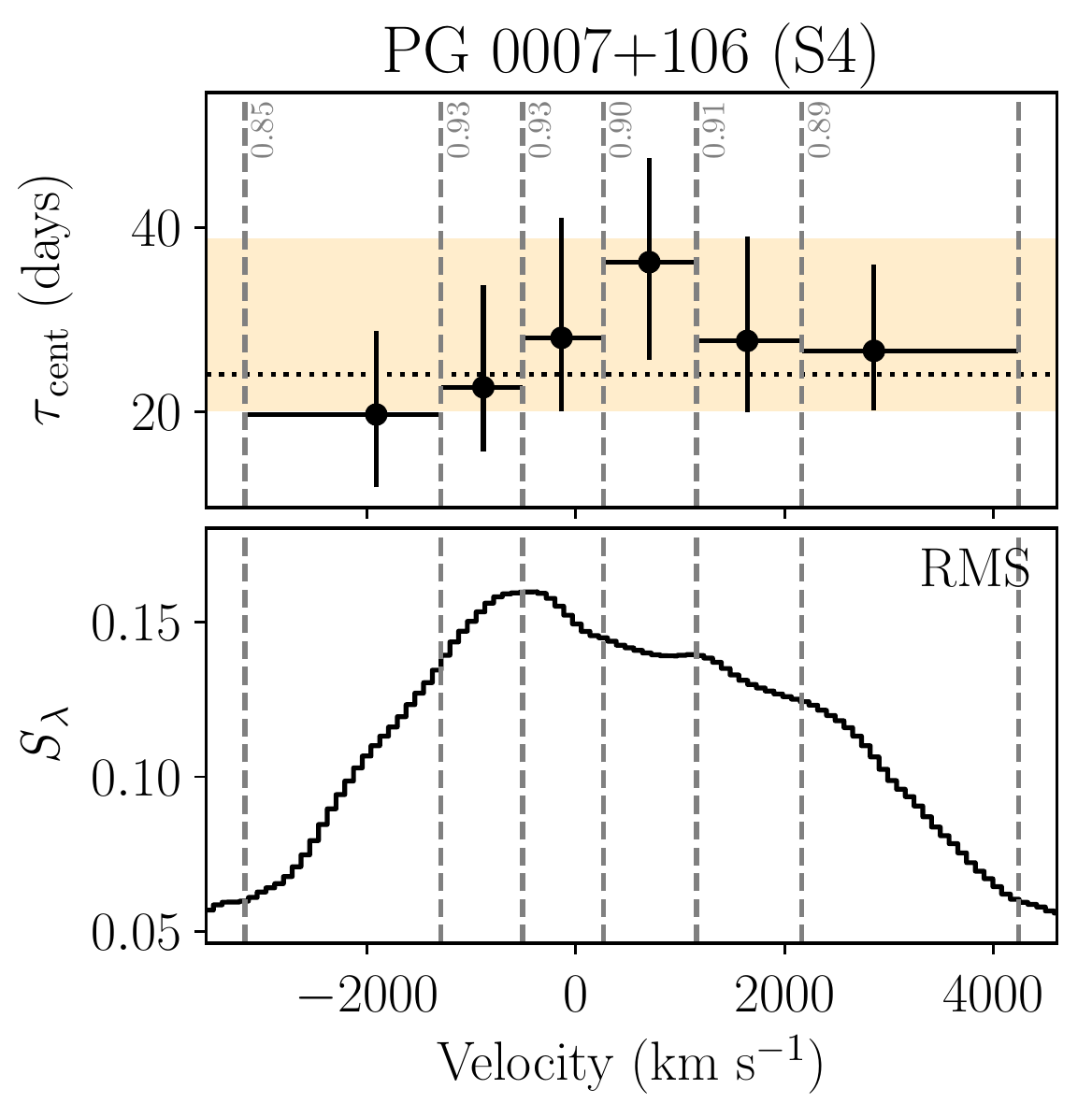}%
    \hspace{0.02in}
    \includegraphics[width=0.3\textwidth, height=0.21\textheight]{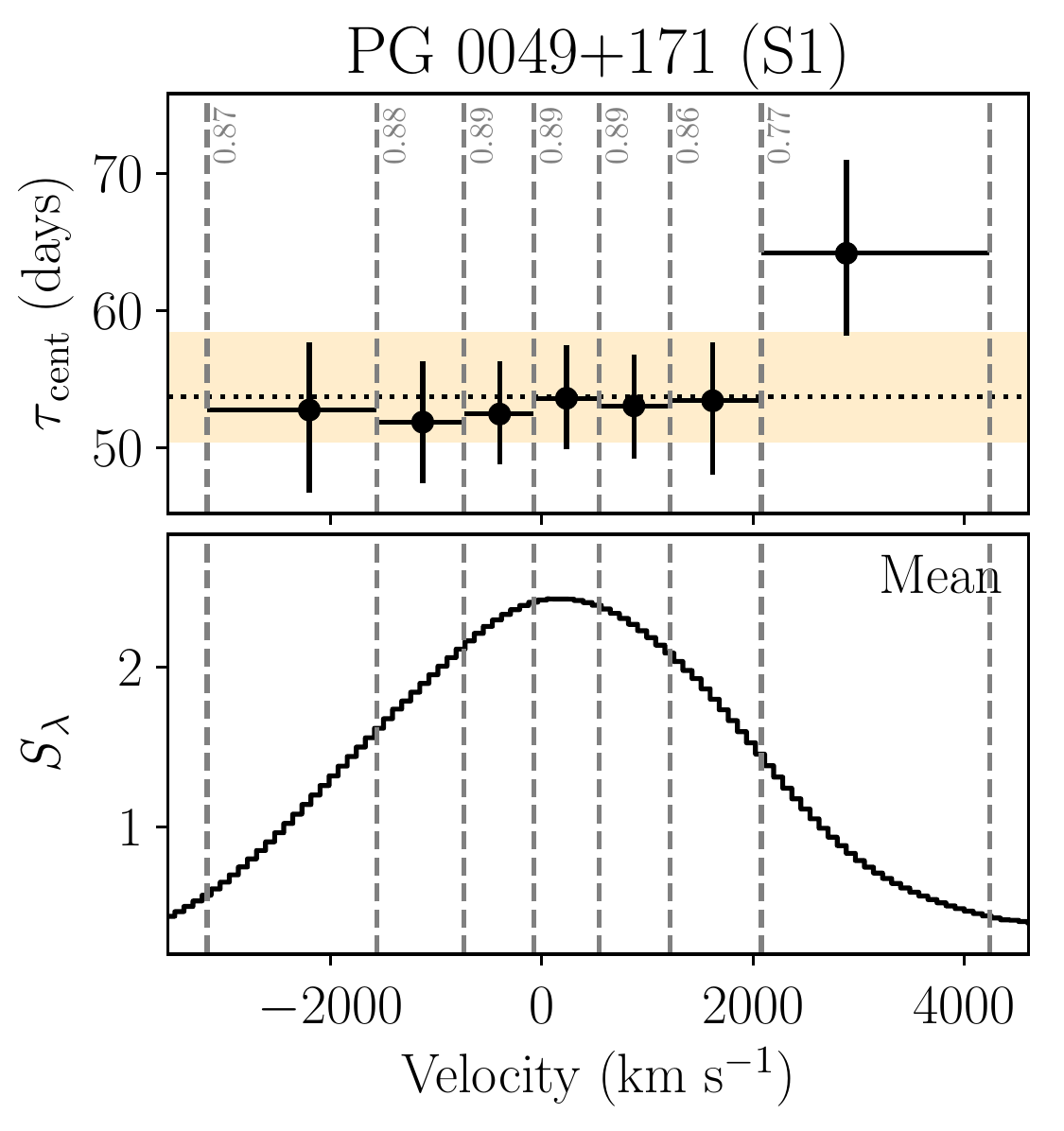}%
    \hspace{0.02in}
    \includegraphics[width=0.3\textwidth, height=0.21\textheight]{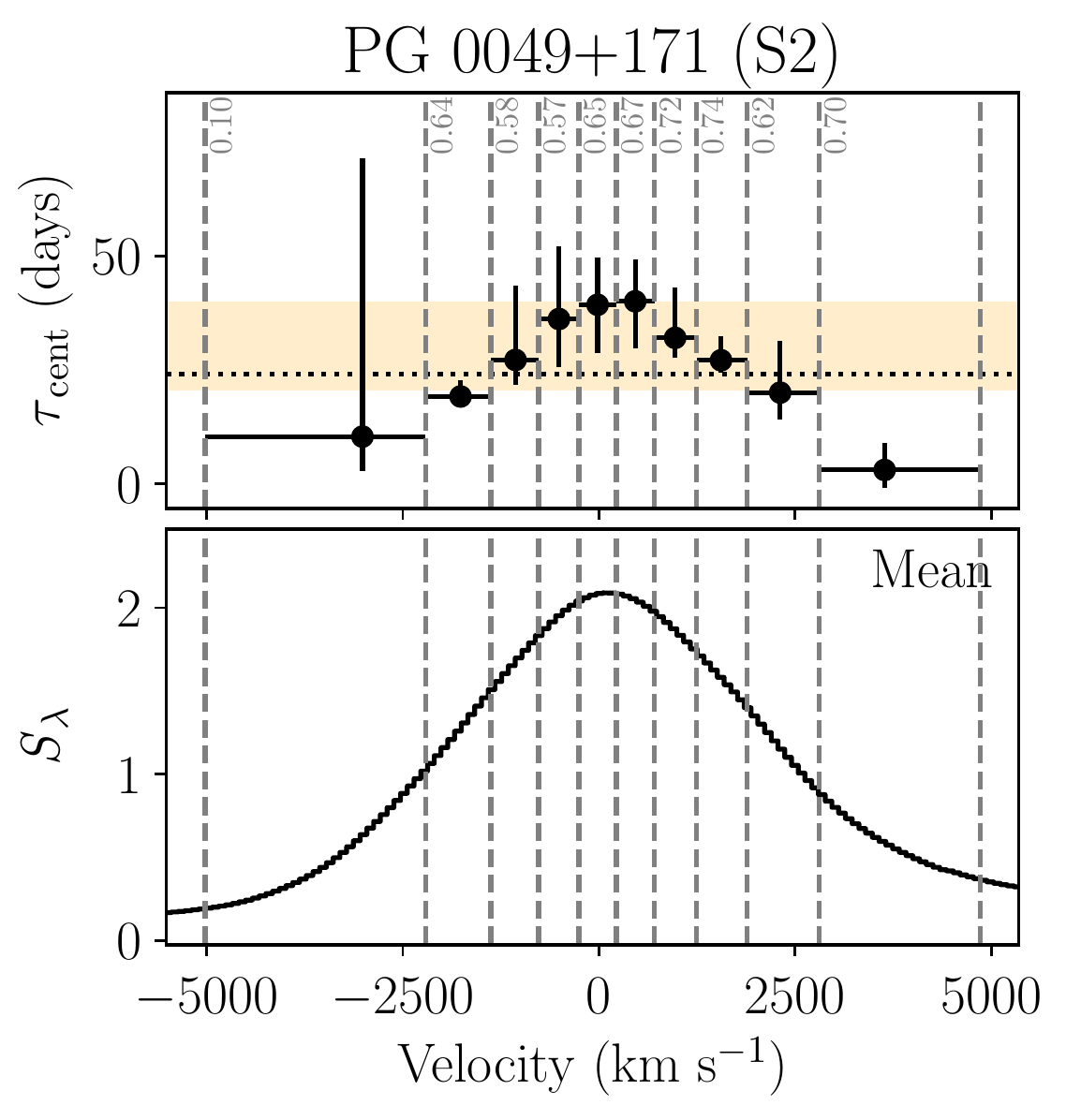}
    \quad \\
    \includegraphics[width=0.3\textwidth, height=0.21\textheight]{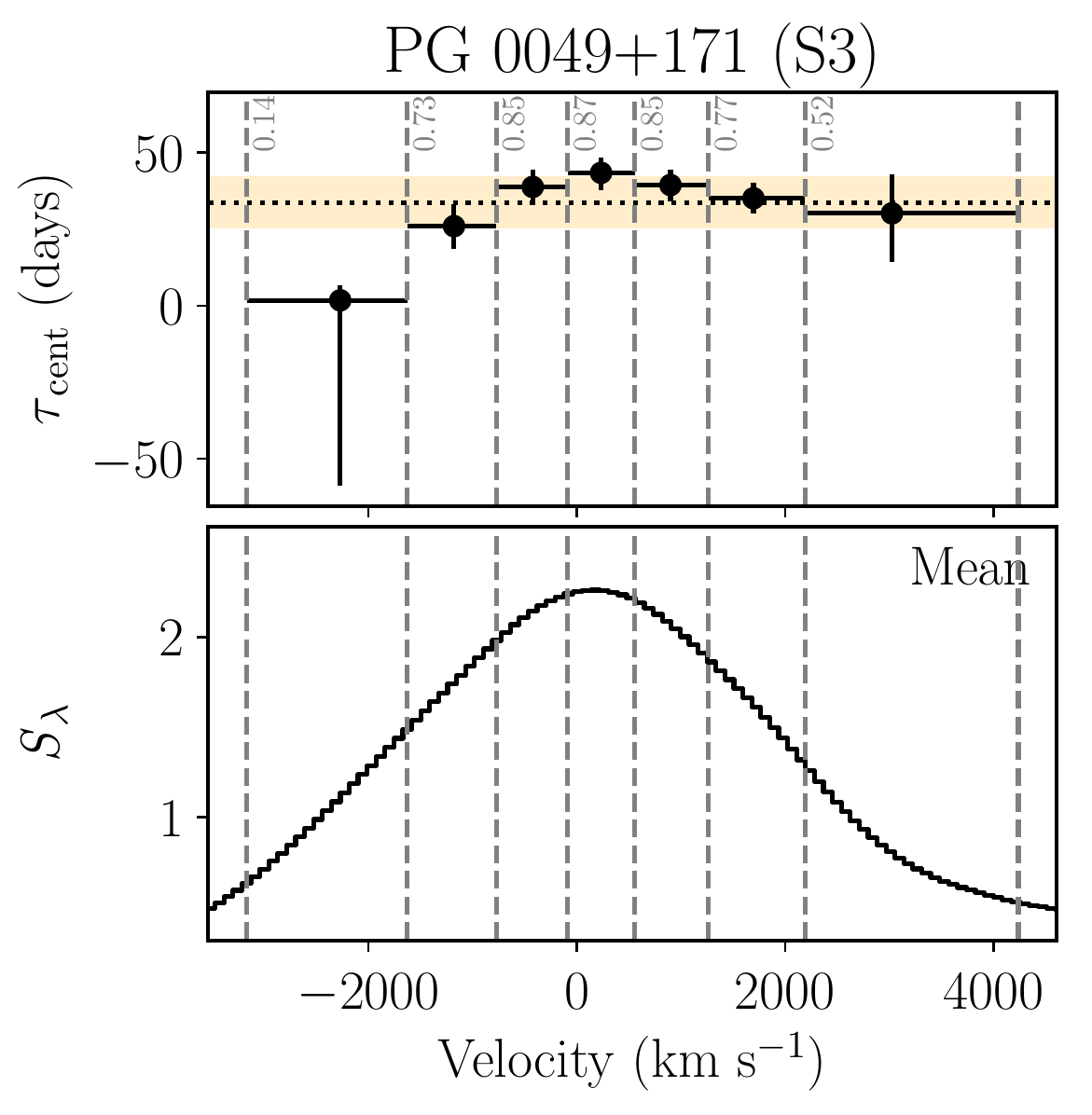}%
    \hspace{0.02in}
    \includegraphics[width=0.3\textwidth, height=0.21\textheight]{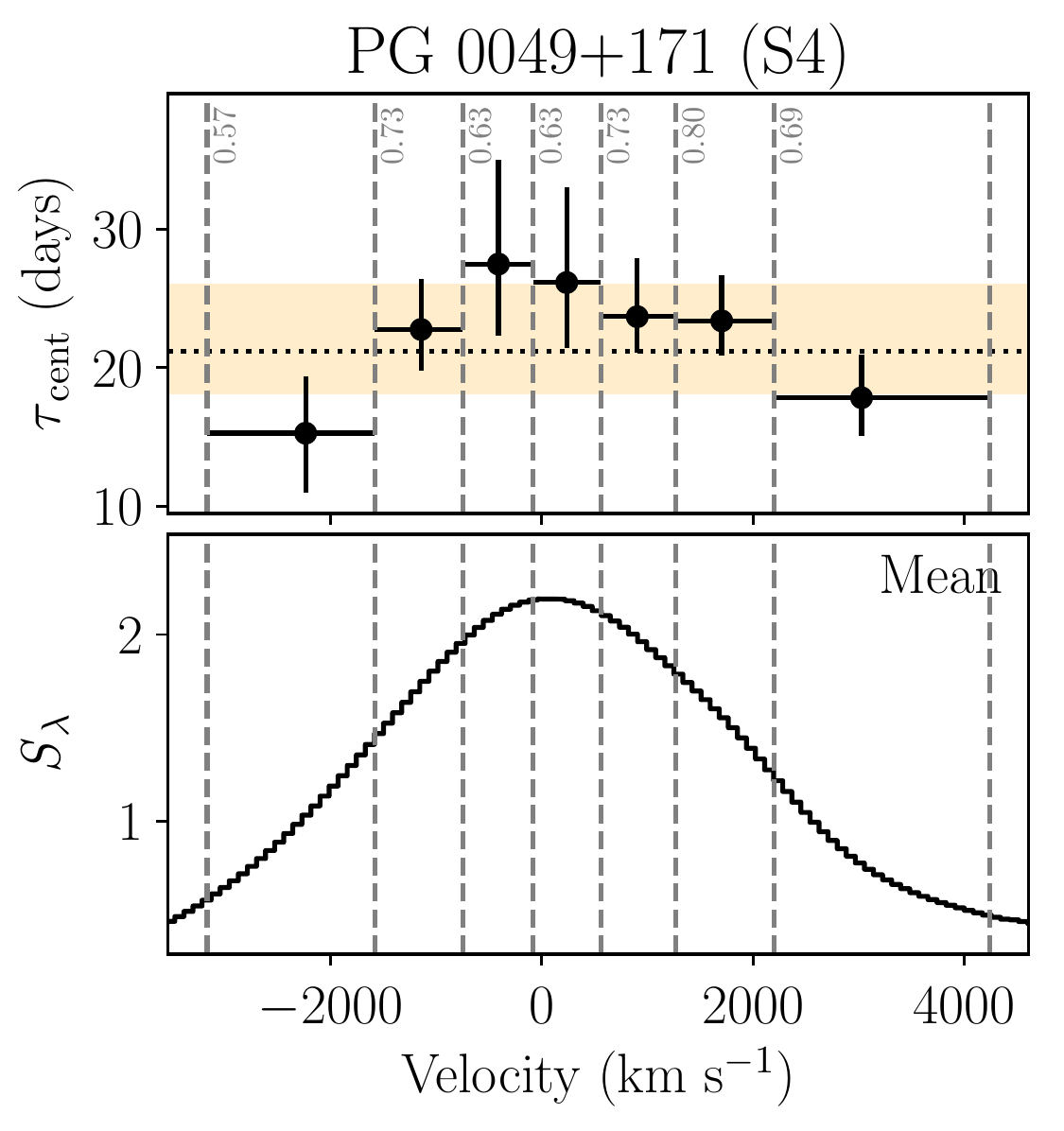}%
    \hspace{0.02in}
    \includegraphics[width=0.3\textwidth, height=0.21\textheight]{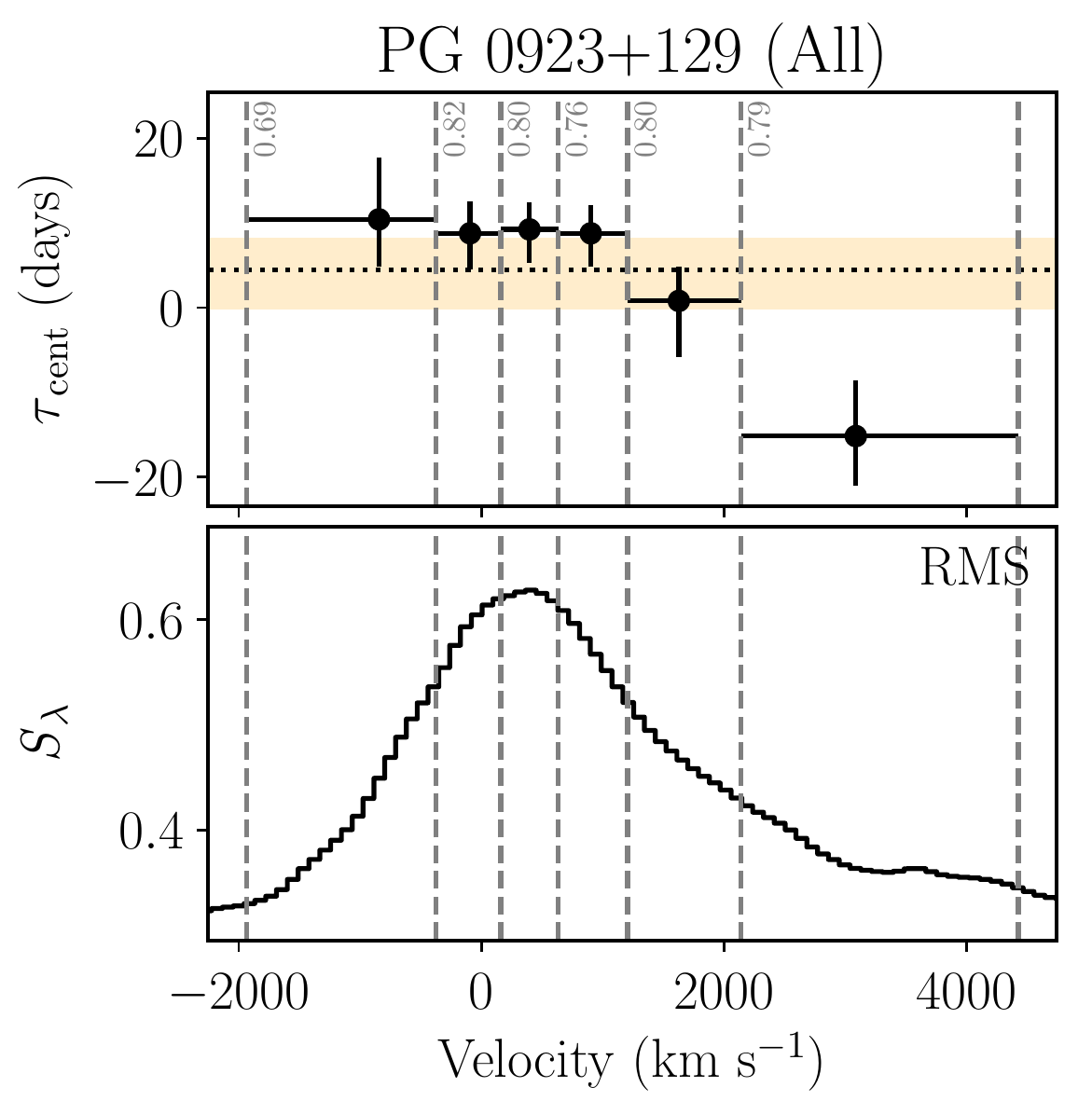}
    \quad \\
    \includegraphics[width=0.3\textwidth, height=0.21\textheight]{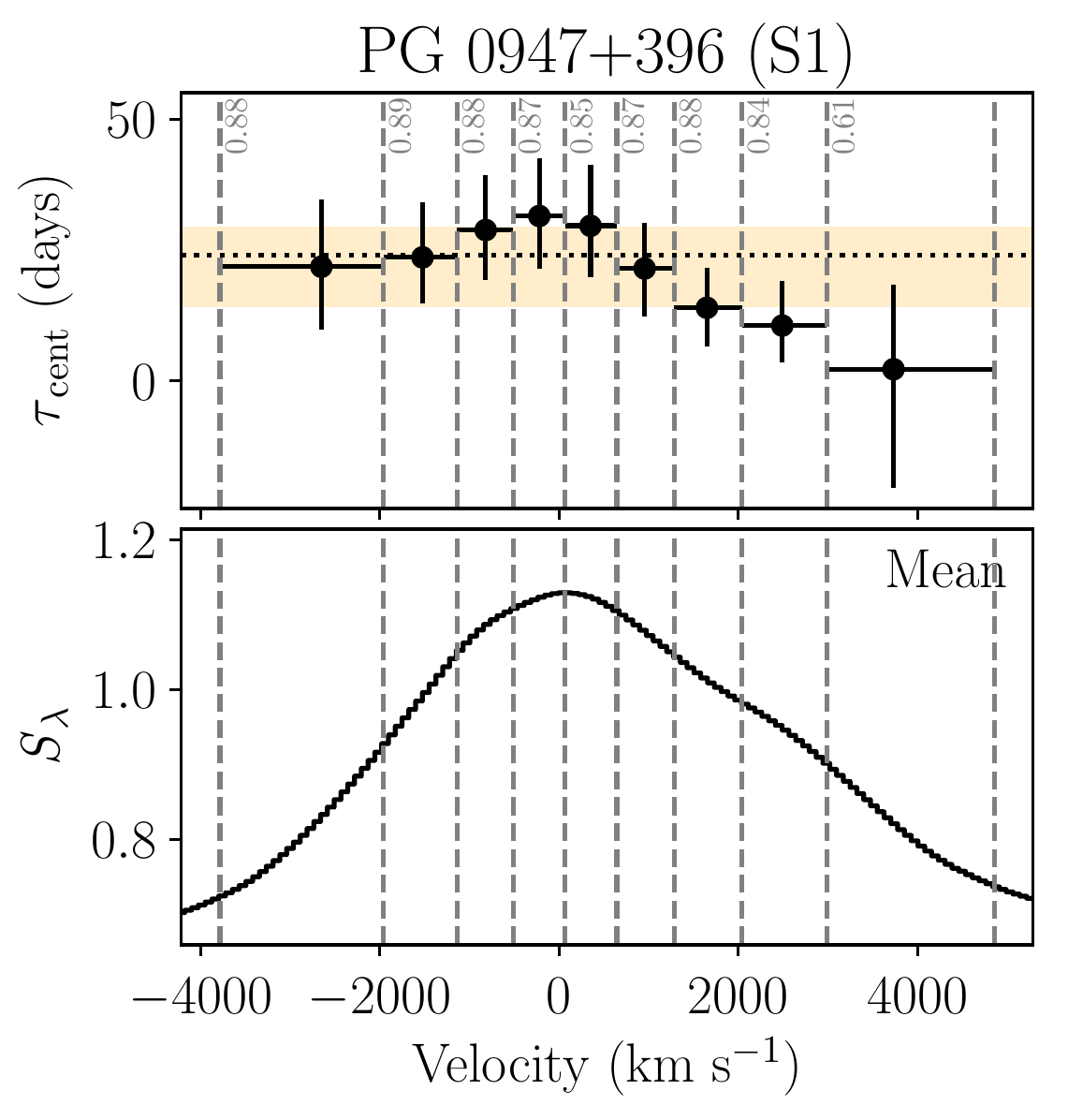}%
    \hspace{0.02in}
    \includegraphics[width=0.3\textwidth, height=0.21\textheight]{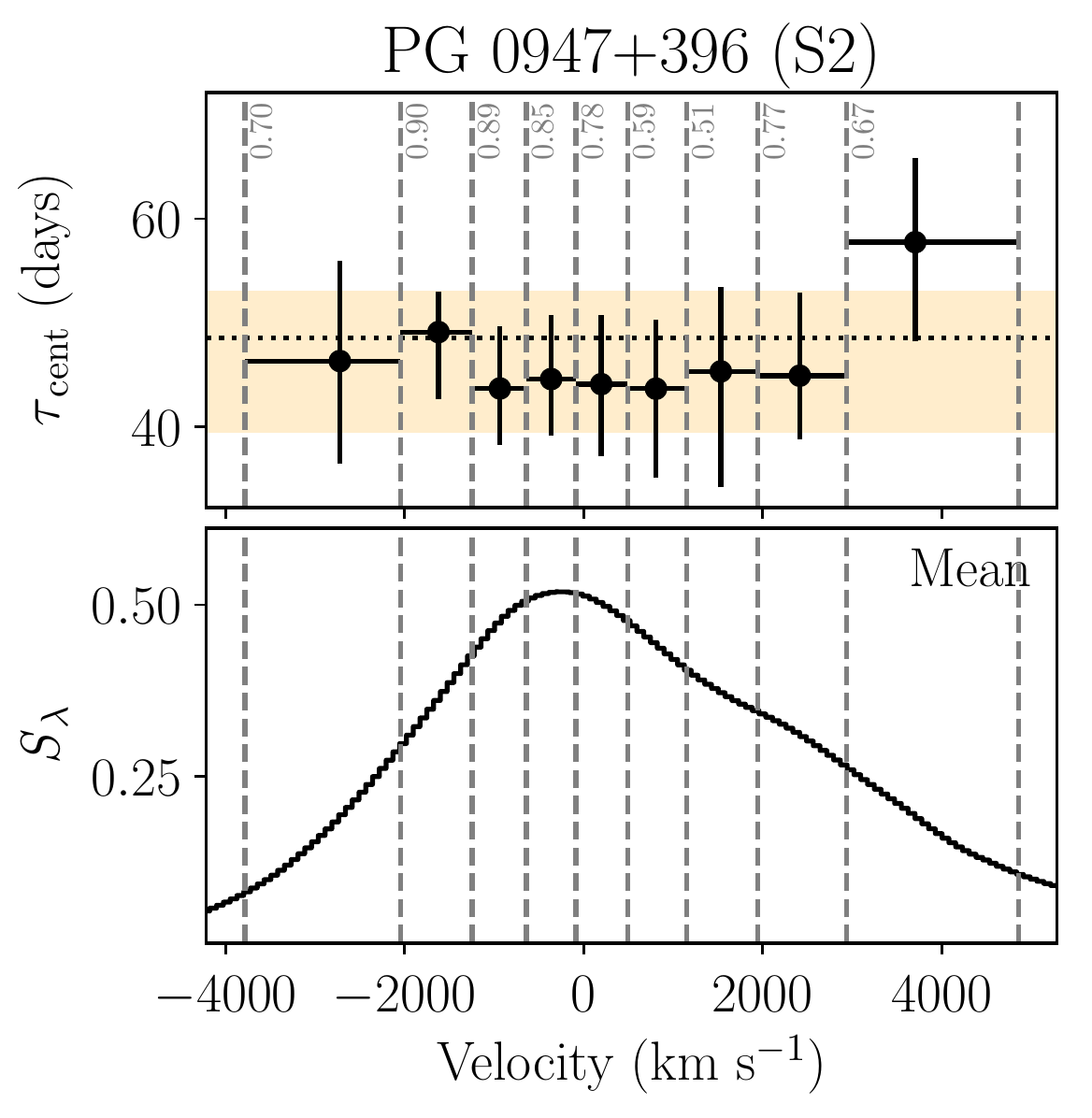}%
    \hspace{0.02in}
    \includegraphics[width=0.3\textwidth, height=0.21\textheight]{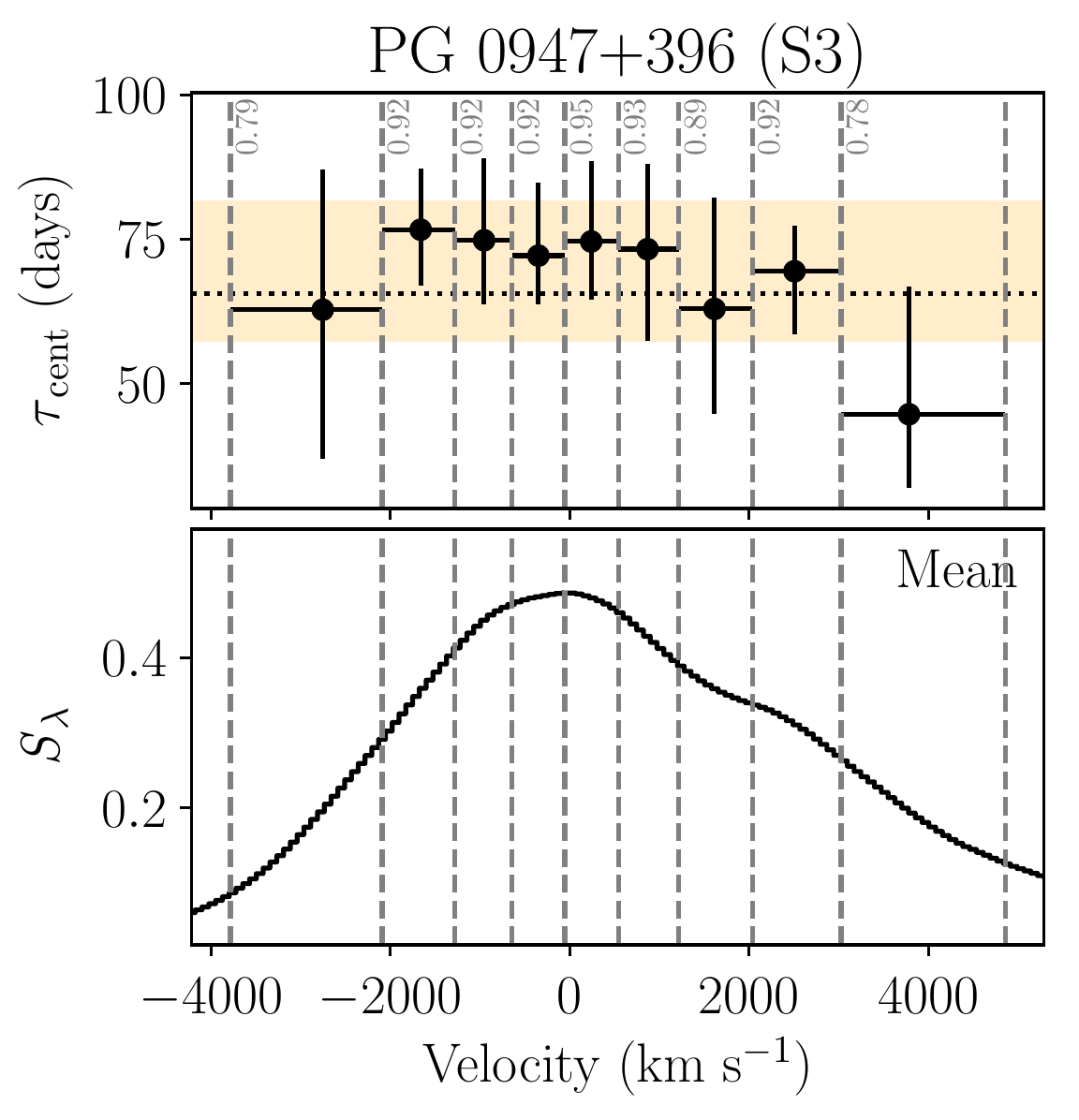}
    \quad \\
    \caption{Velocity-resolved lags. The upper sub-panel in each panel 
    is the velocity-resolved lags and the lower sub-panel is the rms or mean spectrum. 
    The horizontal dash lines and yellow horizontal spans are the 
    mean time lags and their uncertainties.  The vertical dashed lines mark the edges
    of the velocity bins. The peak correlation coefficient is denoted in each bin at the top sub-panels.
    \label{fig:velocity resolved}}
\end{figure*}

\begin{figure*}
    \flushleft
    \figurenum{\ref{fig:velocity resolved}}
    \includegraphics[width=0.3\textwidth, height=0.21\textheight]{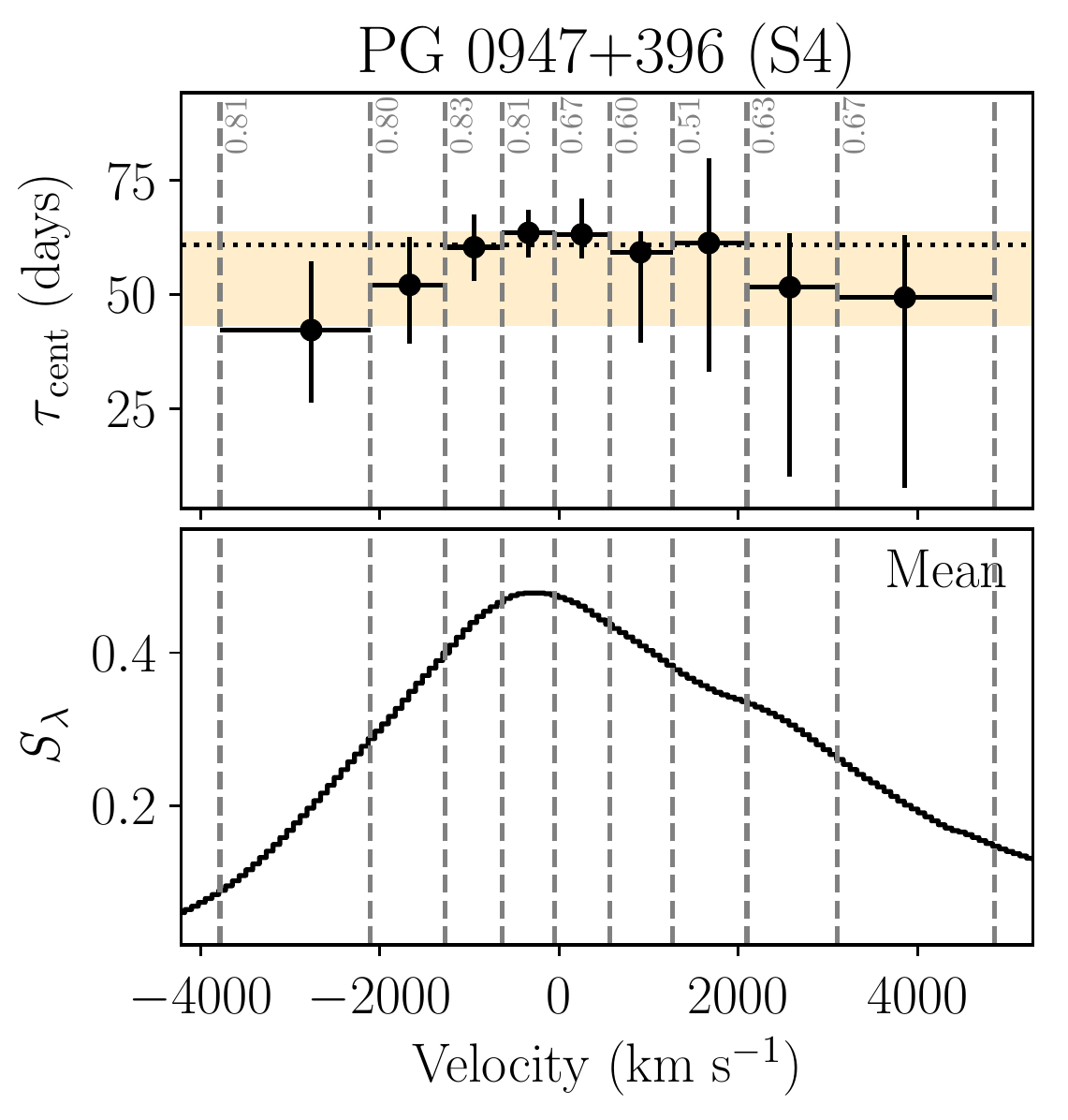}%
    \hspace{0.02in}
    \includegraphics[width=0.3\textwidth, height=0.21\textheight]{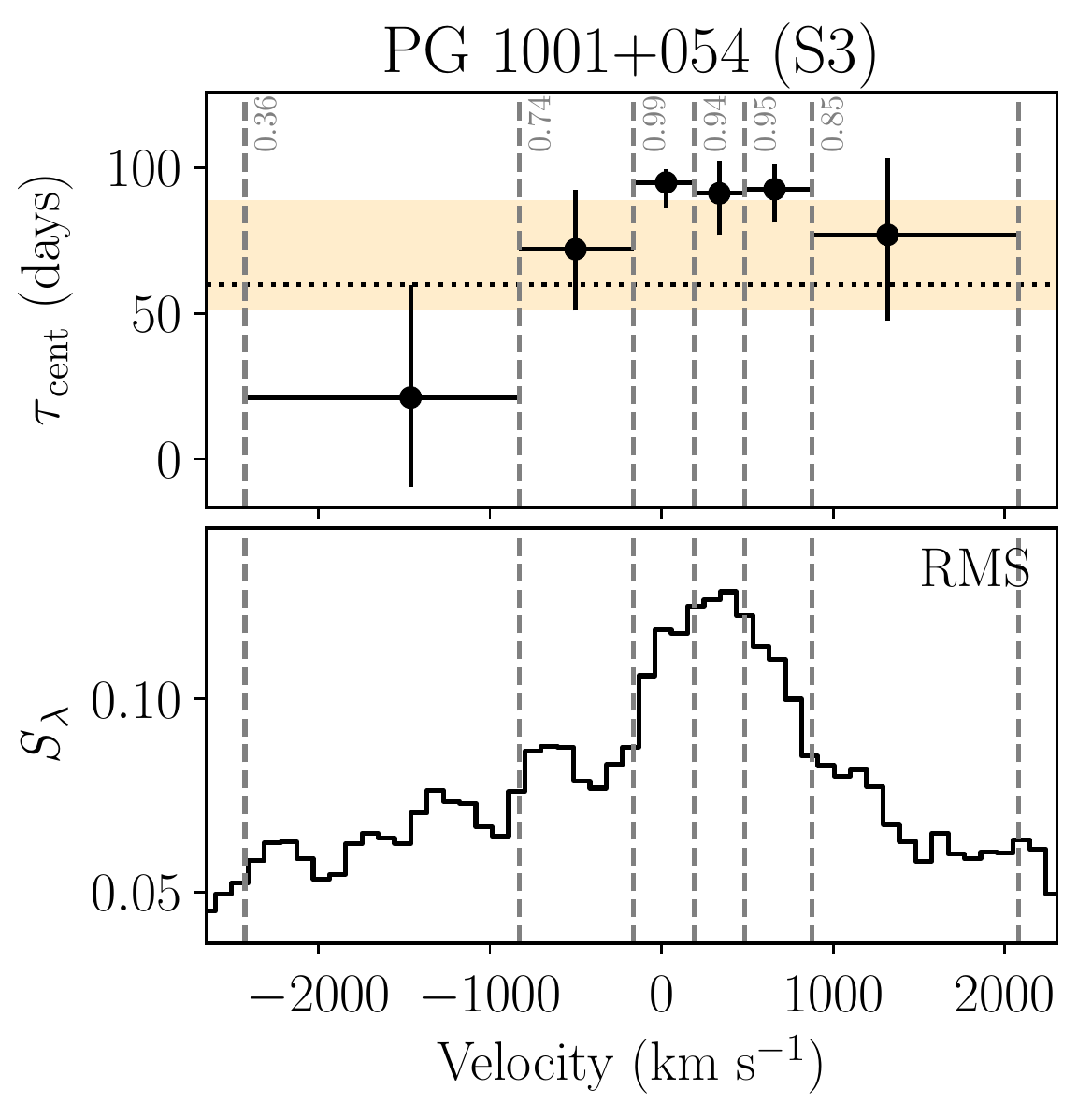}%
    \hspace{0.02in}
    \includegraphics[width=0.3\textwidth, height=0.21\textheight]{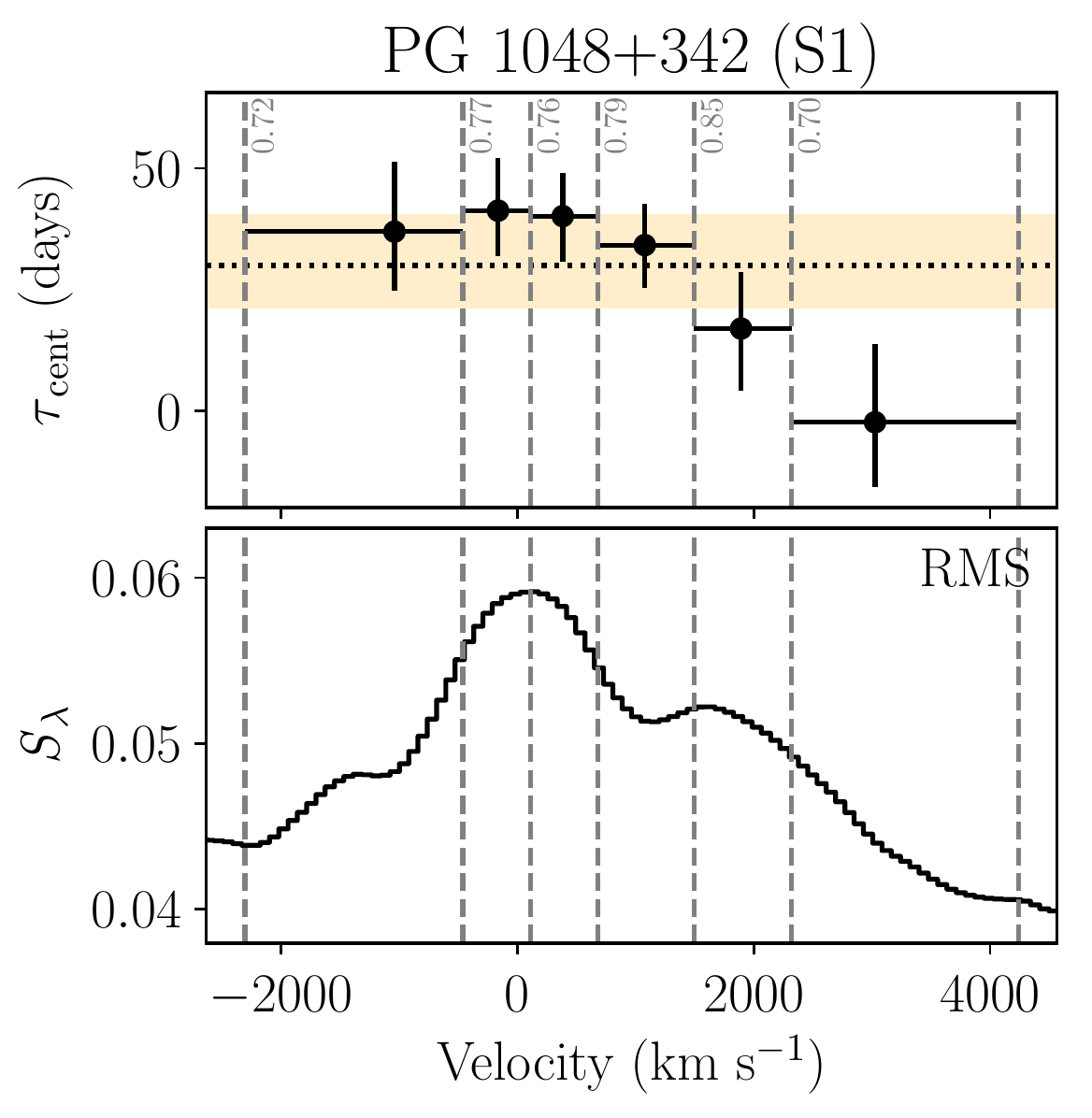}
    \quad \\
    \includegraphics[width=0.3\textwidth, height=0.21\textheight]{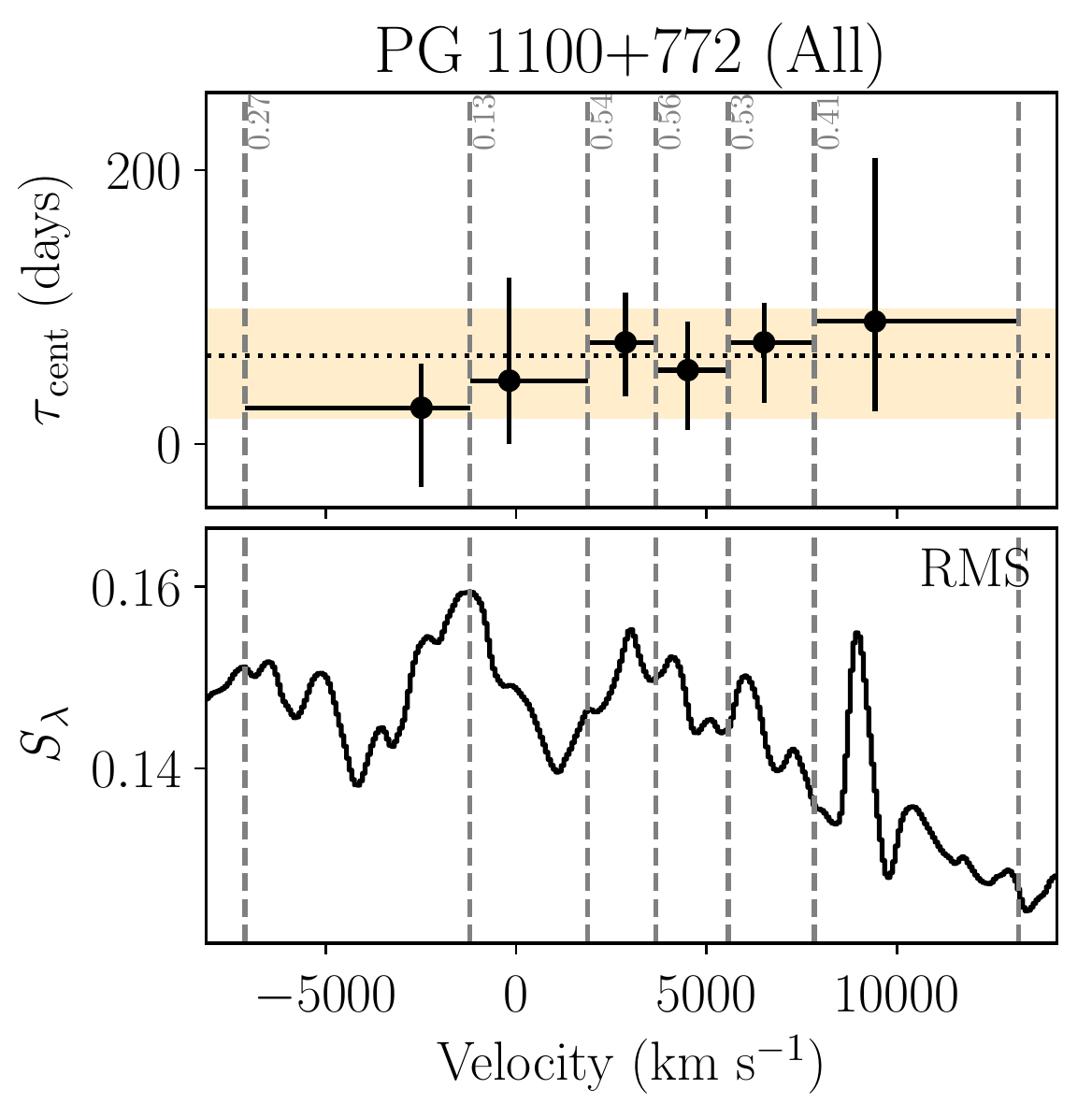}%
    \hspace{0.02in}
    \includegraphics[width=0.3\textwidth, height=0.21\textheight]{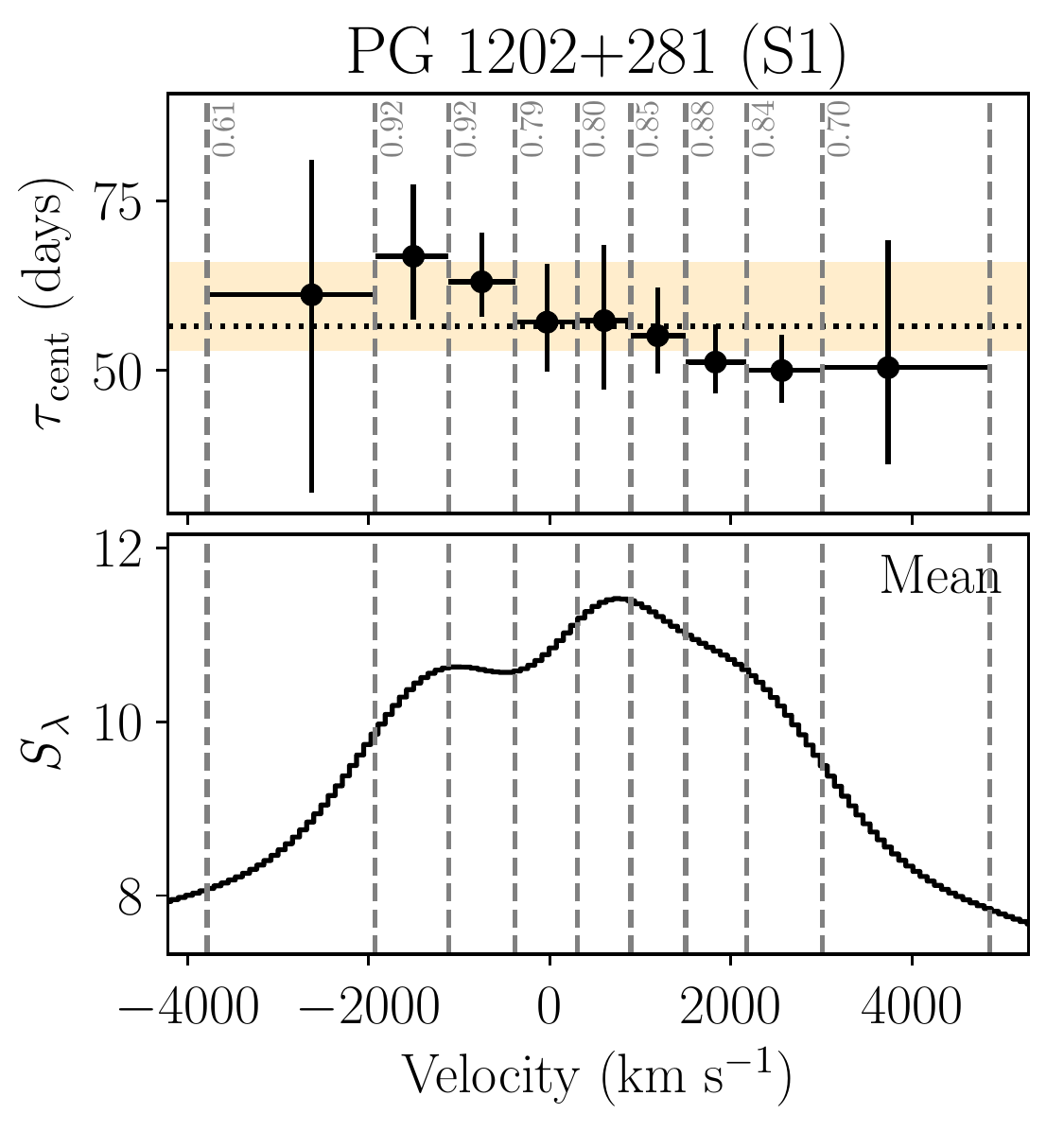}%
    \hspace{0.02in}
    \includegraphics[width=0.3\textwidth, height=0.21\textheight]{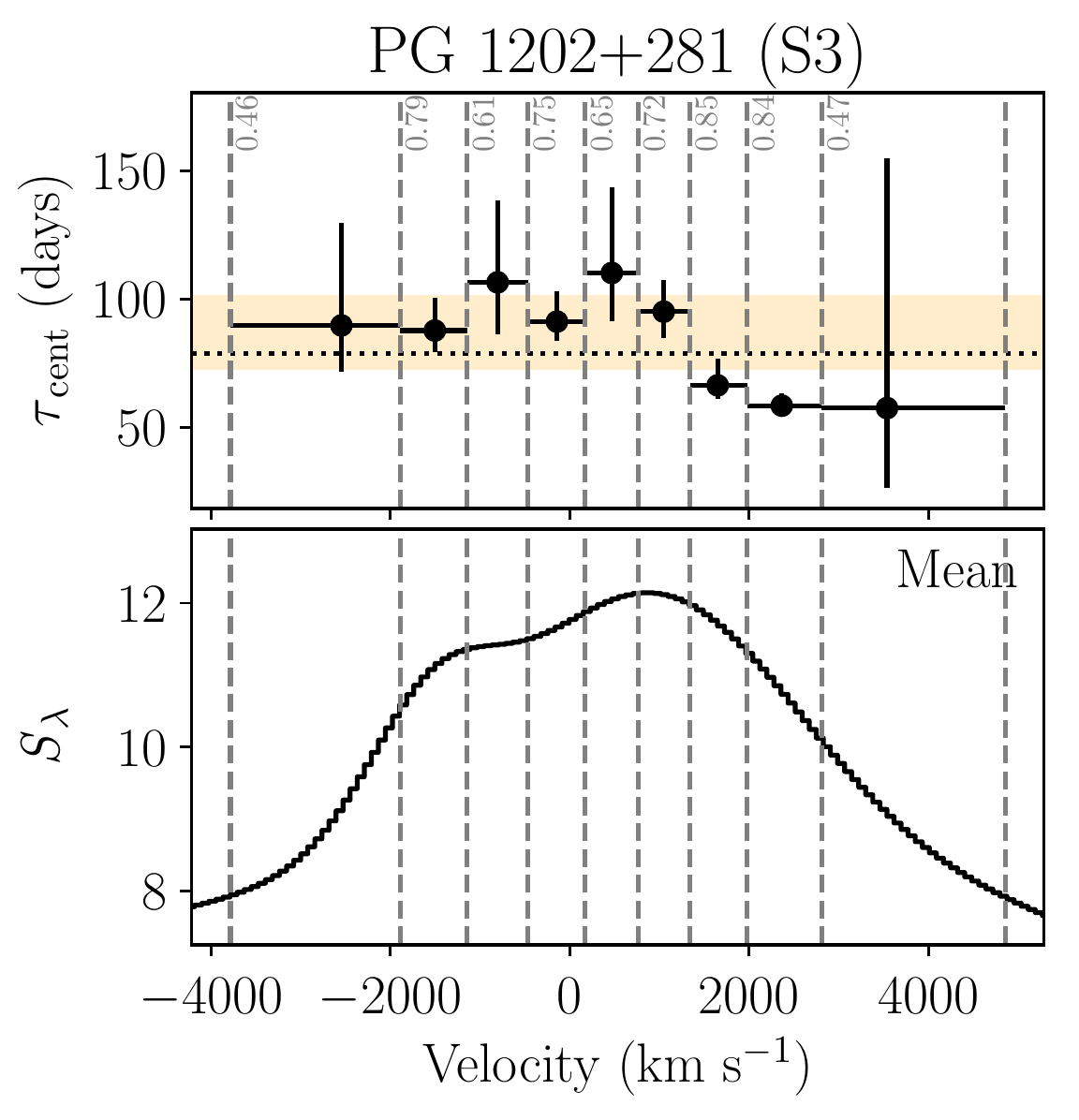}%
    \hspace{0.02in}
    \includegraphics[width=0.3\textwidth, height=0.21\textheight]{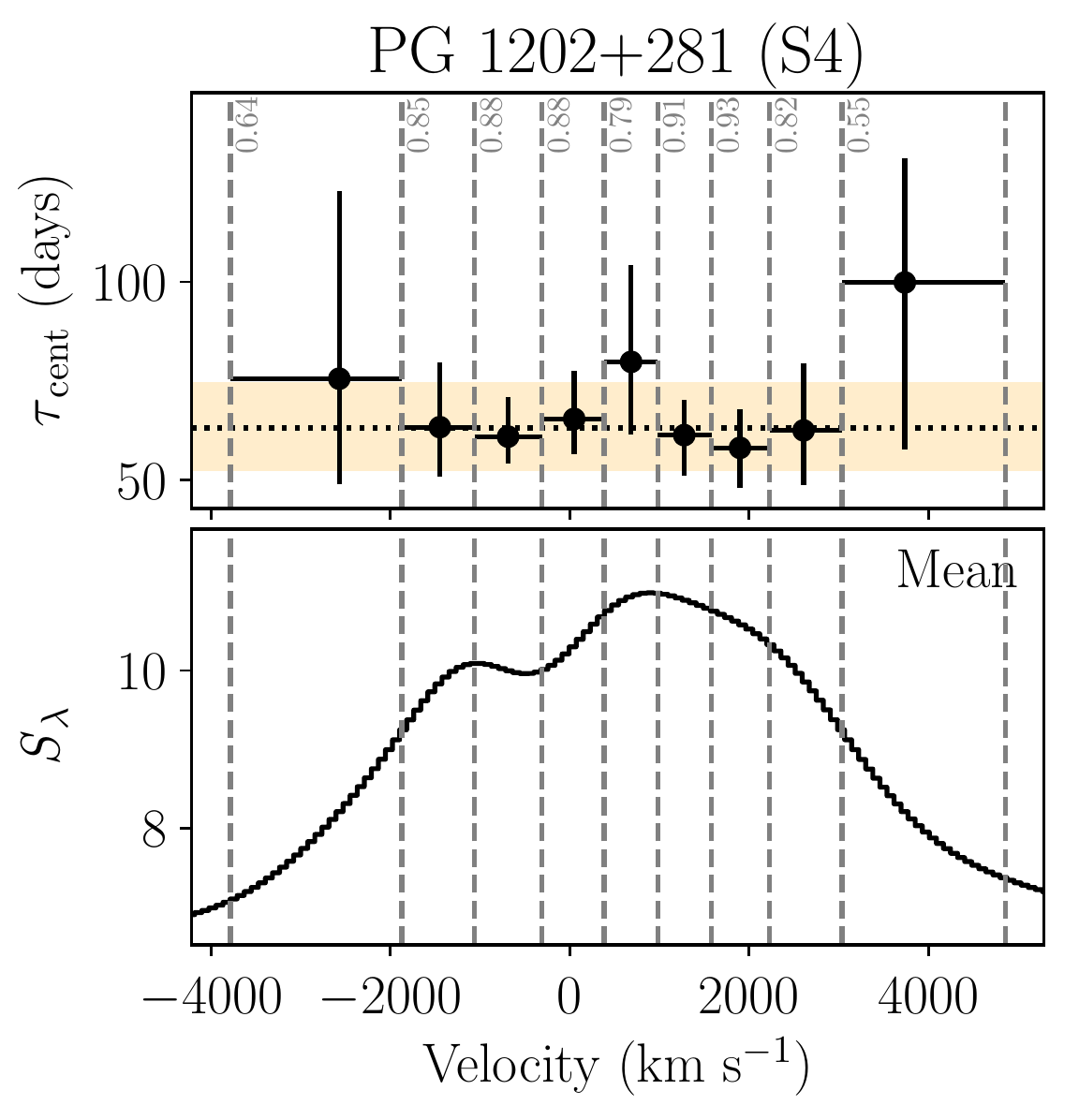}%
    \hspace{0.02in}
    \includegraphics[width=0.3\textwidth, height=0.21\textheight]{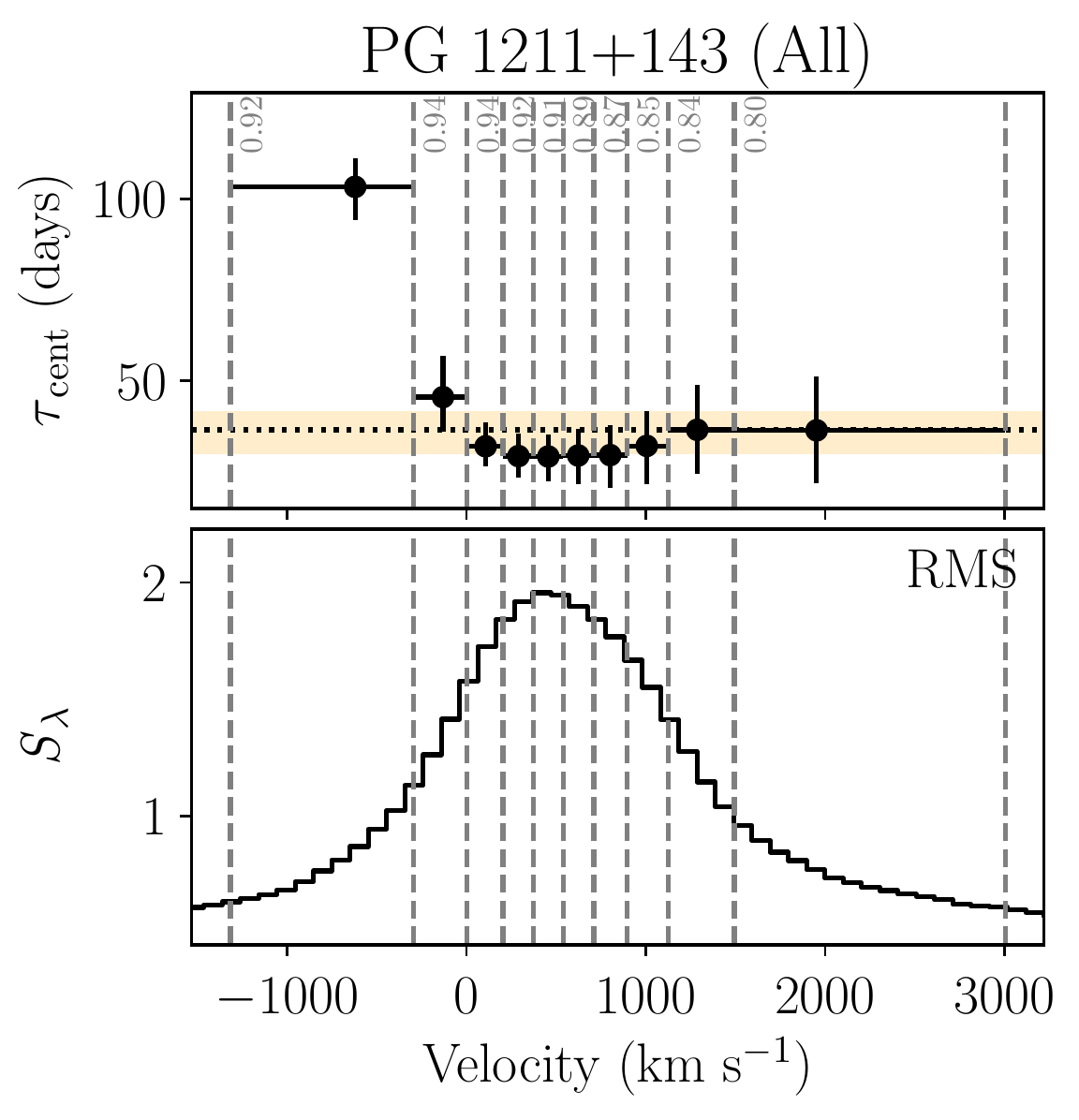}
    \hspace{0.02in}
    \includegraphics[width=0.3\textwidth, height=0.21\textheight]{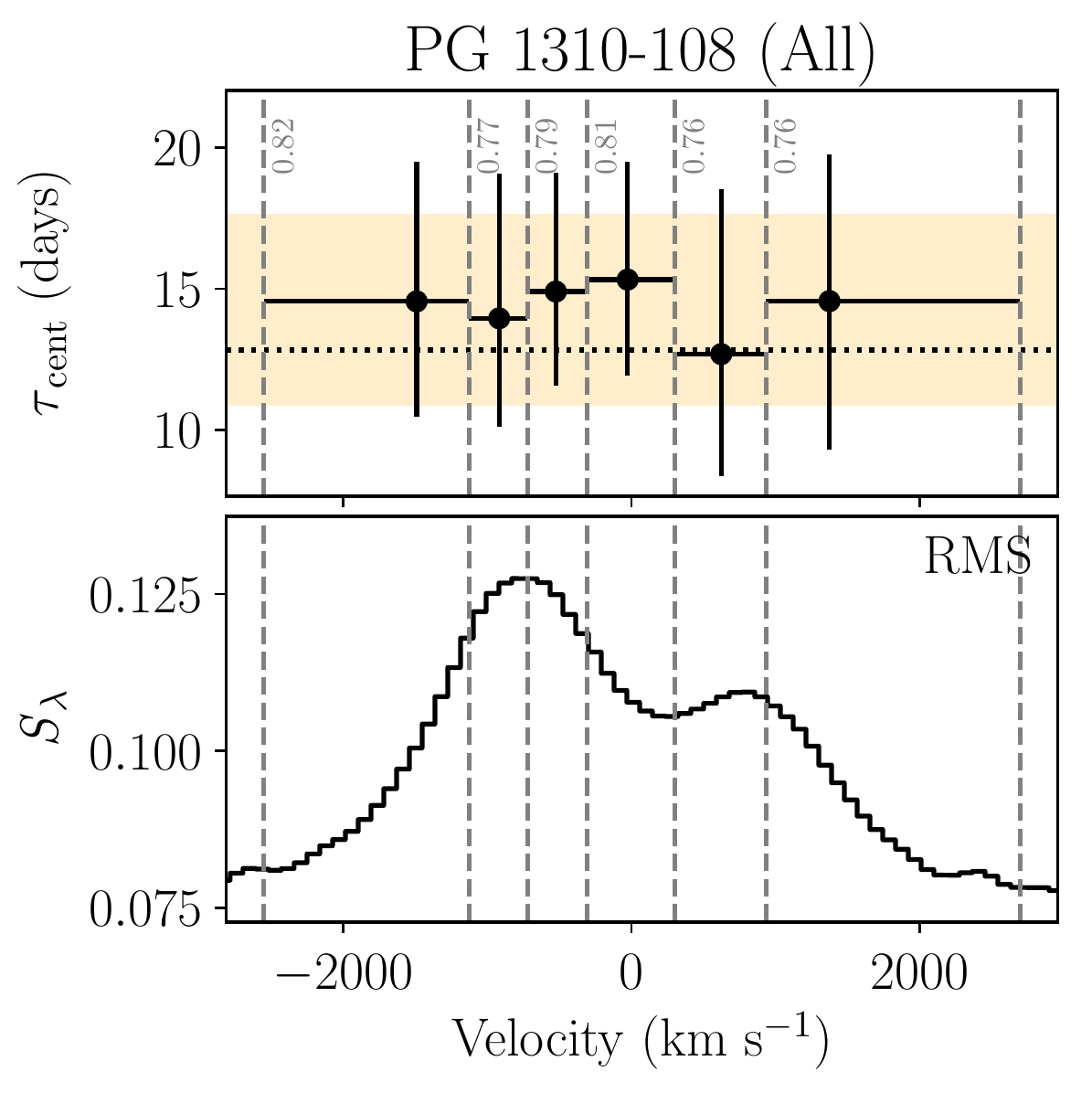}%
    \quad \\
    \includegraphics[width=0.3\textwidth, height=0.21\textheight]{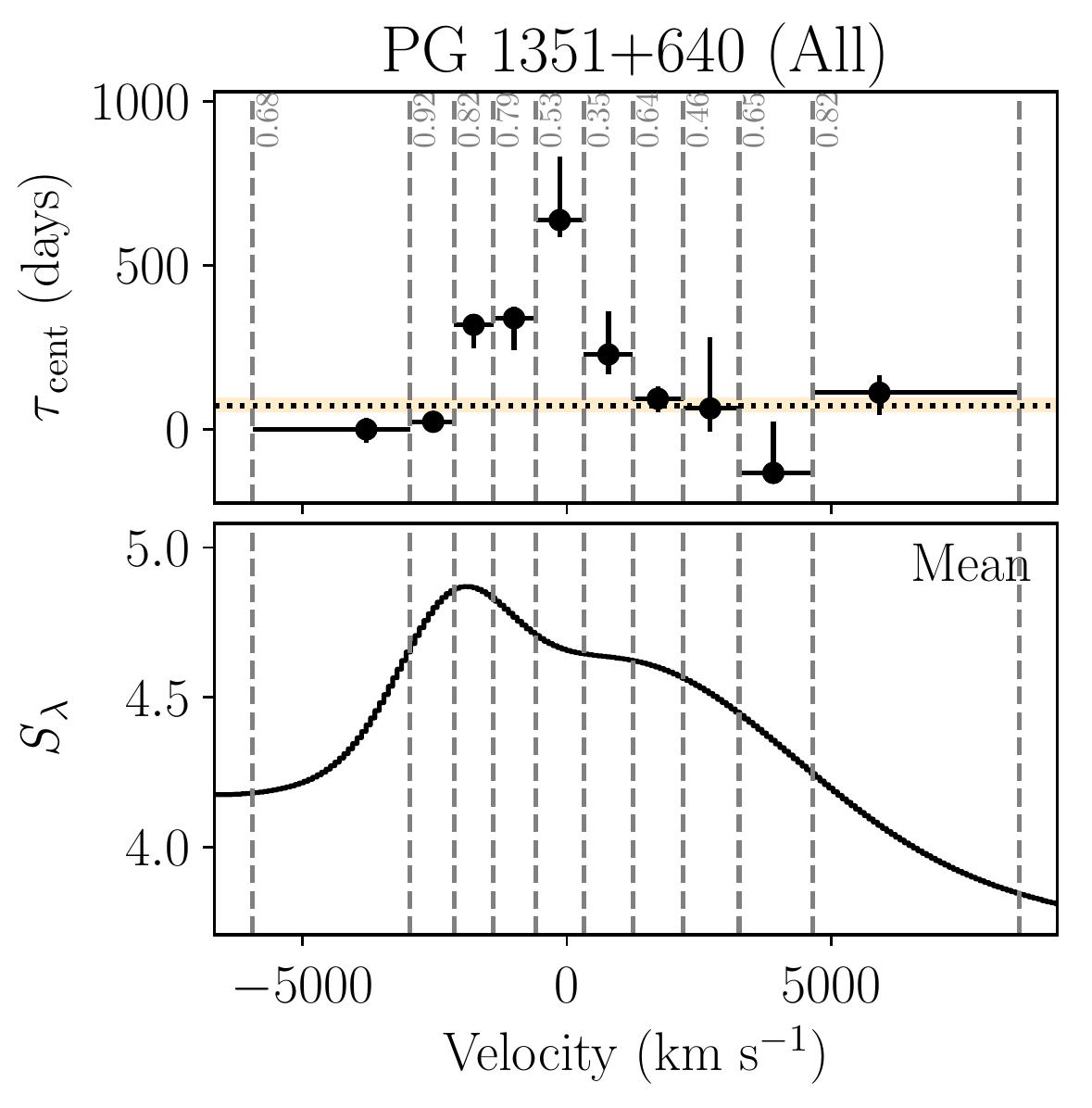}%
    \hspace{0.02in}
    \includegraphics[width=0.3\textwidth, height=0.21\textheight]{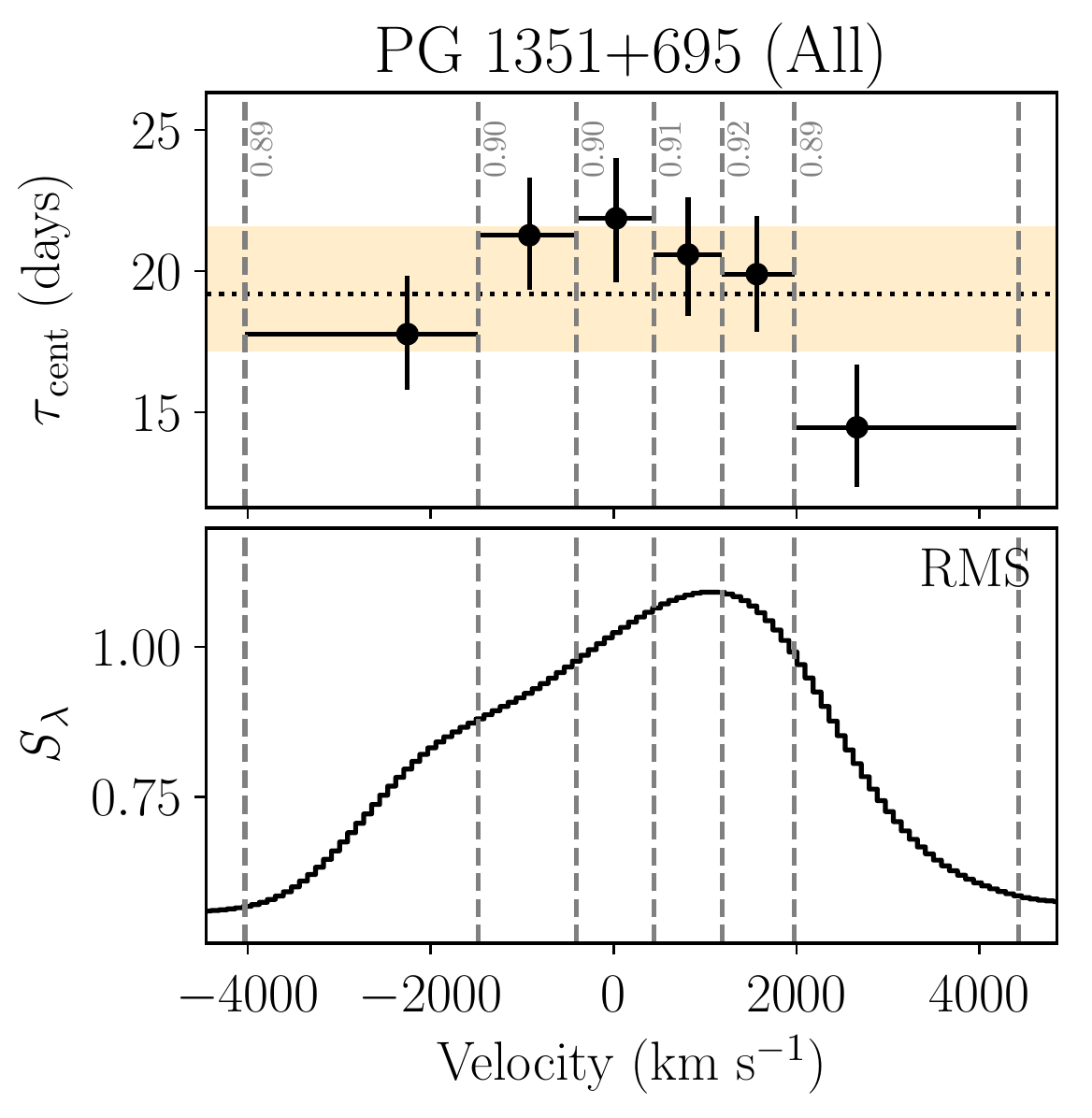}%
    \hspace{0.02in}
    \includegraphics[width=0.3\textwidth, height=0.21\textheight]{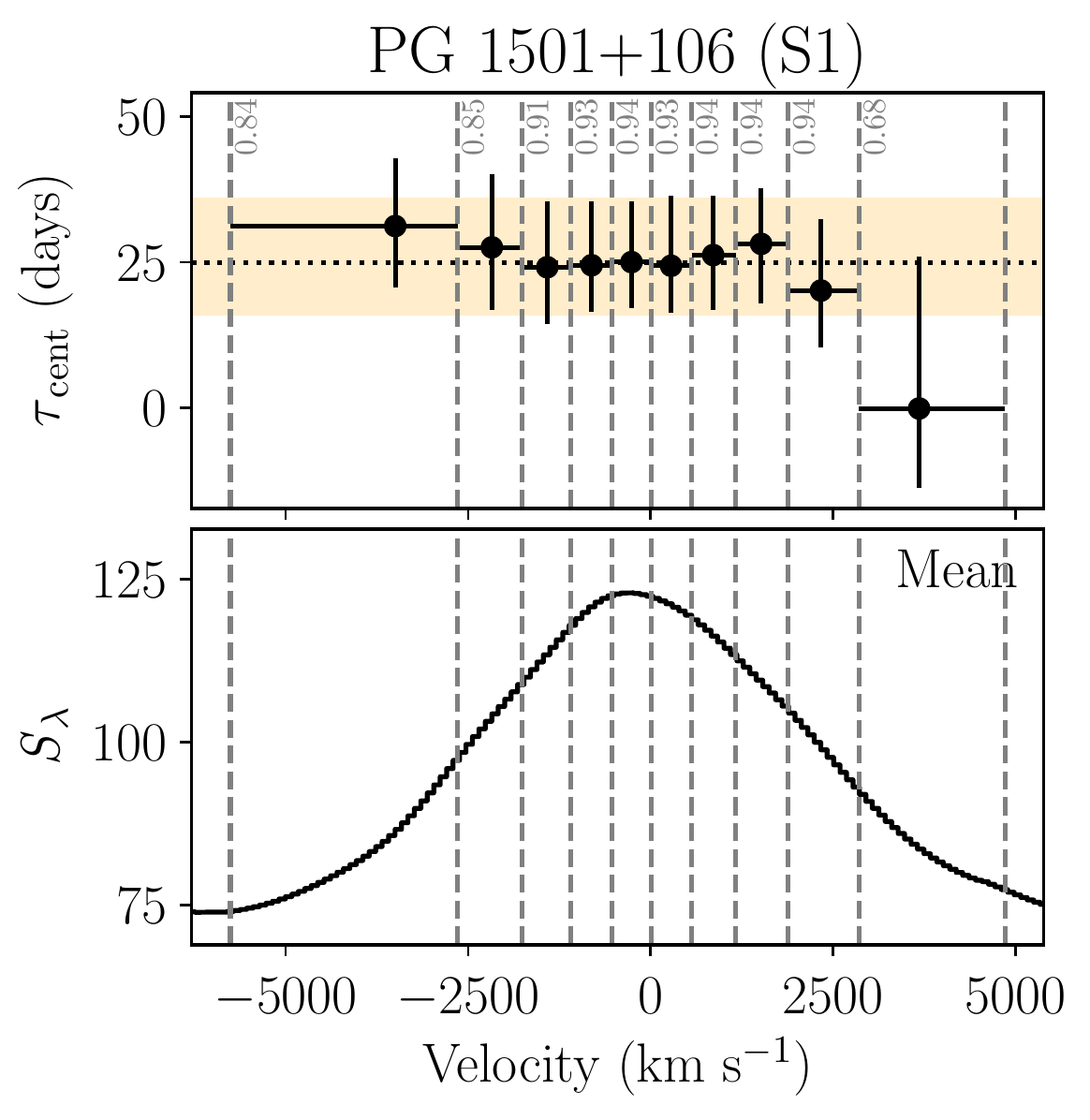}
    \caption{Continued.}
\end{figure*}

\begin{figure*}
    \flushleft
    \figurenum{\ref{fig:velocity resolved}}
    
    \includegraphics[width=0.3\textwidth, height=0.21\textheight]{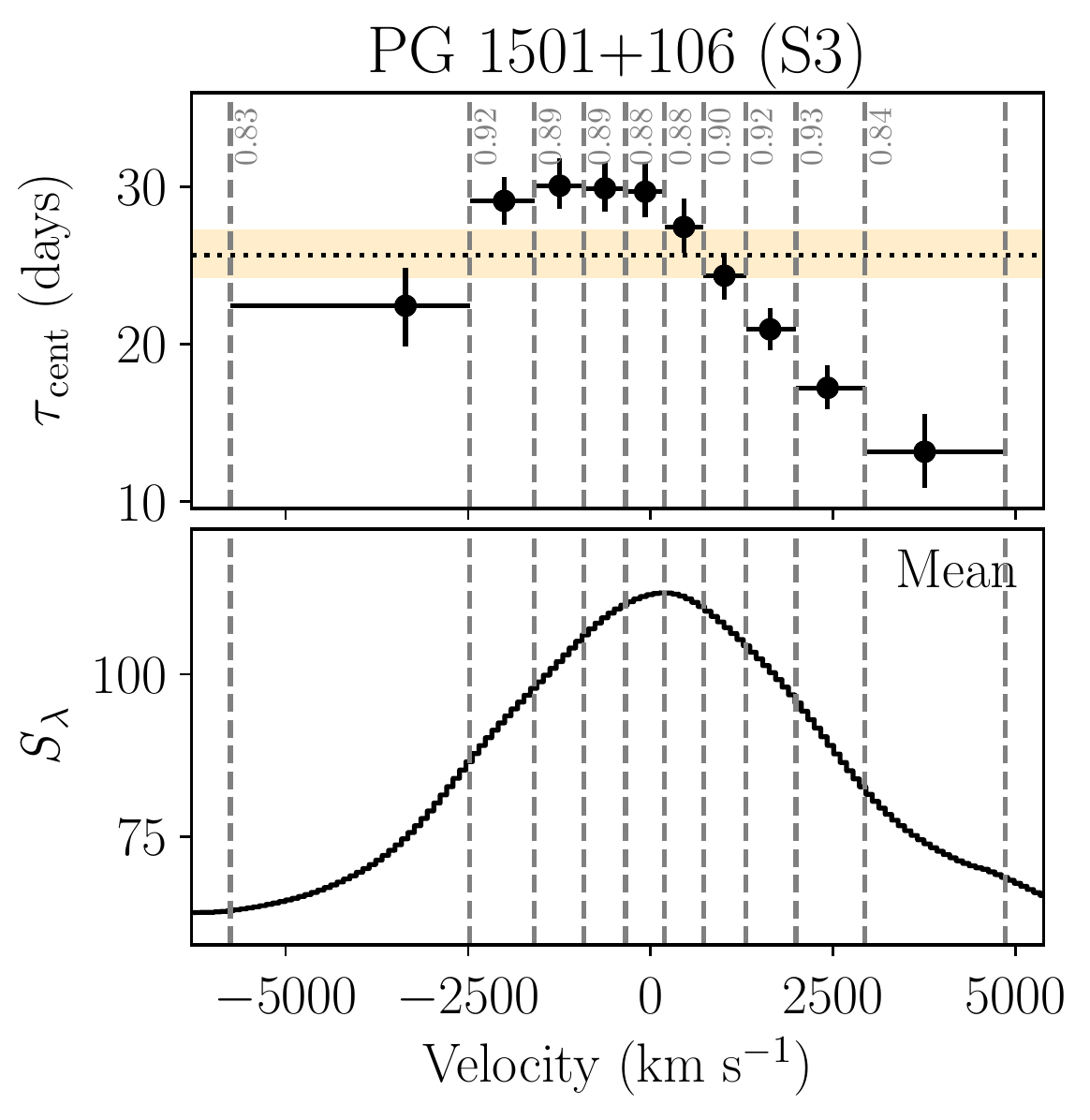}%
    \hspace{0.02in}
    \includegraphics[width=0.3\textwidth, height=0.21\textheight]{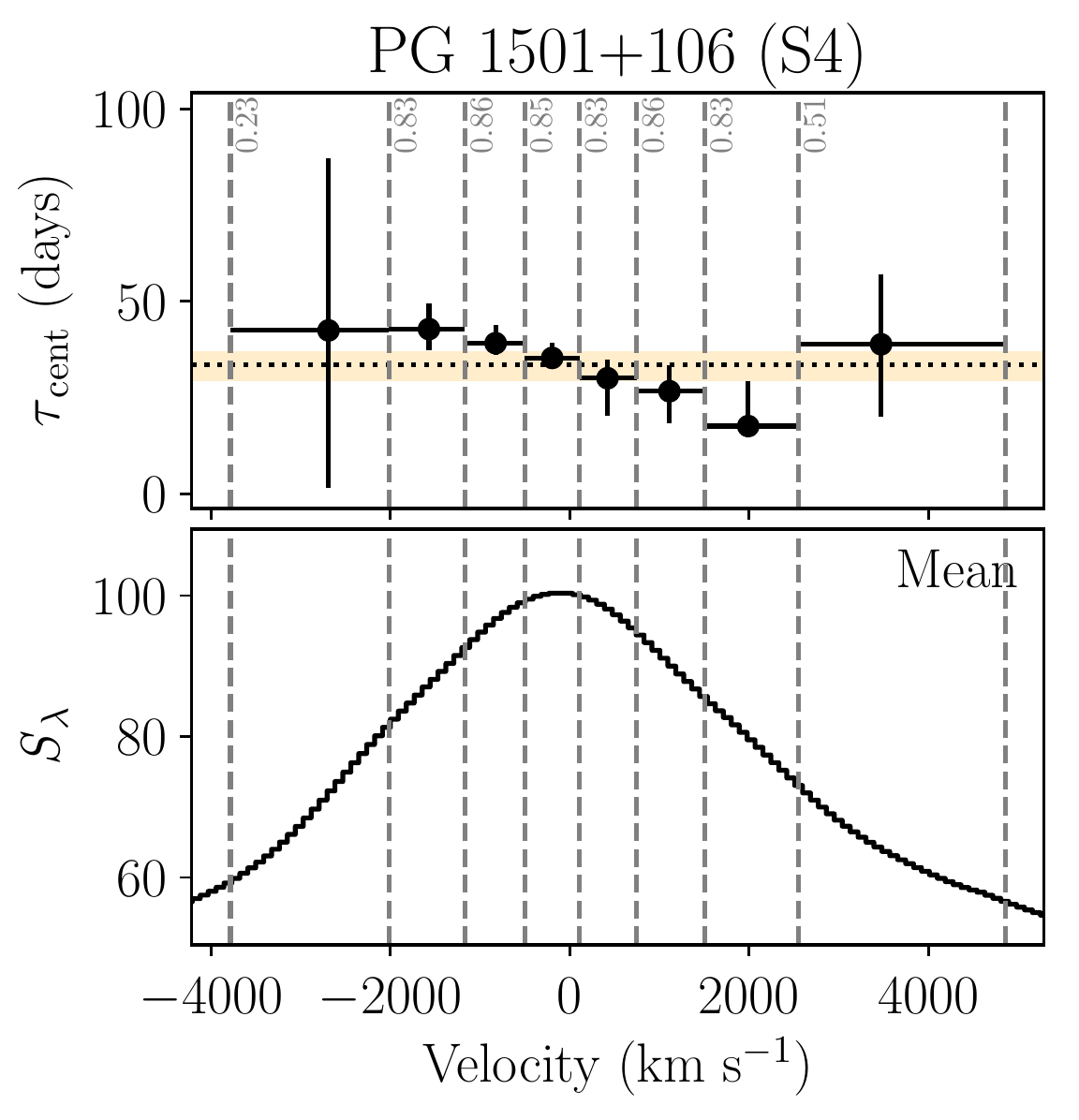}
    \hspace{0.02in}
    \includegraphics[width=0.3\textwidth, height=0.21\textheight]{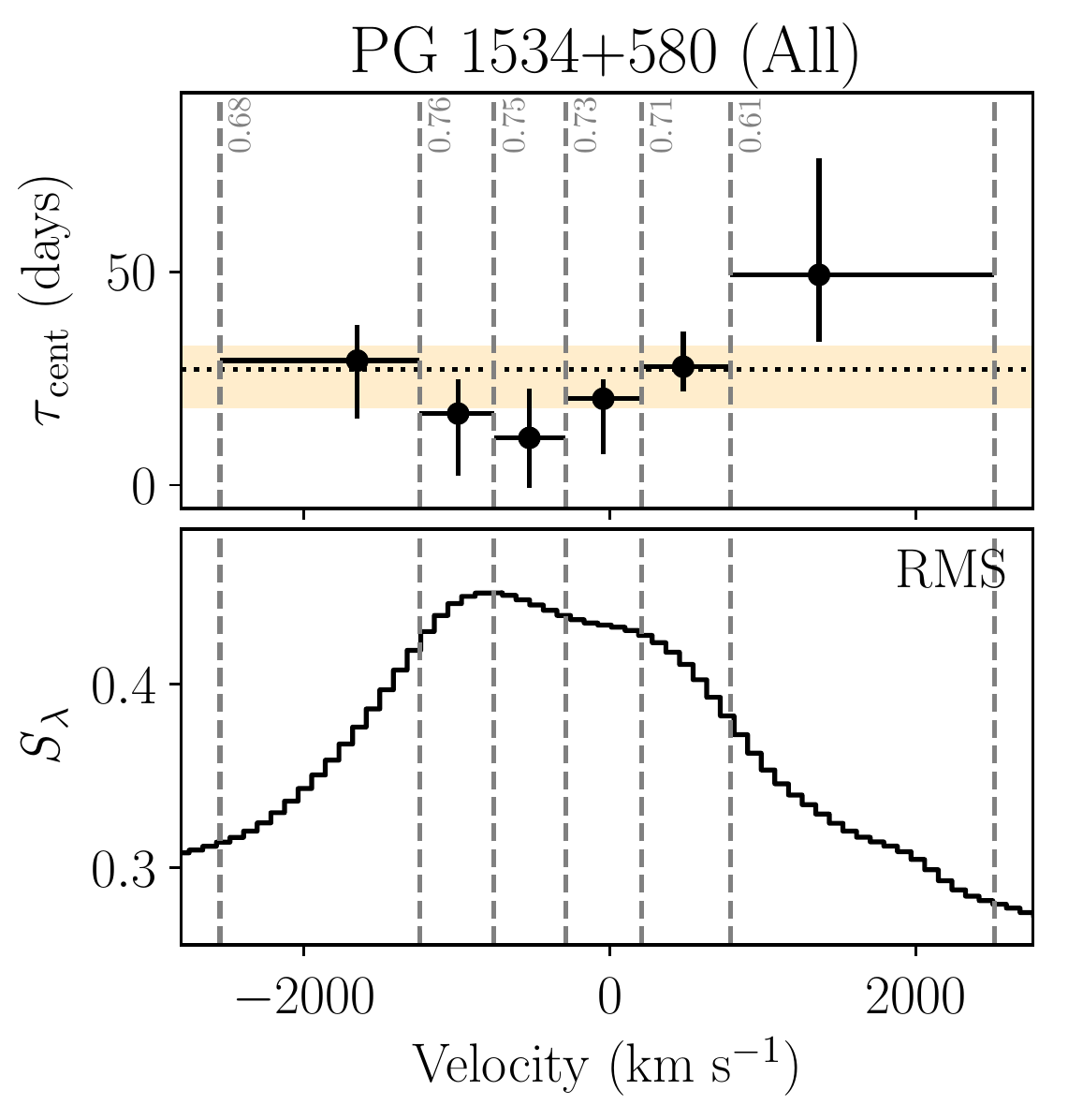}%
    \quad \\
    \includegraphics[width=0.3\textwidth, height=0.21\textheight]{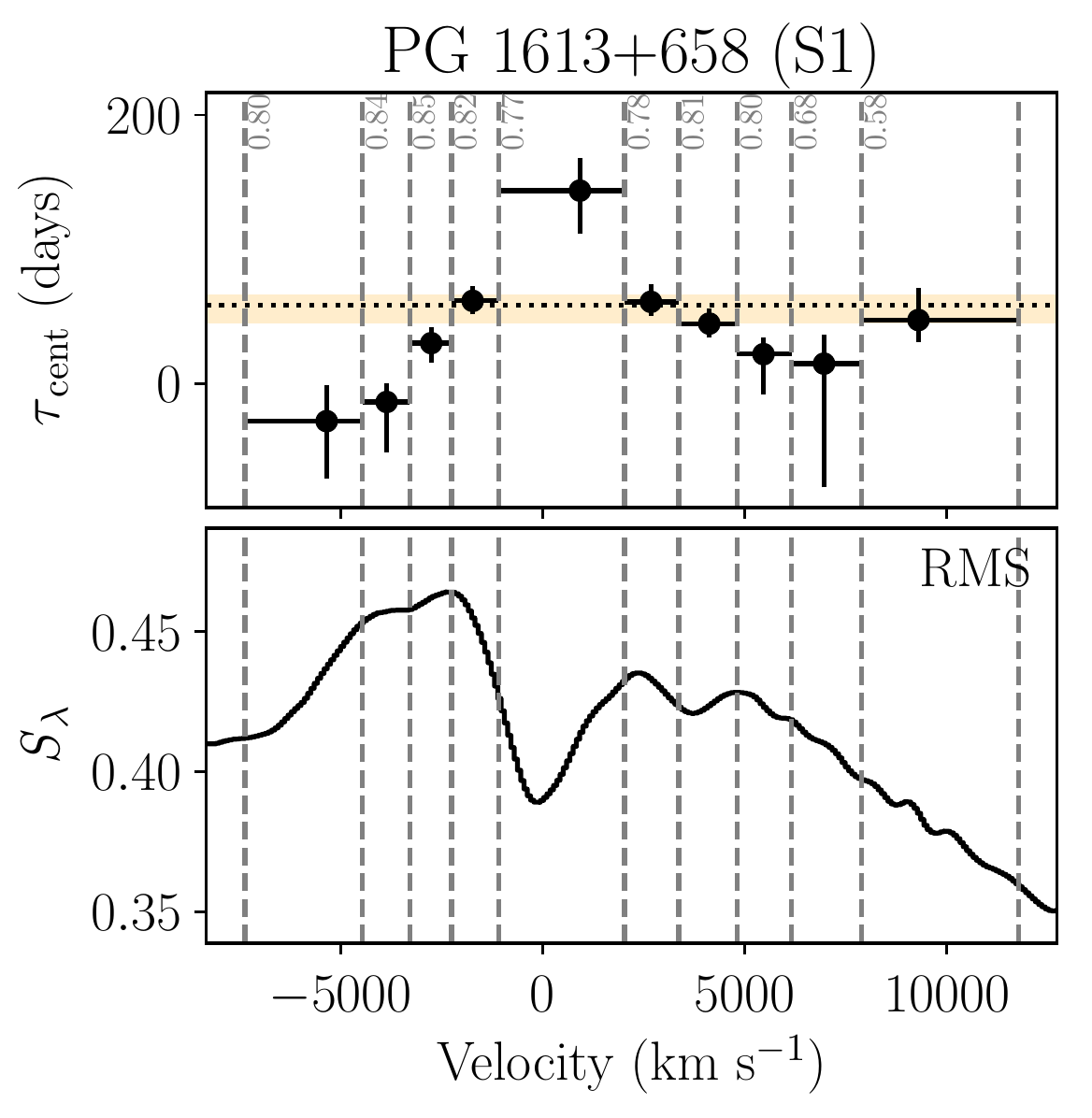}%
    \hspace{0.02in}
    \includegraphics[width=0.3\textwidth, height=0.21\textheight]{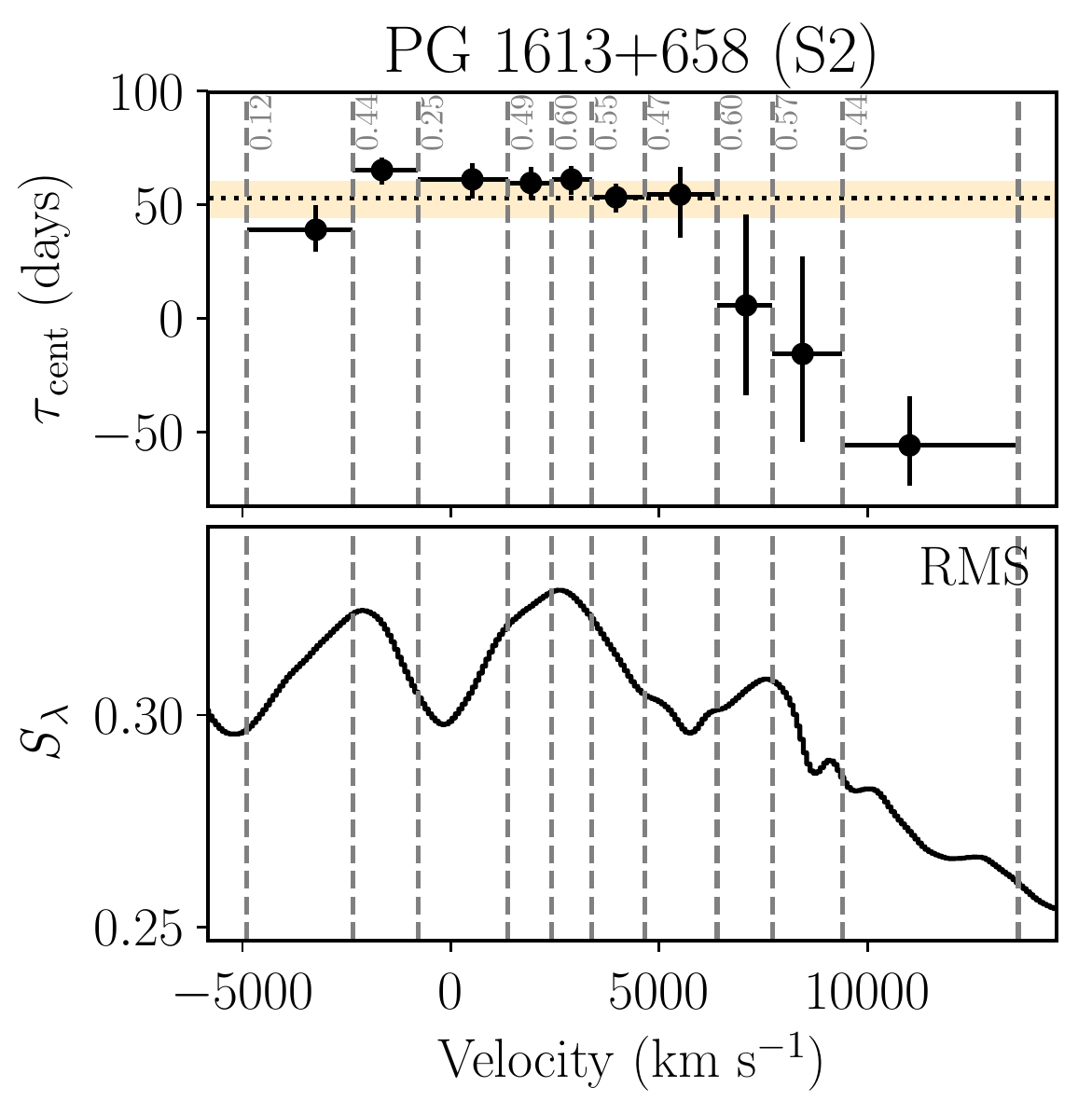}
    \caption{Continued. }
\end{figure*}

\section{DISCUSSION} 
\label{sec:discussion}

One of the primary goals of the MAHA project is to investigate the BLR geometry
and kinematics as well as their potential evolution in AGN BLRs with asymmetric
H$\beta$ lines, which requires both long-term monitoring and high cadence.  More
than half of the present sample (9 objects) have been monitored for 4 -- 5 years
with cadences of $\sim$ 3 -- 8 days. Among them, 6 objects (see Table
\ref{tab:BHmasses}) show clear variations in more than one season and can be
used to investigate the potential evolution in their BLR responses. Here we
discuss the measurements of the individual objects and compare them with the
previous results in the literature.

\subsection{Notes for Individual Objects}

{\it PG~0007+106 (Mrk~1501, III Zw 2):} It is a radio-loud AGN and showed a 5.1
yr quasi-periodicity in its radio light curve which can perhaps be explained by
the helical motion of a jet \citep{Terasranta2005, Li2010}. In our RM campaign,
H$\beta$ time lags of $\sim$ 14 -- 25 days (MICA) were measured for different
seasons. From the 4 seasons and the combined light curve, BH mass measurements
of 4 -- 7$\times10^7M_{\odot}$ were obtained using the line dispersion for the
velocity measurement ($7.03_{-0.31}^{+0.34}\times10^7M_{\odot}$ is preferred).
The previous RM campaign of this object \citep{Grier2012} gave a time lag of
$15.5_{-1.8}^{+2.2}$ days and a BH mass of $\log(\bhm/M_{\odot}) =
7.9_{-0.2}^{+0.2}$ was measured from BLR dynamical modeling
\citep{grier2017mcmc}. Our BH mass is in excellent agreement with that from
\cite{grier2017mcmc}. In all of the four seasons, the velocity-resolved lags in
Figure \ref{fig:velocity resolved} show longer lags at small velocities and
shorter lags at high velocities, which indicates that its BLR is dominated by
virialized motions or a Keplerian disk (the data quality in Seasons 1 and 4 is
relatively poorer). In Seasons 2 and 3, the velocities corresponding to the blue
wing of the line have slightly longer lags than the red wing, implying a
potential contribution of inflowing velocity besides the Keplerian/virialized
motion in its BLR. Similarly, the BLR modeling in \cite{grier2017mcmc} suggested
that its BLR kinematics is a combination of near-circular elliptical and
inflowing orbits. However, the velocity-resolved lags in \cite{Grier2013} showed
a stronger inflowing signature. In our campaign, the H$\beta$ profile is almost
the same as that in \cite{Grier2012} and also similar to those seen in much
earlier single-epoch spectra \citep{DeRobertis1985, Boroson1992,
Marziani2003atlas}. Note that the excess red emission in the H$\beta$ profile
for this object is not well characterized by the De Robertis asymmetry parameter
$A$. The $A$ parameter measures the blue or red extension of line wing. The red
wing of the H$\beta$ in PG~0007+106 does not extend too much with respect to its
line core. Instead the flux excess can be quantified by a systematic velocity
shift (e.g., ``H$\beta$ shift'' in Table 2 of \citealt{Boroson1992}). The
relation between the BLR kinematics and emission-line profiles (including the
velocity shift) will be discussed in details in a forthcoming paper.

{\it PG~0049+171 (Mrk~1148):} The profile of its broad H$\beta$ is slightly red
asymmetric and showed no significant change during our campaign. Its profile
remains  similar to that in \cite{DeRobertis1985} and \cite{Boroson1992}. The
H$\beta$ response in Season 4 is the best among the seasons. It gives a BH mass
of $2.95_{-0.31}^{+0.37}\times10^7M_{\odot}$. Considering that the line signal
in the rms spectrum of Season 4 is not very significant (it is better in the
continuum-cleaned rms spectrum), we prefer to use the mean spectrum to determine
the velocity bins in the velocity-resolved analysis. In the velocity-resolved
analysis, the plateau of the H$\beta$ light curve at the end of Season 1 is too
short to give a good constraint in some of the velocity bins, thus the lags at
different velocities in Season 1 are not very well resolved. The
velocity-resolved lags for the other seasons (Seasons 2, 3, 4) are almost the
same, and show longer lags in low-velocity bins and shorter lags at high
velocities. Similar to PG~0007+106, which is the signature of a Keplerian disk
or virialized motion. Moreover, compared to Season 3, the velocity-resolved lags
in Seasons 2 and 4 look more symmetric. The lags for the blue wing for Seasons 3
are shorter than those for its red wing, which indicates a potential
contribution from outflow in this season. The differences between Seasons 2, 3,
and 4 may imply that the response region of the BLR in PG~0049+171 is undergoing
some minor changes.

{\it PG~0923+129 (Mrk~705, Ark~202):} We have only data of one season for this
object, and it varied strongly only toward the end of our campaign. An H$\beta$
time lag is reported here for the first time. The time lag measured from MICA is
$6.2_{-1.8}^{+3.2}$ days and the corrsponding preferred BH mass is
$0.81_{-0.23}^{+0.42}\times10^7M_{\odot}$. The broad H$\beta$ profile is
slightly red asymmetric and neither the Fe II or [O III] lines are particularly
strong.  Its velocity-resolved lags are clearly longer at blue velocities and
shorter at red velocities, which is the signature of inflow (see Figure
\ref{fig:velocity resolved}). 

{\it PG~0947+396:} Its H$\beta$ profile shows a red asymmetry and has no obvious
changes compared with previous spectra published by \cite{Boroson1992} and
\cite{Shang2007}. Time lags can be detected for each of its four seasons,
although the uncertainties for the second season are the smallest because of its
stronger continuum variation and clear H$\beta$ response (See Sec
\ref{sec:meanrms}). The MICA measurement from the entire light curve is
consistent with the single-season result from Season 2 ($39.5_{-1.7}^{+3.8}$
days vs. $41.8_{-1.1}^{+1.3}$ days). The lag in Season 2 yields a preferred BH
mass of $1.02_{-0.03}^{+0.04}\times10^8M_{\odot}$. Its velocity-resolved lags in
Seasons 1, 3, and 4 are generally symmetric with longer lags at small velocities
and shorter lags at high velocities, which is the signature of a Keplerian disk
or virialized motion (similar to PG~0007+106 and PG~0049+171). The lags at blue
velocities are a little longer than those at red velocities in Seasons 1 and 3,
while the opposite is the case in Season 4. While this effect is not very
pronounced, it may imply weak contributions from inflow and outflow,
respectively. It is a little strange that the lags at different velocities in
Season 2 are not fully resolved, although the uncertainties of the average lag
measurement is the smallest of the four seasons, which may result from its
relatively small variation amplitude of the light curves in this season. 

{\it PG~1001+054:}  The H$\beta$ profile shows significant blue asymmetry (see
Table \ref{tab:basic_info} and Figure \ref{1001lc}). It has stronger Fe {\sc ii}
emission lines (see Figure \ref{fig:mean_spec}) compared to the other objects,
which is consistent with the positive correlation between the asymmetry
parameter $A$ and the relative strength of Fe {\sc ii} reported by
\cite{Boroson1992}. We prefer the BH mass measured from the entire combined
light curve. The BH mass measured in our campaign is
$1.07_{-0.12}^{+0.12}\times10^8M_{\odot}$. Its velocity-resolved lag measurement
in Season 3 suggests an outflow signature (shorter lags at blue velocities and
longer at red).

{\it PG~1048+342:} The profile of its broad H$\beta$ shows a more significant
red asymmetry compared to the profiles in \cite{Boroson1992} and
\cite{Kaspi2000}, but it displayed no significant changes over the four seasons
in our campaign. \cite{Kaspi2000} did not manage to sample this object
sufficiently to successfully measure the time lag of its H$\beta$ line. The
clear variation (especially in Season 1 and in the entire light curve) enables
us to give a reliable measurement of its H$\beta$ time lag for the first time.
The time lag measured from the entire light curve ($36.8_{-3.4}^{+2.4}$ days)
using MICA has smaller uncertainties than that from Season 1
($28.0_{-4.8}^{+5.6}$ days), and is thus preferred for the BH mass
determination, which we calculate to be
$4.44_{-0.42}^{+0.31}\times10^7M_{\odot}$. The longer lags at blue velocities
and shorter lags at red velocities measured from Season 1 show the signature of
inflow.

{\it PG~1100+772 (3C~249.1):}  Although the variability of H$\beta$ ($F_{\rm
var}=1.6\%$) is much smaller than that of the 5100\AA\ continuum flux ($F_{\rm
var} = 8.9\%$, see Table \ref{tab:lc_info}), we can still measure a H$\beta$
time lag using MICA. The ICCF and $\chi^2$ methods cannot give reliable
measurements to the lags because of the small line variation amplitude. The
profile of its broad H$\beta$ shows clear red asymmetry. The time lag measured
from the entire light curve is $55.9_{-1.4}^{+3.0}$ days and the BH mass is
$7.81_{-0.47}^{+0.54}\times10^8M_{\odot}$. It is a radio-loud object
(Fanaroff-Riley II) with asymmetric radio lobes and has an extended
emission-line region. Its jets and its extended emission-line region were
suggested to originate from the merger of the host galaxy of a gas-poor quasar
and a large late-type galaxy \citep{Stockton1983, Gilbert2004, Fu2009}. Because
of the small variation of H$\beta$ flux, the profile of the rms spectrum is
poorly constrained. The lags at different velocities are only marginally
resolved. On average, the lags at blue velocities are shorter than those in the
red, which may indicate an outflow (not on a significant scale because of the
small variation amplitude).

{\it PG~1202+281 (GQ~Com):} The H$\beta$ time lag is reported here for the first
time. The profile of its broad H$\beta$ shows a red asymmetry. The peak of
H$\beta$ was blueshifted in previous spectra from \cite{boroson1985,
Boroson1992, Kaspi2000, Shang2007}, however such a blue shift does not seem so
evident during our campaign. The light curve in Seasons 1 and 3 (and the entire
light curve) show clear variations and can give reliable lag measurements.
However, the lag measured from the entire light curve has the smallest
uncertainties and is preferred for the BH mass determination. This yields a BH
mass of $9.80_{-0.46}^{+0.44}\times10^7M_{\odot}$. Similar to PG~0049+171, we
use the line profiles in the mean spectra to determine the velocity bins in the
velocity-resolved analysis because of the relatively poor quality of the rms
spectra. Its velocity-resolved lags generally show the signature of inflow (see
Seasons 1 and 3, the lags at blue velocities are longer than the red ones).

{\it PG~1211+143:} The X-ray and UV observations suggest that this object has
ultra fast outflows \citep{Pounds2003, Danehkar2017}. It is therefore
interesting to investigate the kinematics of its BLR through RM. As a
narrow-line Seyfert 1 galaxy, this object was monitored from 1991 to 1998 by
\cite{Kaspi2000} and showed a H$\beta$ time lag of $93.2_{-29.9}^{+19.7}$ days
\citep{Kaspi2005}. Because the variation of its light curve in the previous
campaign \citep{Kaspi2000} was slow and the cadence was also not very high, the
past result has relatively large uncertainties. Given the higher cadence in our
campaign ($\sim4$ days), the time lag becomes better defined and we find it to
be significantly shorter ($53.0_{-5.8}^{+5.1}$ days). The BH mass of
$\log(\bhm/M_{\odot}) = 7.87_{-0.19}^{+0.19}$ given in \cite{Peterson2004} is
larger than the value reported here ($2.14_{-0.24}^{+0.21}\times10^7M_{\odot}$
or $2.25_{-0.28}^{+0.25}\times10^7M_{\odot}$ from the FWHM or $\sigma_{\rm
line}$ of the rms spectrum). The longer lags at blue velocities and the shorter
at red (see Figure \ref{fig:velocity resolved}) suggest an inflowing BLR. This
is the first determination of the BLR kinematics in this object. 

{\it PG1310$-$108:} The H$\beta$ time lag is reported here for the first time.
This object historically showed an H$\beta$ profile with a strong and extended
red wing \citep{Boroson1992}.  
The H$\beta$ light curve shows clear response to the varying continuum with a
time lag of $12.8_{-1.7}^{+1.7}$ days. The BH mass measured from our campaign is
$1.33_{-0.22}^{+0.20}\times10^7M_{\odot}$. Its lags at different velocities are
not successfully resolved. 

{\it PG~1351+640:} \cite{Kaspi2000} monitored this object in 1991--1998 but did
not find a reliable H$\beta$ lag measurement because of the relatively low
cadence and large scatter of points in the light curve. Our data demonstrate
significant variations and clear responses. The ICCF and MICA results are
consistent with each other. The rms spectrum shows some residual signal around
the [O {\sc iii}] wavelengths, which may originate from the variations in the
contribution of the broad He {\sc i}$\lambda$4922,5016 lines
\citep[e.g.,][]{Jackson1989} or a broad component of [O {\sc iii}]
\citep[e.g.,][]{Zakamska2016}. The Fe {\sc ii} lines in this object are weak
(see Table \ref{tab:asym_kinematics}), so this residual signal is less likely
from Fe {\sc ii}$\lambda$4924,5018 lines. Because of the long-term variation
timescale, we did not separate the light curves into different seasons. The time
lag measured from the entire light curves in our campaign is
$74.8_{-2.3}^{+2.3}$ days and the BH mass is
$1.52_{-0.06}^{+0.07}\times10^8M_{\odot}$. Similar to PG~0049+171, we used the
line profile in the mean spectrum to determine the bins of velocity-resolved
analysis because of the relatively lower S/N ratio of the rms spectrum. The
inferred BLR kinematics is Keplerian/virialized motion.

{\it PG~1351+695 (Mrk~279):} Its H$\beta$ variation amplitude in the present
campaign is around $F_{\rm var}=27\%$, which is stronger than the continuum
variability ($F_{\rm var}=12\%$), probably because the continuum flux is diluted
by the contribution from its host galaxy. This object was first monitored from
Dec. 1987 to Jul. 1988 with 39 points by \cite{Maoz1990}, who reported a
H$\beta$ time lag of $12\pm3$ days. After that, it was monitored again from Jan.
1996 to Jul. 1996 by \cite{Santos-Lleo2001}, giving a lag of
$16.7_{-5.6}^{+5.3}$ days. More recently, \cite{Barth2015} reported a new RM
measurement for this object from Mar. 2011 to Jun. 2011, with a time lag
consistent with the previous measurements (see also \citealt{williams2018}). The
time lag in the present paper is $19.9_{-1.0}^{+1.0}$ days and the derived BH
mass is $4.35_{-0.23}^{+0.24}\times10^7M_{\odot}$. Its velocity-resolved lags
show a Keplerian disk or virialized motion of the BLR with probable
contributions from inflow (see Figure \ref{fig:velocity resolved}).

{\it PG~1501+106 (Mrk~841):}  We monitored this object for 4 years (from 2017 to
2020). The light curve of the first season was published in Paper
\citetalias{Brotherton2020}. In the present paper, we slightly adjusted the
window for measuring the H$\beta$ fluxes in order to make sure that the
variation signals in the rms spectra of all four seasons are covered. We used
PyCALI \citep{Li2014} to perform the inter-calibration of the spectroscopic and
photometric continuum light curves, which is different from the simple linear
regression method in Paper \citetalias{Brotherton2020}. This also makes the time
lag measured from Season 1 slightly different but within 1$\sigma$ uncertainties
with respect to the value provided in Paper \citetalias{Brotherton2020}. The
variation of Season 2 is too weak to give a good constraint to the time lag,
however Seasons 3 and 4 show clear and strong H$\beta$ responses. It should be
noted that the peak around JD $\sim$ 2458700 days in the H$\beta$ light curve in
Season 3 and the trough around JD $\sim$ 2459000 days in Season 4 are both
narrower than their corresponding features in the continuum light curve (see
Figure \ref{1501lc}). This phenomenon makes the transfer function calculated
through MICA for Season 3 and 4 have a second very broad component in addition
to the primary narrow peak (see Figure \ref{1501lc}). Although Season 3 has a
very broad component in the transfer function from MICA, we still prefer to use
the lag from this season in the BH mass measurement because its variability is
the strongest during the campaign. The three methods (ICCF, $\chi^2$, and MICA)
yield generally consistent time lags for Season 3. The preferred BH mass
measurement is $7.17_{-0.79}^{+1.66}\times10^7M_{\odot}$, which is slightly
larger than the measurement in Paper \citetalias{Brotherton2020}. \cite{U2021}
monitored this object one year before our campaign although their light curves
are of shorter duration. They obtained a BH mass measurement
($4.7_{-1.6}^{+2.6}\times10^7M_{\odot}$) slightly smaller than ours in the
present paper (but within $1\sigma$ uncertainties). The velocity-resolved lags
in Season 1 are not clear, as reported in Paper \citetalias{Brotherton2020}, but
both of the velocity-resolved lags in Seasons 3 and 4 show definite inflow
signatures. This is consistent with the BLR kinematics reported by
\cite{U2021}.

{\it PG~1534+580 (Mrk~290):} This object was monitored before by
\cite{denney2010} who reported a time lag of 8.72 days in the rest frame. We
measured a time lag of $25.4_{-1.4}^{+2.0}$ days, which is much longer than the
result reported in \cite{denney2010}. It is not unexpected because, with a
similar spectroscopic aperture, the fluxes in our campaign
($\sim3.9\times10^{-15}\ {\rm erg\ s^{-1}\ cm^{-2}\ \AA^{-1}}$) are much higher
than those ($\sim0.9\times10^{-15}\ {\rm erg\ s^{-1}\ cm^{-2}\ \AA^{-1}}$) in
\cite{denney2010}. The BH mass obtained from our campaign is
$2.89_{-0.19}^{+0.25}\times10^7M_{\odot}$ and is almost the same as that
determined by \cite{denney2010}. \cite{denney2010} did not resolve the lags at
different velocities. In our campaign, the data also do not allow us to give
high-quality velocity-resolved lag measurements. However, the general structure
of velocity-resolved lags implies complicated BLR geometry or kinematics. 

{\it PG~1613+658 (Mrk~876):} This object was monitored during 1991--1998
\citep{Kaspi2000} and a H$\beta$ time lag of $40.1_{-15.2}^{+15.0}$ was reported
\citep{Kaspi2005}. The trough of the H$\beta$ light curve is in the gap in
Season 1, which gives a poorer constraint to the time lag than from Season 2.
The H$\beta$ time lag measured from Season 2 (with better data quality) is
$48.3_{-3.8}^{+5.0}$ days, which is similar to the value in \citep{Kaspi2005}
but much better constrained. The profile of broad H$\beta$ shows a strong red
asymmetry and does not show significant changes compared to that of
\cite{Boroson1992}, \cite{Erkens1995}, and \cite{Kaspi2000}. However, the
H$\beta$ profile plotted by \cite{DeRobertis1985} shows a much stronger red wing
and a slightly blueshifted peak. The radius of the innermost part of its dusty
torus ($334.1_{-37.0}^{+42.4}$ days) was measured by infrared reverberation
mapping \citep{Minezaki2019} and is larger by a factor of $\sim 7$ compared with
the BLR size in the present paper. Similar to the average lag determination, the
gap in Season 1 makes the velocity-resolved lag measurement somewhat unreliable.
The velocity-resolved lags of Season 2 indicate that its BLR is dominated by
inflow.

\begin{deluxetable*}{llll}
    \tablecaption{H$\beta$ Asymmetry vs. BLR kinematics\label{tab:asym_kinematics}}
    \setlength{\tabcolsep}{10pt}
    \tablehead{
    \colhead{Target}  & \colhead{$R_{\rm Fe}$} &  \colhead{H$\beta$ Asymmetry}   & \colhead{BLR kinematics} 
    }
    \startdata
    PG~0007+106 & 0.48 & Symmetric & Keplerian/Virialized + weak inflow          \\
    PG~0049+171 & 0.13 & Red       & Keplerian/Virialized + weak outflow         \\
    PG~0923+129 & 0.53 & Red       & Inflow                                      \\
    PG~0947+396 & 0.33 & Red       & Keplerian/Virialized + weak inflow/outflow  \\
    PG~1001+054 & 0.89 & Blue      & Outflow                                     \\
    PG~1048+342 & 0.28 & Red       & Inflow                                      \\
    PG~1100+772 & 0.05 & Red       & Outflow                                     \\
    PG~1202+281 & 0.19 & Red       & Inflow                                      \\
    PG~1211+143 & 0.51 & Blue      & Inflow                                      \\
    PG~1310-108 & 0.23 & Red       & Unresolved                                  \\
    PG~1351+640 & 0.20 & Red       & Keplerian/Virialized                        \\
    PG~1351+695 & 0.47 & Symmetric & Keplerian/Virialized + inflow               \\
    PG~1501+106 & 0.26 & Red       & Inflow                                      \\
    PG~1534+580 & 0.21 & Red       & Complicated                                 \\
    PG~1613+658 & 0.57 & Red       & Inflow                                       
    \enddata
    \tablecomments{$R_{\rm Fe}$ is the flux ratio of Fe {\sc ii} and H$\beta$
    emission lines measured from our campaign (from an
    individual exposure with high S/N ratio).
    }
    \end{deluxetable*}

\subsection{H$\beta$ Asymmetry and BLR kinematics}

To investigate if there is any correlation between the H$\beta$ asymmetry and
BLR kinematics, we make a short summary in Table \ref{tab:asym_kinematics}.
Although the size of the present sample is limited, it is obvious that the
kinematics of Keplerian/virialized motion and inflow is more common than
outflow, especially in the objects with broader H$\beta$ (e.g., FWHM (H$\beta$)
$\gtrsim$ 4000 km/s, corresponding to Population B in \citealt{Marziani2003}),
similar to the cases reported in the literature \citep[e.g.,][]{Bentz2009,
denney2010, Grier2013, Du2016VI}. It appears that the asymmetry of the emission
line does not directly correlate with the BLR kinematics (e.g., red-asymmetric
lines can be associated with inflow, outflow, or Keplerian/virialized BLR
kinematics). This is consistent with the fact that the emission-line profile is
the integration of the clouds in BLR and has relatively stronger degeneracy than
the velocity-resolved lags for the BLR geometry and kinematics. The flux ratios
of Fe {\sc ii} (from 4434\AA\ to 4684\AA) and H$\beta$ lines ($R_{\rm Fe}$) are
also listed in Table \ref{tab:asym_kinematics}.

The parameter $A$ listed in Table \ref{tab:basic_info} is measured from an
individual exposure with high S/N ratio. We have checked that the variation of
$A$ is relatively weak (although not zero) for each object during the campaign.
The ``blue'' or ``red'' asymmetry did not change in our observations. It is the
same as expected because the varying part of the emission line is only a small
portion of the entire profile. This can be justified from the much weaker
emission lines in the rms spectra with respect to those in the mean spectra (see
Figures \ref{0007lc}-\ref{1613lc}). Therefore, it is enough to list the
parameter $A$ measured from one individual exposure for exhibiting the blue or
red asymmetry of the emission-line profiles for the present sample.

\section{SUMMARY} 
\label{sec:summary}

In this third paper of the series, we present the RM measurements of 15 PG
targets from the MAHA project. Our campaign has both long-term duration (spans
from 1 to 5 years for different objects) and high cadence. We successfully
measure reverberation time lags between the continuum and H$\beta$ light curves
for individual seasons using three different methods (ICCF, $\chi^2$, and MICA).
ICCF and MICA show more consistent results, while the $\chi^2$ method
demonstrates slightly larger scatter. The BH masses of PG~0049+171, PG~0923+129,
PG~0947+396, PG~1001+054, PG~1048+342, PG~1100+772, PG~1202+281, PG~1310$-$108,
PG~1351+640 are reported for the first time. The velocity-resolved lags of the
objects are also measured and show very diverse kinematics (virialized, inflow,
and outflow signatures). The results from the present sample suggest that the
BLR kinematics of Keplerian/virialized motion and inflow is more common than
outflow. Future BLR modeling will investigate their BLR geometry and kinematics
in more detail.

\section*{ACKNOWLEDGEMENTS}


We thank the anonymous referee for valuable comments that improved the
manuscript. We thank WIRO engineers James Weger, Conrad Vogel, and Andrew Hudson
for their indispensable and invaluable assistance. We also acknowledge the
precious support from the staff of the Lijiang 2.4 m telescope, CAHA 2.2m
telescope, Asiago 1.82m telescope, and SAAO 1.9 m telescope. Funding for the
Lijiang 2.4 m telescope has been provided by CAS and the People’s Government of
Yunnan Province. This work is partly based on observations collected at the
Centro Astron\'omico Hispanoen Andaluc\'ia (CAHA) at Calar Alto, operated
jointly by the Andalusian Universities and the Instituto de Astrof\'isica de
Andaluc\'ia (CSIC). This research is based in part on observations collected at
Copernico telescope (Asiago, Italy) of the INAF - Osservatorio Astronomico di
Padova. We thank the South African Astronomical Observatory for the allocation
of telescope time, and Francois van Wyk for obtaining some of the spectra. We
acknowledge the support by the National Science Foundation of China through
grants {NSFC-12022301, -11991051, -11991054, -11873048, -11833008}, by National
Key R\&D Program of China (grants 2016YFA0400701), by Grant No. QYZDJ-SSW-SLH007
from the Key Research Program of Frontier Sciences, Chinese Academy of Sciences
(CAS), by the Strategic Priority Research Program of CAS grant No.XDB23010400,
and by the International Partnership Program of CAS, Grant
No.113111KYSB20200014. LCH was supported by the National Science Foundation of
China (11721303, 11991052) and the National Key R\&D Program of China
(2016YFA0400702). MB enjoyed support from the Chinese Academy of Sciences
Presidents International Fellowship Initiative, grant No. 2018VMA0005. YRL
acknowledges financial support from the National Natural Science Foundation of
China through grant NSFC-11922304 and from the Youth Innovation Promotion
Association CAS. HC acknowledges financial support from grant NSFC-12122305. We
also acknowledge support from a University of Wyoming Science Initiative Faculty
Innovation Seed Grant. This work is supported by the National Science Foundation
under REU grant AST 1852289. T.E. Zastrocky acknowledges support from NSF grant
1005444I.

This work is also based on observations obtained with the Samuel Oschin 48-inch
Telescope at the Palomar Observatory as part of the Zwicky Transient Facility
project. ZTF is supported by the National Science Foundation under Grant No.
AST-1440341 and a collaboration including Caltech, IPAC, the Weizmann Institute
for Science, the Oskar Klein Center at Stockholm University, the University of
Maryland, the University of Washington, Deutsches Elektronen-Synchrotron and
Humboldt University, Los Alamos National Laboratories, the TANGO Consortium of
Taiwan, the University of Wisconsin at Milwaukee, and Lawrence Berkeley National
Laboratories. Operations are conducted by COO, IPAC, and UW. This work makes use
of the public data from ASAS-SN project. ASAS-SN is supported by the Gordon and
Betty Moore Foundation  through  grant  GBMF5490  to  the  Ohio  State
University, and NSF grants AST-1515927 and AST-1908570.Development of ASAS-SN
has been supported by NSF grant AST-0908816, the Mt. Cuba Astronomical
Foundation, the Center for Cosmology and AstroParticle Physics at the Ohio State
University,  the  Chinese  Academy  of  Sciences  South America Center for
Astronomy (CAS- SACA), the Villum Foundation, and George Skestos.

\software{DASpec (\url{ https://github.com/PuDu-Astro/DASpec}), PyCALI \citep{Li2014}, 
MICA \citep{Li2016}.}

\clearpage

\bibliography{ref}
\bibliographystyle{aasjournal}


\end{document}